\documentclass[a4paper,twocolumn,11pt,accepted=2019-12-03]{quantumarticle}
\pdfoutput=1
\usepackage{amsmath}
\usepackage{amsfonts}
\usepackage{amssymb}
\usepackage{graphicx}
\usepackage{sidecap}
\usepackage{braket}

\usepackage{blindtext}
\usepackage{subfigure}
\usepackage{lipsum}
\usepackage{fullpage}
\newcommand*\diff{\mathop{}\!\mathrm{d}}
\usepackage{appendix}
\usepackage{subfigure}
\usepackage{color}
\usepackage[dvipsnames]{xcolor} 
\usepackage{physics}

 \usepackage{amsthm}

\newtheorem*{conj}{Conjecture}

\newcommand{\ee}{\mathrm{e}}
\usepackage[numbers,sort&compress]{natbib}

\begin{document}

\title{Exact Markovian and non-Markovian time dynamics 
in waveguide QED:
 collective interactions, bound states in continuum, superradiance and subradiance}

\author{Fatih Dinc}
\orcid{0000-0003-0921-0162}
\email{fdinc@stanford.edu}
\affiliation{Perimeter Institute for Theoretical Physics, Waterloo, Ontario, N2L 2Y5, Canada}
\affiliation{Department of Applied Physics, Stanford University, Stanford, CA 94305, USA}
\author{\.{I}lke Ercan}
\orcid{0000-0003-1339-9703}
\affiliation{Electrical \& Electronics Engineering Department, Bo\u{g}azi\c{c}i University, Istanbul, 34342, Turkey}
\author{Agata M. Bra\'nczyk}
\orcid{0000-0001-9681-4671}
\affiliation{Perimeter Institute for Theoretical Physics, Waterloo, Ontario, N2L 2Y5, Canada}
\maketitle

\begin{abstract}
We develop a formalism for modelling \emph{exact} time dynamics in waveguide quantum electrodynamics (QED)  using the real-space approach. The formalism does not assume any specific configuration of emitters and allows the study of Markovian dynamics \emph{fully analytically} and non-Markovian dynamics semi-analytically with a simple numerical integration step. We use the formalism to study subradiance, superradiance and bound states in continuum. We  discuss new phenomena such as subdivision of collective decay rates into symmetric and anti-symmetric subsets and non-Markovian superradiance effects that can lead to collective decay stronger than Dicke superradiance. We also discuss possible  applications such as pulse-shaping and coherent absorption. We thus broaden the range of applicability of real-space approaches beyond steady-state photon transport.
\end{abstract}

\section{Introduction}
Careful control of interactions between single-photons and multiple quantum systems within waveguide arrays is an important ingredient in the development of quantum networks \cite{sipahigil2016integrated,kimble2008quantum,chang2007single}. In such structures the waveguides  would  operate  as  quantum channels that efficiently transport information in the form of quantum light between the quantum systems, while the quantum systems themselves would store and  process the quantum information.

Waveguides that, to a good approximation, confine light to one spatial dimension (1D) have two advantages over three-dimensional waveguides. The first is practical: scattering in 1D makes photon transport more efficient. The second is theoretical: 1D waveguides are much simpler to model. Understanding  time evolution of single-photon states in 1D waveguide quantum electrodynamics (QED) will therefore benefit  development of quantum networks and quantum information science and technologies more broadly \cite{roy2017colloquium}.

Various approaches have been explored to describe interactions between light confined in 1D waveguides and two-level systems  such as atoms, cavities, resonators, superconducting qubits etc. Some examples include the real-space approach \cite{zheng2010waveguide,shen2007strongly,bermel2006single,cheng2016single,shen2018exact}, diagrammatic approaches \cite{roulet2016solving}, computational methods for energy-non-conserving systems \cite{fischer2018scattering}, the input-output formalism  \cite{caneva2015quantum,rephaeli2013dissipation,rephaeli2013dissipation,fan2010input}, generalized master equations \cite{shi2015multiphoton,baragiola2012n}, the LSZ Reduction formula \cite{shi2009lehmann,shi2011two} and the more-recently-developed SLH Framework \cite{combes2018two,brod2016two}. Each method offers unique intuition and can be preferable over another depending on the problem of interest. 

Of particular interest in 1D waveguide systems is the identification of  excitation probabilities of quantum systems and photon scattering amplitudes. These have been computed using various methods \cite{song2018photon,das2018photon,song2017photon,ruostekoski2017arrays,liao2016dynamical,liao2015single,zhou2008controllable,mirza2016multiqubit,cheng2017waveguide}, but often semi-analytically (that is, through a combination of analytical and numerical methods). While semi-analytical approaches 
are quite reliable for calculating excitation probabilities of atoms and single-photon scattered pulse shapes \cite{liao2016photon,liao2015single,wang2011efficient}, a fully analytical treatment of quantum networks would deepen intuition and enhance  understanding by making clear the general relationship between the system's parameters and its  behaviour. We thus choose  the real-space approach, whose strength lies in its ability to yield exact solutions---as long as the light-matter interactions take the form of delta functions located at positions of quantum emitters---particularly for single- or few-photon scattering problems (the approach is less suited to multi-photon scattering with many photons \cite{roy2017colloquium}, for which numerical approaches such as  \cite{shi2015multiphoton,baragiola2012n} are more appropriate). 

In this paper, we make several fundamental and applied contributions. {We extend the applicability of the theory of single-excitation time dynamics of photon-mediated interactions---in both the Markovian and non-Markovian  \cite{fang2018non,zheng2013persistent,gonzalez2013non} regimes---by incorporating bound states in continuum (BIC) for a general system and performing complex analysis.} We describe the connection between the poles of the scattering parameters and collective decay rates, opening the way to analytic results in a variety of scenarios. We study time-delayed coherent quantum feedback for a system of three qubits, and introduce the notion of super-superradiance (SSR). We also identify a connection between collective decay modes and the symmetry/anti-symmetry of the system. On the applied side, we introduce a recursive method for finding the scattering parameters, making tractable the study of time-dynamics in  systems with an unprecedented number of qubits, as well as non-identical qubits. We  also consider applied aspects such as Fano minima (transparent frequencies), {a method for} pulse shaping and nearly-perfect coherent absorption.

In doing so, we make---what we believe to be---a strong case that 
the real-space approach should become the method-of-choice when dealing with single-excitation subspaces in 1D waveguide QED systems. 

\section{Outline}

In Section \ref{sec:formalism}, we  review previous work on the real-space formalism while clarifying some steps that were omitted in the literature, and construct the time evolution operator for the single-excitation subspace. We review the scattering problem in the real-space formalism \cite{fang2018non,tsoi2008quantum} and extend the analysis to multi-qubit systems coupled to a 1D waveguide. 
We also touch on some interesting many-body physics phenomena such as bound states in continuum (BIC) {\cite{facchi2016bound,calajo2019exciting,gonzalez2014generation,longhi2007bound,tanaka2007electron,tufarelli2013dynamics}}---{ which had only been studied for $N=2$ qubits \cite{gonzalez2013non} using the real space approach, and had not been considered in time evolution}---and the related dark and bright states. We show that while this scenario may be challenging to model exactly using numerical techniques, we can take advantage of the analytical nature of the real-space approach to study it. More broadly, we provide a fully analytical description for the time evolution of any \emph{regular} single-photon states and system observables, such as atomic excitation probabilities. {For the scope of this paper, we define a regular state $\ket{\psi}$, as a state such that $\braket{k}{\psi}$ has analytic expansion in the complex plane and has vanishing behavior as its energy $E_k$ tends to complex infinity. As an example, a superposition of singly excited states such as $\ket{\psi}= \Sigma_m \alpha_m \ket{e_m}$ is always regular, whereas a Gaussian scattering state is not.}

In Section \ref{sec:coldecrat}, we discuss collective decay behaviour such as  subradiance \cite{albrecht2019subradiant,asenjo2017exponential} and superradiance \cite{zhou2017single,goban2015superradiance}.
In the Markovian limit
the photon-mediated interactions between  qubits occur instantly. Consequently, the collective behavior of the multi-qubit network gains immense importance
through the formation of so-called ``collective decay rates" of the network. 
In the non-Markovian regime\footnote{ In the literature, one finds two types of non-Markovianity: one that already exists at the level of a single qubit due to coupling to the environment and another that becomes important at the level of collective behavior due to time-delayed feedback within the system. The single-qubit level non-Markovianity was considered in \cite{fang2018non,valente2016non} and we re-derive the findings of \cite{valente2016non} in Appendix \ref{sec:appendixa}. Throughout this paper, we study only the non-Markovianity introduced by the many-body behavior of the system, i.e. by the macroscopic separation of qubits and corresponding time-delayed coherent interactions.}, the qubits are separated far enough such that the interactions between the light and the qubits remain effectively isolated. 
This  leads to a slightly different definition of collective decay rates, as non-Markovian processes introduce new types of decay modes \cite{zheng2013persistent} and individual interactions between distant qubits can now be observed. 

In Sections \ref{sec:Markovian} and  \ref{sec:nonmarkgen}, we apply the formalism to study time-dependent dynamics of various observables. 
We first model spontaneous decay of an initially excited system, then explore pulse scattering.
In the Markovian limit (Section \ref{sec:Markovian}) we demonstrate how to apply various complex analysis tricks to obtain highly intuitive and simple results. 
In the non-Markovian regime (Section \ref{sec:nonmarkgen}), we show that  modeling time-evolution  boils down to computing a single numerical integral, which is much simpler than existing approaches such as the finite-difference time-domain (FDTD) method \cite{liao2015single,liao2016photon}{  or solving coupled delay differential equations \cite{sinha2019non}.}

In Section \ref{sec:beyond}, we sketch out various, more exotic,  scenarios in which we expect the real-space approach to shine. In subsection \ref{sec:symmetry}, we conjecture that the set of collective decay rates is divided into two subsets: symmetric and anti-symmetric.  In subsection \ref{sec:pulseshaping}, we  introduce {a method towards} pulse shaping by exploiting collective decay rates. Finally, in subsection \ref{sec:beast-mode-on},  we consider  time dynamics of a photon interacting with a $500$-qubit system,  and show that the system approaches near-perfect coherent excitation with increasing number of qubits,
demonstrating the full power of the real-space approach in terms of scale-ability.

We conclude in Section \ref{sec:sum} with a discussion of the results.

\section{System model} \label{sec:formalism}

In the real-space formalism, the light-matter interactions almost always take the form of delta functions located at positions of quantum emitters (although, at least one example of non-delta interactions exists \cite{UVcutoff}). Delta interactions simplify the model significantly---the light propagates as a free field inside the waveguide, its amplitude changing only at the positions of interaction. From a mathematical point of view, this results in scattering energy eigenstates that are plane waves outside the system. Consequently, the problem of scattering from quantum emitters becomes analogous to delta-function scattering taught in introductory quantum mechanics courses \cite{griffiths2010introduction}. This property of the real-space formalism makes it simple to obtain exact eigenstates for single and few-photon scattering \cite{liao2016photon}. 

{In this section,} we demonstrate the real-space approach for a linear chain of { identical} $N=3$ qubits inside a one-dimensional waveguide { (we generalize the formalism to multi-qubit systems in Appendix \ref{sec:appendixgeneral} and to non-identical qubits in Appendix \ref{sec:appendixnonidentical})}. We show how to identify the single-photon eigenstates (both scattering and bound) for the system  using the Bethe ansatz. We set up the formalism for modelling time-evolution using the eigenstates. We then show that, in the Markovian limit, the time evolution of any regular state can be described via a contour integration. {Finally, we show that the real-space approach can be preferable for modelling exact time evolution in the presence of BICs, since analytical methods only require knowing the scattering eigenstates (without BICs).}

\subsection{Hamiltonian and energy eigenstates} \label{sec:energyeigenstate}

Consider the interaction of a single qubit (such as a two-level atom or a quantum emitter) with light in 1D.  The real-space Hamiltonian \cite{shen2005coherent,tsoi2008quantum} that describes the qubit-light system is $H = H_{\textsc{q}}+H_{\textsc{f}}+H_{\textsc{i}}$, where 
\begin{align}
     H_{\textsc{q}}={}& \sum_{m\in\{\rm qubits\}} \sigma_m^\dag \sigma_m \Omega
     \end{align}
     is the free qubit Hamiltonian, where $\sigma_m$ is the de-excitation operator for the $m^{th}$ qubit and  $\Omega$ is its energy separation, 
     \begin{align}
   H_{\textsc{f}}={}& i \hbar v_g \int_{-\infty}^\infty \diff x\\\nonumber
           & \times\left( C_L^\dag(x)\frac{\partial}{\partial x} C_L(x) -C_R^\dag(x)\frac{\partial}{\partial x} C_R(x)  \right) 
           \end{align}
is the free field Hamiltonian,  where $v_g$ is the group velocity of photons inside the waveguide and $C_{R/L}(x)$ are annihilation operators for right/left moving photons,  and
           \begin{align}
           \begin{split}
         H_{\textsc{i}}={}&\sqrt{J_0} \sum_{m\in\{\rm qubits\}} \int_{-\infty}^\infty \diff x \delta(x-mL)\\
           &\times\left( (C_R^\dag(x)+C_L^\dag(x)) \sigma_m + \text{H.c.} \right)\,,
           \end{split}
\end{align}
is the qubit-field interaction Hamiltonian with coupling energy  $J_0$. {Here, $H_{\textsc{f}}$ is obtained assuming a linear dispersion relation for the photon such that $E_k=\hbar v_g |k|$ and that $E_k$ is much larger than the cut-off frequency of the waveguide \cite{shen2009theory}, and $H_{\textsc{i}}$ is in the rotating-wave approximation. Hence, our approach focuses on energy scales $\Omega \pm O(J_0)$ where $J_0 \ll \Omega$. With our definition of the coupling energy, the population decay rate of a single emitter is $\gamma_0=2J_0$.} For the remainder of the paper, we set $\hbar = v_g =1$ to simplify notation {and use $J_0$ as the main units of energy, distance and time; while measuring collective decay rates in units of $\gamma_0$.}

In the real-space formalism, the stationary states of the Hamiltonian can be found via the Bethe Ansatz approach. For a single qubit, scattering eigenstates are the only type of stationary states, whereas for multi-qubits, the stationary states consist of both scattering eigenstates and BIC (\cite{tsoi2008quantum} {considers a similar problem and} identifies the scattering eigenstates, but does not discuss the existence of BIC). 

\begin{figure}[h!]
    \centering
    \includegraphics[width=\columnwidth]{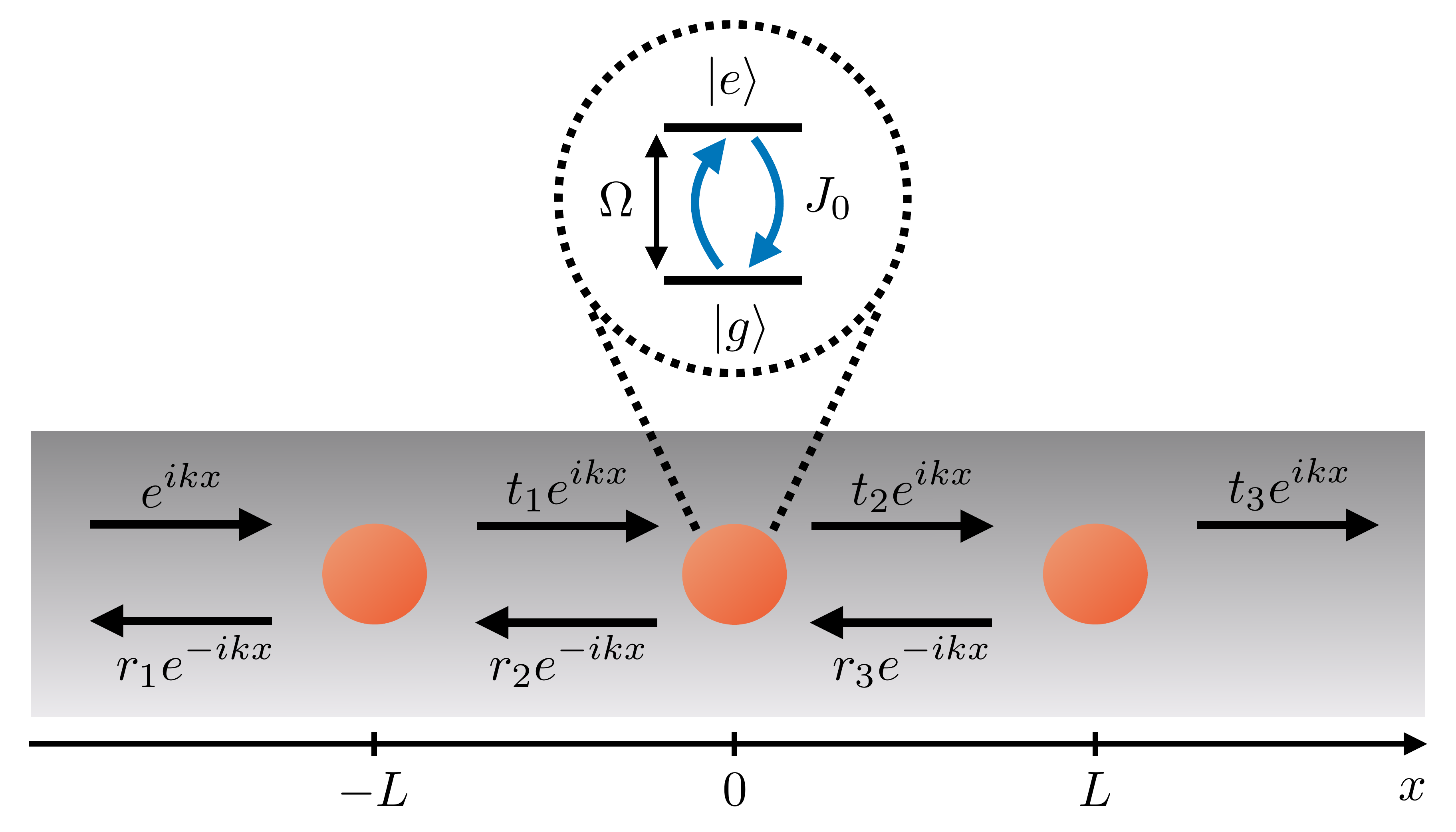}
    \caption{The system model for $N=3$ qubits in a linear chain with arrows depicting photon-scattering coefficients.}
    \label{fig:main}
\end{figure}

{We now review how to find the scattering eigenstates and find the condition of BIC for the linear chain of $N$ qubits.} An example scattering energy-eigenstate incident from far left is illustrated in Fig.~\ref{fig:main}.
Here, three identical equidistant qubits are $L$ apart from each other. {To preserve mirror symmetry in the system, the position of the center of the atom chain is taken arbitrarily as $x=0$.}

The Bethe Ansatz for a scattering eigenstate for a photon incident from far left can be written as
\begin{equation}\label{eq:ba}
\begin{split}
       \ket{E_{k}}=& \sum_{\Xi \in \{L,R\}}\int_{-\infty}^\infty \diff x \phi_\Xi(x) C_\Xi^\dag(x) \ket 0 \\ &+ \sum_{m\in \{\rm qubits\}} e_m(k) \ket{e_m},  
\end{split}
\end{equation}
where $\ket 0$ is the vacuum state, $\ket{e_m}=\sigma_m^\dag \ket 0$,  $e_m(k)$ is the excitation coefficient for the $m^{th}$ atom and $\phi_R(x)$/$\phi_L(x)$ are right/left moving field amplitudes, that contain the transmission and reflection coefficients, $t_m$ and $r_m$ respectively, as illustrated in Fig. \ref{fig:main}. As already discussed in \cite{tsoi2008quantum}, the parameters $t_m$, $r_m$ and $e_m(k)$ can be found by solving the eigenvalue equation $H \ket{E_k} = E_k \ket{E_k}$. 
As a next step, we define $\Delta_k = E_k - \Omega$ and {assume $J_0\ll \Omega$, consistent with the rotating-wave approximation.} This is particularly important for constructing the time evolution operator, as we assume that the eigenstates are continuum normalized. This approximation becomes exact in the limit $J_0/\Omega \to 0$. Nonetheless, the final result obtained using this approximation is more general and applies to the case of finite $J_0/\Omega$ as well. We leave the discussion to a future work. For practical purposes, the delta-interaction Hamiltonian assumption is only reasonable under the condition $J_0/\Omega \to 0$ anyway \cite{UVcutoff}.

The example scattering state illustrated in Fig. \ref{fig:main} does not constitute the complete basis for the scattering eigenstates as identified by \cite{tsoi2008quantum}. In fact, we denote $k<0$ values for scattering eigenstates, where the field is initially incident from right. For $\ket{E_{-k}}$, where the photon is initially incident from far left, one can simply mirror the state $\ket{E_k}$ w.r.t. the center of the multi-qubit system (which is $x=0$ for the $3$ qubit system in Fig. \ref{fig:main}). The operator corresponding to this operation is called the parity operator and is discussed in Appendix \ref{sec:appendixparity}. Using the parity operator, finding the scattering eigenstates incident from one side only is sufficient to form the complete scattering eigenbasis.

{Defining $\theta=\Omega L$ as the phase acquired by a plane wave of frequency $\Omega$ travelling between two adjacent qubits, when $\theta=n\pi$, the photon gains a pre-factor of $(-1)^n$ while travelling between qubits.} For this highly special condition, the scattering eigenstates do not constitute a complete basis for the single-photon subspace. To form a complete basis for this case, one must also include BICs  (which are equivalent to dark states in the absence of non-radiative decay). {BICs have previously been found and used for $N=2$ qubits coupled to a 1-D waveguide \cite{gonzalez2013non}. Here we identify BICs in a linear chain of $N$ atoms coupled to a 1D waveguide.} A Bethe ansatz for the BIC can be written as
\begin{equation} \label{eq:darkf}
    \ket{D} = \sum_{m\in \rm \{qubits\}} e_m \ket{e_m} +  \ket S,
\end{equation}
where $\braket{x}{S}\propto \sin(\Omega x)$ describes a stationary field (since BICs are non-radiating) confined within the system boundary. { The frequency of the photonic component ($E_k=\Omega$) is a direct consequence of the eigenvalue equation $\hat H \ket{D}=\Omega \ket{D}$, where the out-radiating photonic parameters in $\ket{D}$ are constrained to zero by the definition of BIC.}

The quantity $\braket{x}{S}$  effectively vanishes in the Markovian limit (which will be discussed in Section \ref{sec:mlim}) and is not relevant for the discussion of collective qubit subspaces. Solving the eigenvalue equation $H \ket{D}=\Omega \ket{D}$ for both Markovian and non-Markovian cases, we obtain the following mutual conditions\footnote{These conditions  for  BIC were also found in \cite{afterver1}  using a different approach.}
\begin{align} \label{eq:dark}
    \theta=n \pi, \quad  \sum_{m \in {\rm \{qubits\}}} (-1)^{nm} e_m &=0.
\end{align}
Thus, for $\theta=n\pi$, the dimensionality of the BIC subspace is $N-1$ where $N$ is the total number of qubits. 
This result shows an important property in waveguide QED: the dark and bright states, i.e. states that couple to light, manifest differently for odd and even $n$ (in (\ref{eq:dark})). For example, in a two-qubit  system, the symmetric state is dark for odd $n$ \cite{muller2017nonreciprocal}, whereas it is bright for even $n$  \cite{van2013photon}. This clarifies why the seemingly conflicting results of \cite{muller2017nonreciprocal} and \cite{van2013photon} are not actually conflicting. 

From (\ref{eq:darkf}-\ref{eq:dark}), we find the BICs (i.e. dark states) $\ket{D_i}$ ($i=1\dots N$). From these, we identify the superradiant bright state $\ket{B}$, which is orthogonal to the dark states. {Collectively, these form a basis for the single-excitation subspace of the atomic effective Hamiltonian that can be obtained by  adiabatically eliminating the photonic degrees of freedom in Markov approximation.} For $N=3$, this basis is
\begin{subequations}\label{eq:DDB}
\begin{align}
    \ket{D_1}&= \frac{1}{\sqrt{2}} \left( \ket{e_{-1}} {-} \ket{e_1} \right), \\
    \ket{D_2}&= \frac{1}{\sqrt{6}} \left( \ket{e_{-1}} {-} 2 (-1)^n \ket{e_0} {+} \ket{e_1}  \right), \\
    \ket{B}&= \frac{1}{\sqrt{3}} \left( \ket{e_{-1}} {+}  (-1)^n \ket{e_0} {+} \ket{e_1}  \right)\,.
\end{align}
\end{subequations}
{The basis for the general Hamiltonian can be obtained by including the photonic degrees of freedom. Throughout this paper, we use both definitions interchangeably. The distinction should be made for considerations of  non-Markovian dynamics, where $\ket{B}$ and $\ket{D_i}$ have non-vanishing photonic components in this regime. It is important to note that we have picked the basis states $\ket{D_i}$ that are either symmetric or anti-symmetric w.r.t. the center qubit (for reasons which will become clear  in Section \ref{sec:symmetry}). Any linear combination of $\ket{D_i}$ is still within the bound-state subspace. $\ket{B}$ is orthogonal to the subspace spanned by $\ket{D_i}$.}
Having specified the complete eigenbasis for the single-photon eigenstates, we can now turn our attention to the time-evolution dynamics. 

\subsection{Time evolution}\label{sec:timecoldec}
In this section, we extend the real-space formalism to include time-evolution of arbitrary single-photon states. We also show that, in the Markovian limit, for a broad class of states, which we call regular states, the poles of the stationary states of the system play an important role in the time evolution. 

First, let us consider the case where $\theta\neq n\pi$, where  the scattering states constitute the complete basis with the normalization $\braket{E_k}{E_p}\simeq 2\pi \delta(k-p)$. Here, the negative $k$ values stand for scattering of the photon from far right, for which we use the symmetry considerations (discussed in Appendix \ref{sec:appendixparity}) to construct. The time evolution operator in this case is
\begin{equation} \label{eq:timeevolutionop}
\begin{split}
        U(t) &=  \int_{-\infty}^\infty  \frac{\diff k}{2\pi}\ket{E_k} \bra{E_k} e^{-iE_kt}.
\end{split}
\end{equation}
The state of the system, at time $t$, is 
\begin{subequations}
\begin{align} 
           \ket{\psi(t)} &=  \int_{-\infty}^\infty\frac{ \diff k}{2\pi} \ket{E_k} \braket{E_k}{\psi(0)} \ee^{-iE_kt}, \\\label{eq:arbitrary}
        &= \int_{0}^\infty \frac{\diff k}{2\pi} \ket{g(k)} \ee^{-iE_kt},
\end{align}
\end{subequations}
where $\ket{g(k)}=\ket{E_k} \braket{E_k}{\psi(0)} +\ket{E_{-k}} \braket{E_{-k}}{\psi(0)}$. The case where $\theta\neq n\pi$ was discussed in \cite{tsoi2008quantum}, where they studied time evolution in a linear chain of atoms coupled to a 1D waveguide. Their work, however,  focused on the qubit survival probability rather than evolution of the entire system. Therefore, they did not include BICs. Here we also introduce the  special case where $\theta=n\pi$. In this case, a full basis must also include BICs, so the time-evolution operator is
\begin{align}\nonumber
        U_{\textsc{bic}}(t) &=  \int_{-\infty}^\infty  \frac{\diff k}{2\pi}\ket{E_k} \bra{E_k} \Big|_{\theta=n\pi} e^{-iE_kt} \\\label{eq:timeevol2}
        &+ \sum_{i=1}^{N-1} \ket{D_i} \bra{D_i} e^{-i\Omega t}.
    \end{align}
{Here, $\ket{D_i}$ denote the bound states of the full Hamiltonian, rather than of the atomic effective Hamiltonian (in contrast to Eq. (\ref{eq:DDB})).} The state of the system, at time $t$, is 
\begin{align}\nonumber  \ket{\psi_{\textsc{bic}}(t)}  ={}&\frac{1}{2\pi} \int_{0}^\infty \diff k \ket{g(k)} \ee^{-iE_kt}\Big|_{\theta=n\pi}\\\label{eq:arbitrary3}
&+\sum_{i=1}^{N-1}\ket{h_i}e^{-i\Omega t},
\end{align}
where $\ket{h_i}=\ket{D_i} \braket{D_i}{\psi(0)}$. 

Equations \eqref{eq:arbitrary} and \eqref{eq:arbitrary3} can be used to compute the dynamics of various observables, which we do in Section \ref{sec:nonmarkgen}. But first, we consider the Markovian limit, and show that in this limit, one can drastically simplify the expressions by taking advantage of complex analysis. 

\subsubsection{The Markovian limit and the power of complex analysis}\label{sec:mlim}
{ The Markovian limit considered here }is the limit where the qubits are separated by microscopic distances such that $L\sim O(\Omega^{-1})$. Following the approach by \cite{tsoi2008quantum}, we linearize the phase shift $e^{ikL}\simeq e^{i \theta}$, where {$ k L = (\Delta_k + \Omega) L \simeq \Omega L=\theta$} is the phase that light acquires when travelling between adjacent qubits. This linearization process{---which is the real space equivalent of the Wigner-Weisskopf approximation in  Laplace space---}ignores the time delay caused by inter-system propagation of photons and is accurate as long as $L \sim O(\Omega^{-1})$. This is because the characteristic time for the time evolution of states in the Markovian limit is $\sim O(J_0^{-1})$ (consequently $|\Delta_k| \leq O(J_0)$), as we show in Section \ref{sec:Markovian}. Since the propagation time of photons between qubits is $\sim O(\Omega^{-1})$, it is neglected in this limit (hence the name Markovian). 

For $\theta\neq n\pi$, we perform the substitution $\Delta_k = k-\Omega$, and rewrite \eqref{eq:arbitrary} as
\begin{subequations}
\begin{align}
           \ket{\psi(t)}        &= \int_{-\Omega}^\infty  \frac{\diff \Delta_k}{2\pi}\ket{g(k)} \ee^{-i\Delta_kt},\\
        &\simeq  \int_{-\infty}^\infty  \frac{\diff \Delta_k}{2\pi}\ket{g(k)} \ee^{-i\Delta_kt},  \\ \label{eq:arbitrary2}
        &= \sum_p \underset{\Delta_k = p}{\text{Res}} \left[ \ket{g(k)} \ee^{-i\Delta_kt} \right], 
\end{align}
\end{subequations}
where $p$ are the lower-half plane (LHP) poles of the collective system such that $\lim_{\Delta_k \to p} \ket{E_k}$ diverges, and $\text{Res}$ stands for residue. Here,  we made use of the assumption $J_0 \ll \Omega$ and $L \sim O(\Omega^{-1})$  in the first step, and  invoked the residue theorem in the second step. The regularity condition of the state $\ket{\psi}$ has been used when completing the contour through a circle including the lower half plane. When applicable, we neglect the global phase. The state $\ket{g(k)}$ has the same lower half plane poles as the scattering parameters $t_m,r_m,e_m$ (see equations (13-15) from \cite{tsoi2008quantum}). In fact, for a linear chain of qubits, all poles are in the LHP. Therefore finding the poles of any of the scattering parameters is the same as finding the poles of $\ket{E_k}$ (except for rare occasions where poles are cancelled by an introduced zero for a specific scattering parameter, in which case one shall pick another one that has $N$ poles for $N$ qubits).

The key result here is that, according to \eqref{eq:arbitrary2}, the time evolution of any regular state can be described via a contour integration and poles $p$. As an analytical approach, this is often much simpler to do than solving the integrals in \eqref{eq:arbitrary} directly.

For $N=3$ qubits, we identify the three poles of the scattering parameters as
\begin{subequations}\label{eq:poles}
\begin{align}
    p_1 &=-\frac{i}{2}J_0  \left(
   e^{2 i \theta}+2 + e^{ i \theta}\sqrt{8  + e^{2 i \theta}}\right), \\
    p_2 &=-\frac{i}{2}J_0 \left( e^{2
   i \theta}+2 -e^{ i \theta}\sqrt{8+ e^{2 i \theta}} \right), \\
    p_3 &= -i J_0 \left(1-e^{2 i \theta}\right).
\end{align}
\end{subequations}

What about when $\theta= n \pi$? Taking the limit $\theta \to n \pi$ (in, e.g., \eqref{eq:poles} for $N=3$)  results in $N-1$ of the poles converging at the origin and getting cancelled by a zero introduced in the numerator of the scattering parameters. The cancellation of these poles corresponds to loss of information about the system. In this case, the BIC terms in the time-evolution operator would be needed to supply this lost information. In fact, one would first need to find the BICs and then evolve the state according to \eqref{eq:arbitrary3}. But we now show that if the problem is treated analytically, we can get around this apparent loss of information as long as we are clever about it.

The scattering eigenstates contain all the information needed to model time evolution of the system, even in the limit $\theta\rightarrow n\pi$. But the limit must be taken at the right step in the calculation. The correct strategy is to \emph{first} evaluate the residues at the positions of the poles $p$ and \emph{then} take the limit $\theta \to n \pi$. Reversing the order (i.e. first taking the limit and then computing the residues) does not result in the same expression due to pole cancellation. Concretely,
\begin{equation} \label{eq:inequality}
\begin{split}
        &\lim_{\theta \to n \pi} \sum_p \underset{\Delta_k = p}{\text{Res}} \left[ \ket{g(k)} \ee^{-i\Delta_kt} \right] \neq \\ &\sum_p \underset{\Delta_k = p}{\text{Res}} \left[ \ket{g(k)} \ee^{-i\Delta_kt} \Big|_{\theta=n\pi} \right] .
\end{split}
\end{equation}

A natural question may arise: why is the residue corresponding to $N-1$ subradiant poles not zero in the limit $\theta = n \pi$? We discuss this in Appendix \ref{sec:appendixfinal}.

This analysis highlights two important properties of the real-space formalism. First, that the time evolution as described in this section is only exact in the limit {$J_0/\Omega \to 0$}, otherwise the analytical expressions found for the poles are approximate due to the linearization $kL \simeq \Omega L$ and the exact poles are described by a transcendental equation.
Second, in the regimes that it is valid, the method can be preferable for modelling time evolution in the presence of BICs, since with the analytical methods the time evolution can be obtained through only the scattering eigenstates.

In Sections \ref{sec:Markovian} and \ref{sec:nonmarkgen}, we will use the time-evolved state to compute dynamics of various observables. But first, we will discuss another interesting property of multi-qubit systems: their collective decay rates. 

\section{Collective decay rates}\label{sec:coldecrat}
Collective behavior of many-qubit systems can best be probed by considering the collective decay rates, as they reveal interesting physical phenomena such as subradiance and superradiance. In this section, we propose a strategy for finding the collective decay rates. 

Collective excitations of two qubits have been investigated perturbatively in the Markovian limit using the SLH formalism and a general master equation \cite{muller2017nonreciprocal}. Time evolution via the real-space formalism offers further insight by providing \emph{exact} collective decay rates. In this limit, our strategy provides collective decay rates that  describe the exact time evolution of regular states according to \eqref{eq:arbitrary2}. Non-Markovian collective decay rates for $N=2$ qubits coupled to a 1D waveguide have been investigated in \cite{zheng2013persistent} using a Green's function method. We develop a more general theory applicable to any multi-qubit system. 

Our proposed strategy for computing the collective decay rates is based on the observation that the complete set of decay rates is given by the poles of the scattering parameters via {a rotation and scaling such that $\Gamma = 2i p$.} A similar observation was made in  \cite{tsoi2008quantum},  but they did not explore it further. We do this here. 

The strategy for finding collective decay rates is as follows:
\begin{enumerate}
    \item[1)] Write down the Bethe Ansatz \eqref{eq:ba} and find the scattering parameters.
    \item[2)] Pick a scattering parameter, say $r_1$, and set its denominator to zero. This gives the characteristic equation for the collective decay rates.
    \item[3a)] For the Markovian limit, find analytical expressions  for the poles. 
    \item[3b)] For the non-Markovian regime, apply numerical methods to find the roots.
\end{enumerate}
{This strategy parallels the well-known connection between the resonance poles of a scattering matrix and the decay of unstable states. In the context of waveguide QED, the scattering parameters' poles provide a more general description due to the existence of qubit excitation coefficients ($e_k$), which may not always be eliminated to construct an S-matrix. Consequently, our definition of collective decay rates is analogous to \emph{resonance poles} in quantum scattering theory, \emph{quasi-normal modes} in gravitational wave physics and \emph{scattering poles} of the wave scattering theory \cite{zworski2017mathematical}.}

To illustrate this approach, for both the Markovian and non-Markovian regimes, we consider $N=3$ identical qubits coupled to a 1D waveguide. The characteristic equation for this case is
\begin{align}
\begin{split}
   0={} & 2 i e^{2 i (\Delta_k+\Omega) L} (\Delta_k+i )\\
     &+ e^{4 i (\Delta_k+\Omega)L} (1+i \Delta_k)\\
     &+i (\Delta_k+i )^3,
     \end{split}
\end{align}
where we normalize $\Delta_k$ w.r.t. $J_0$. 

\subsection{Markovian regime}
 
In the Markovian limit,  the qubits are separated by a distance $L \sim O(\Omega^{-1})$, which justifies the linearization of the light propagation phase $k L \simeq \Omega L=\theta$, which results in a polynomial characteristic equation for the collective decay rates. Solving this equation yields following collective decay rates:
\begin{subequations} \label{eq:coldecay}
\begin{align}
      \Gamma_1 &=\frac{1}{2} \left(
   e^{2 i \theta}+2 + e^{ i \theta}\sqrt{8  + e^{2 i \theta}}\right)\gamma_0 , \\
    \Gamma_2 &=\frac{1}{2} \left( e^{2
   i \theta}+2 -e^{ i \theta}\sqrt{8+ e^{2 i \theta}} \right)\gamma_0 , \\
    \Gamma_3 &=   \left(1-e^{2 i \theta}\right)\gamma_0 .
\end{align}
\end{subequations}
where we recall that $\gamma_0=2J_0$ is the single-qubit decay rate.

If we set $\theta=n\pi$, we find that  two of the decay rates become zero, whereas one becomes $3\gamma_0$. This also occurs for an arbitrary $N$-qubit system: as $\theta\rightarrow n\pi$, $N-1$ collective decay rates cluster around zero, while the $N$th one  approaches $N\gamma_0$. The corresponding physical phenomena are called \emph{subradiance} and \emph{superradiance}, respectively.  Subradiance occurs when \emph{destructive} interference \emph{suppresses} spontaneous emission. Conversely, superradiance occurs when \emph{constructive} interference \emph{enhances} spontaneous emission. Consequently, the corresponding collective decay rates are called subradiant or superradiant decay rates of the system. Hence, finding collective decay rates also gives further information on superradiance and subradiance occurring in the multi-qubit system. The $N\gamma_0$ scaling of the superradiant decay rate is well-known and is usually referred to as the Dicke superradiance \cite{andreev1980collective}. We observe the Dicke superradiance in the Markovian limit, however we will observe a new kind of superradiance when we discuss the non-Markovian limit.

\subsection{Non-Markovian regime} \label{sec:coldecratnonmark}

In the non-Markovian regime, which consists of every condition where the Markovian approximation is no longer valid, the qubits are separated by macroscopic distances such that $L\sim O(J_0^{-1})$. The propagation time of light within the network is no longer negligible, consequently, for $L \sim O(J_0^{-1})$, the second term of $k L = \Omega L + \Delta_k L$ can no longer be neglected. Hence, the important distinction in the non-Markovian limit is that the characteristic equation of poles is a transcendental equation that includes complex exponentials in addition to polynomial terms. We must thus use numerical methods.

To use our approach in the  non-Markovian limit, we first expand the exponential as
\begin{subequations}
\begin{align}
    e^{i(\Delta_k+\Omega)L} ={}&e^{i\theta (1+\Delta_k/\Omega)}\\
    \simeq {}&e^{i\theta} \sum_{n=0}^M  \frac{(i\theta \Delta_k/\Omega)^n}{n!},
\end{align}
\end{subequations}
where we truncate this expansion for some large $M$. This approach works as long as the poles have the property $|p|\ll \Omega$, such that $(\Delta_k/\Omega)^M \to 0$ for the region of interest. If this is not the case, one needs to employ more complicated numerical approaches to find the roots. 

\begin{figure}[b!]
    \centering
    \includegraphics[width=\columnwidth]{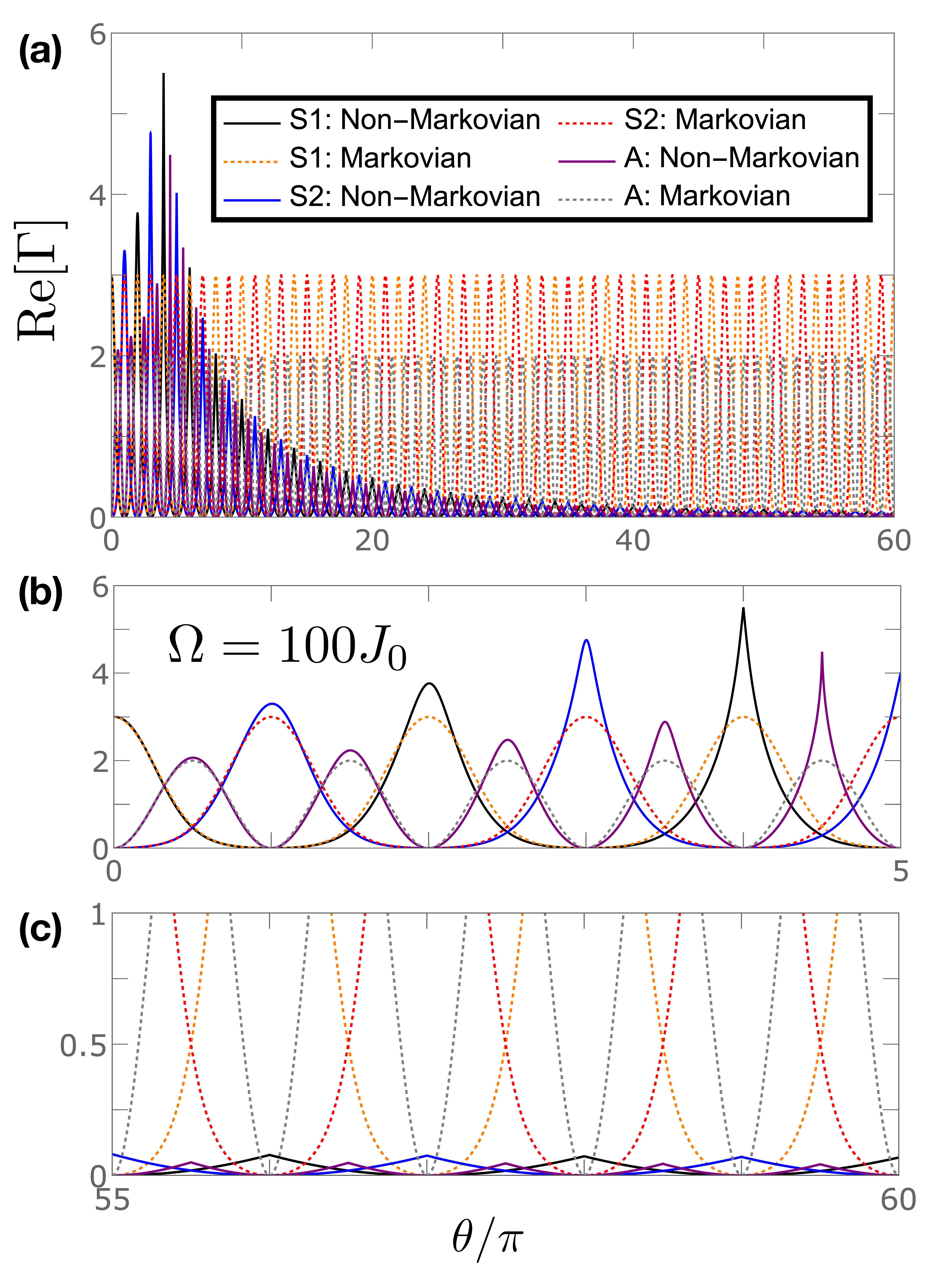}
    \caption{The behavior of three Markovian and non-Markovian collective decay rates (in units of $\gamma_0$) for $N=3$ qubits coupled to a 1D waveguide for  a) $\theta \in [0,60]\pi$, b) $\theta \in [0,5]\pi$, c) $\theta \in [55,60]\pi$. S (A) stand for symmetric (anti-symmetric) collective decay rates. In all figures, $\Omega=100J_0$. }
    \label{fig:non-mark-col-dec}
\end{figure}

For illustration purposes, we pick $\Omega=100J_0$ and plot the Markovian and non-Markovian collective decay rates (more accurately, their real part which is responsible for the decaying behavior) in Fig. \ref{fig:non-mark-col-dec}. Here, we denote the symmetric (anti-symmetric) collective decay rates with S (A). This figure shows many resemblances to the findings of \cite{zheng2013persistent}, where the non-Markovian decay rates become subradiant in the large $\theta$ limit. As explained in \cite{zheng2013persistent}, after the initial emission of a decaying exponential pulse from the first qubit, it gets reflected by the second and third qubits and re-excites the first qubit. This cycle continues for longer times due to light trapping between the qubits, as we shall see in Section \ref{sec:nonmark}. The non-Markovian decay rates describe this light-trapping and slow decay quantitatively. Since in the non-Markovian regime, the interactions are effectively isolated between single qubits and light, increasing $L$ decreases the collective decay rates due to the increased time delay of inter-qubit photon propagation. 

Fig. \ref{fig:non-mark-col-dec} shows another interesting phenomenon. For $\theta \sim 4 \pi$, the superradiant collective decay rate becomes larger than $3\gamma_0$.   Dicke superradiance alone cannot describe this phenomenon, an additional physical mechanism (i.e. time-delayed coherent quantum feedback) must be at play. We will discuss this new mechanism---which we call super-superradiance (SSR)---in Section \ref{sec:nonmarkspontem} when we discuss strong collective spontaneous emission from $N=3$ qubits in the non-Markovian regime. For now, we note the highly counter-intuitive observation that by introducing time delay of photon propagation within the system, the overall decay of the system can be enhanced. This has also been observed for  $N=2$  in Fig. 4(d) of \cite{zheng2013persistent}, but was not discussed in the main text.  Recently, \cite{sinha2019non} discussed this effect for $N=2$ using a different approach. The findings of \cite{zheng2013persistent}, \cite{sinha2019non} and our findings agree perfectly for $N=2$ although all three papers use different methods. Our findings suggest that SSR is a more general phenomenon than the $N=2$ case.

On another note, the non-Markovian processes introduce many more decay modes, in addition to $3$ original decay modes that are present in the Markovian limit. This can be understood by the travel time of light between qubits in two different regimes. In the non-Markovian regime, the delayed feedback mechanism introduces infinitely many poles, as light gets trapped inside the system and is released within intervals of $2L$ with each release time corresponding to a non-Markovian process. For the Markovian limit, all these processes happen at an instant and are confined to an infinitesimal amount of time, since photons propagate between qubits without any delay. Hence, any exclusively non-Markovian process in the Markovian limit has an infinite decay rate. In other words, one can think of the additional non-Markovian collective decay rates as flowing from LHP complex infinity to finite values as the photon travel time between qubits becomes relevant, i.e. as $\theta$ increases. When plotting Fig. \ref{fig:non-mark-col-dec}, we picked the three decay rates with the lowest real part, following \cite{zheng2013persistent}.

\begin{figure}[h!]
    \centering
    \includegraphics[width=\columnwidth]{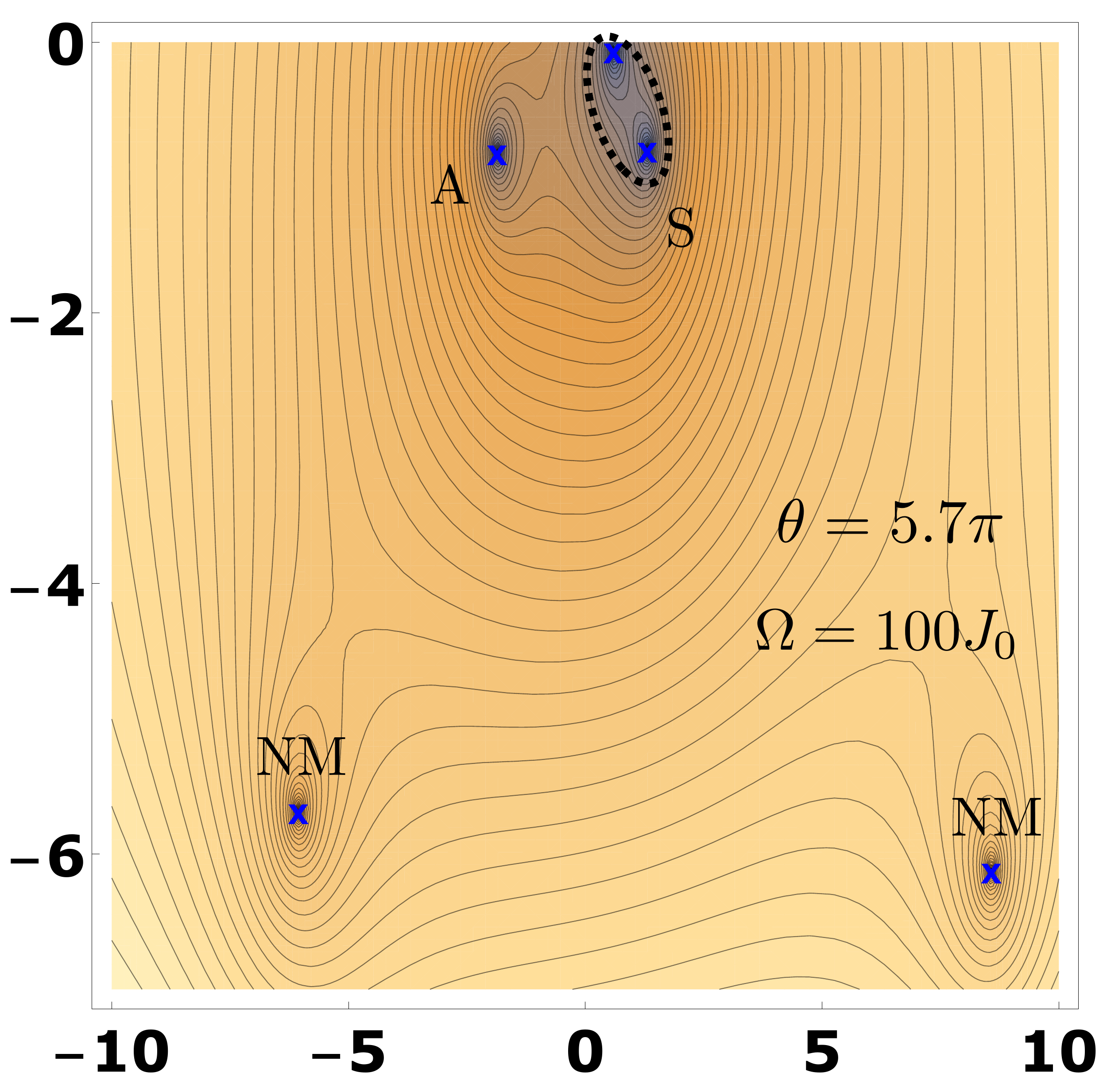}
    \caption{The poles (in units of $J_0$) obtained via the numerical method for $\theta=5.7\pi$, $\Omega=100J_0$ and $N=3$ qubits marked by blue crosses. S (A) stand for symmetric (anti-symmetric) collective decay rates {(See Section \ref{sec:symmetry})}, NM stands for decay modes generated by non-Markovian processes. The color figure shows the logarithmic modulus plot of the full characteristic equation, where the darker regimes correspond to lower values of the function. Since the characteristic equation is analytical, its modulus has a minimum inside a finite region only if it has a zero within that region (from the minimum modulus principle). Thus, it is clear that the poles found via the numerical method are indeed physical, as the marks correspond to the zeros of the plotted function.}
    \label{fig:non-mark-2}
\end{figure}

In Fig. \ref{fig:non-mark-2}, we plot the three Markovian poles (two corresponding to symmetric modes and one to anti-symmetric mode) and additional poles corresponding to non-Markovian processes. For the parameters considered, the {processes corresponding to} non-Markovian poles decay much faster and therefore can be neglected. For larger $\theta$, this is no longer the case and non-Markovian processes gain importance.

The method introduced in this section is more general and can be applied to any multi-qubit system, as the characteristic equation for any general system can be found by simply finding energy eigenstates. Analogously, we also note that a similar approach can be applied to consider Fano minima, i.e. reflection minima, that lead to transparency of the multi-qubit system for certain frequencies. \cite{tsoi2008quantum} finds $N-1$ such frequencies for a linear chain of $N$ qubits. Non-Markovian processes can increase this number, as the characteristic equation describing the minima frequencies become transcendental in this limit. For example, for $N=3$, we have observed several non-Markovian Fano minima (in addition to $2$ that are observed in the Markovian limit \cite{tsoi2008quantum}) in our own explorations. \cite{fanopaper} observes similar effects for a two-qubit system. We believe that this is an interesting phenomenon to study in future work.

We have also developed an  alternative method for finding collective decay rates, which we describe  in Appendix \ref{sec:appendixalternativemethod}.  While this method is less efficient than the one presented here, it provides further insight into the physics of the system, which might be useful for constructing proofs.

\section{Time dynamics of observables in the Markovian limit} \label{sec:Markovian}

In this section, we study time dynamics in the Markovian limit. We can therefore take  advantage of  complex analysis to simplify the analysis.

As an example, we consider a chain of three qubits. We first consider spontaneous emission and study the time dynamics of atom excitation probabilities and emitted single-photon pulses. We then study the scattering of a single photon pulse from a system on multiple qubits. 

\subsection{Spontaneous emission} \label{sec:spontemission}
Here we consider spontaneous emissions with the initial condition where the central qubit ($m=0$) is excited. We begin by  defining the relevant observables.

\subsubsection{Survival probability}
The survival probability (introduced in \cite{tsoi2008quantum}) is defined as the probability that an initially excited atom remains in its excited state:
\begin{equation} \label{eq:main}
    P_e(t) = \left|  \int_{-\infty}^\infty  \frac{\diff k}{2\pi}|e_0(k)|^2 e^{-i \Delta_k t} \right|^2.
\end{equation}
After performing the complex analysis, this quantity takes the  form
\begin{equation}
    P_e(t) = \left|  \sum_p \underset{\Delta_k = p}{\text{Res}} \left[2 |e_0(k)|^2 e^{-i \Delta_k t}\right] \right|^2.
\end{equation}

\subsubsection{Side-atom excitation probability} The side-atom excitation probability is the probability that a side atom (either $m=-1$ or $1$) is excited at time $t$, given that the center atom is initially prepared in an excited state. To find this probability, we first note that the initial state of the system can be projected onto energy eigenstates via a resolution of identity
\begin{equation}
    \ket{\alpha(t)} =  \int_{-\infty}^\infty  \frac{\diff k}{2\pi}\ket{E_k} \braket{E_k}{\alpha}  e^{- i E_k t},
\end{equation}
where $\ket{\alpha}= \sigma_0^\dag \ket 0$ is the initial state of the system. We realize that $\braket{E_k}{\alpha} = e_0^*$, where $^*$ denotes the complex  conjugate. We  find that the probability, $P_s(t)=|\braket{e_1}{\alpha(t)}|^2=|\braket{e_{-1}}{\alpha(t)}|^2$, that a side atom is excited at time $t$ is
\begin{equation} \label{eq:side}
    P_s(t) = \left| \int_{-\infty}^\infty  \frac{\diff k}{2\pi}e_{\pm 1}(k) e_0^*(k) e^{-i\Delta_k t} \right|^2.
\end{equation}
After performing the complex analysis, this quantity takes the  form
\begin{align}
\begin{split}
    P_s(t) ={}& \Big|  \sum_p \underset{\Delta_k = p}{\text{Res}} \Big[ e_0^*(k) (e_1(k) \\
    &~~~~~~~~~~~~~~+ e_{-1}(k)) e^{-i \Delta_k t}\Big] \Big|^2,
    \end{split}
\end{align}
where we recall that $e_1(k)=e_{-1}(-k)$ due to symmetry.

\subsubsection{Emitted photon probability density} The emitted photon probability density is defined as $\mathcal{P}(x,t)=|\psi (x,t)|^2$  where $\psi(x,t)=\braket{x}{\alpha (t)}$ is the emitted photon waveform. Following a  derivation similar to that of  the side-atom excitation probability, the emitted photon probability density takes the form
\begin{equation} \label{eq:appendixa}
      \mathcal{P}(x,t) = \left| \int_{-\infty}^\infty  \frac{\diff k}{2\pi}\braket{x}{E_k} e_0^*(k) e^{-i\Delta_k t} \right|^2.
\end{equation}
After performing the complex analysis, this quantity takes the  form
\begin{equation} \label{eq:emittedphoto}
      \mathcal{P}(x,t) = \left| \underset{\Delta_k = p}{\text{Res}} \left[(t_3 + r_1 ) e_0^*(k) e^{-i\Delta_k (t-|x|)}\right] \right|^2,
\end{equation}
where we recall that in the Markovian limit, the complete system is effectively situated at $x=0$ and $\braket{x}{E_k}=t_3 e^{ikx} \Theta(k) + (r_1 e^{ikx}+e^{-ikx}) \Theta(-k)$ for $x>0$. The $e^{-ikx}$ term becomes zero since $e_0^*(k)$ has no upper-half plane poles and the emitted photon probability density is symmetric w.r.t. the origin. 

For completeness, we define the probability of photon emission as $P_w = \int_{-\infty}^\infty \mathcal{P}(x,t) \diff x$. We also include an illustrative example of how these formulae can be applied to a single atom inside a waveguide in Appendix \ref{sec:appendixa} for comparison with existing literature.  

\subsubsection{Numerical results}

\begin{figure*}
\centering
\includegraphics[width=16cm]{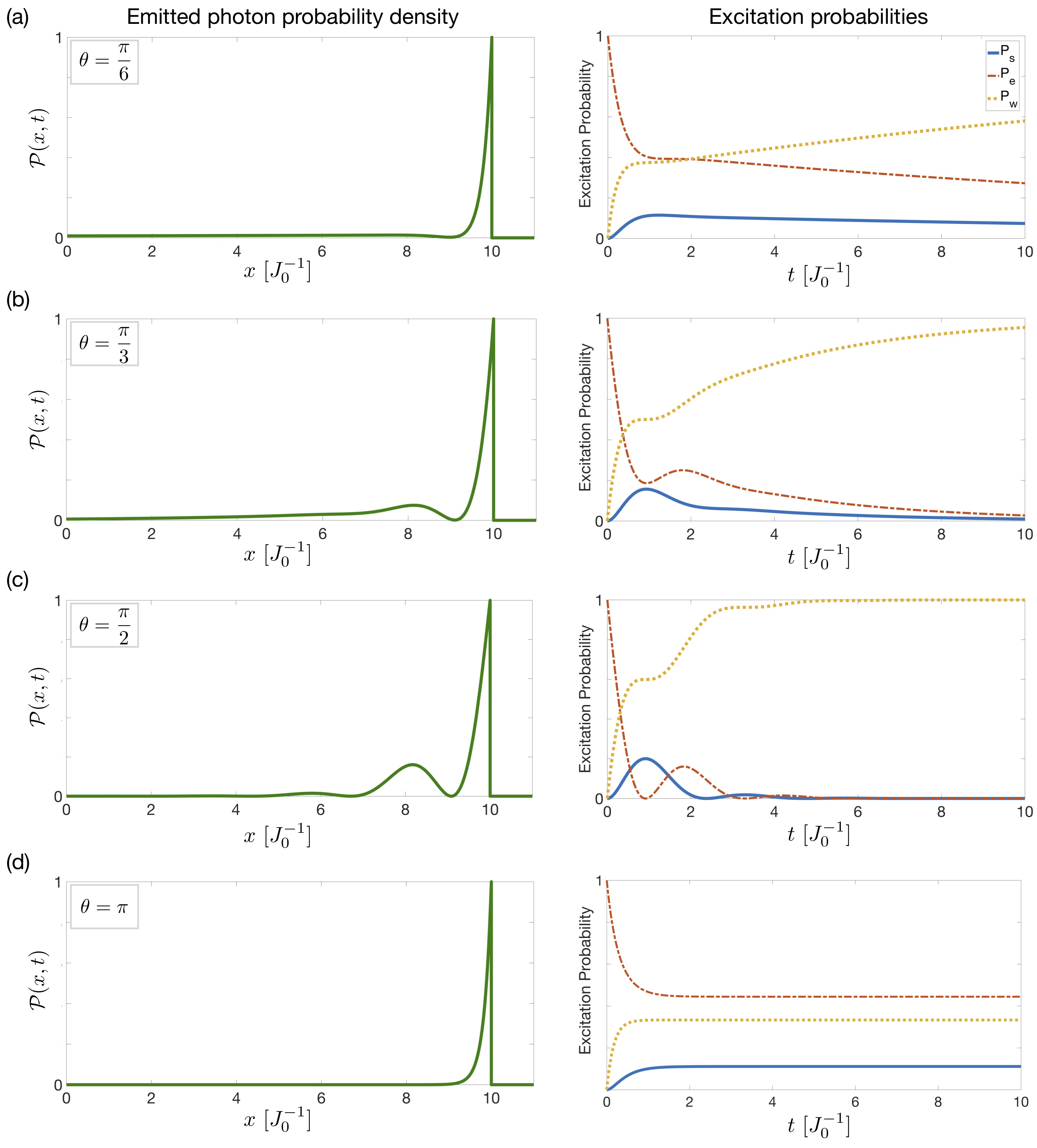}
\caption{Spontaneous emission in a system of three qubits in the Markovian limit. Emitted photon probability densities  $\mathcal{P}(x,t)=|\psi(x,t)|^2$ (normalized w.r.t. $J_0$, for $t=10/J_0$), and atomic excitation probabilities $P_e$ and $P_s$ and emission  probability $P_w$, for different values of $\theta$ in (a)-(d). The corresponding poles are: (a) $p_1/J_0 =-0.794-i0.133 $ and $p_2/J_0=1.66-i1.37 $, (b) $p_1/J_0 =-1.323-i0.5  $ and $p_2/J_0=1.323-i0.5  $, (c) $p_1/J_0 =-1.323-i0.5  $ and $p_2/J_0=1.323-i0.5  $, and (d)  $p_1/J_0 =-3i  $ and $p_2/J_0=0$. The three-qubit systems is located at $x=0$, the distance between the atoms become effectively zero for the scales $x \sim O(J_0^{-1})$, as $J_0/\Omega \to 0$. The system is symmetric w.r.t. the origin, therefore only $x>0$ is illustrated. The qubits at later times decay corresponding to the pole with lowest imaginary component, as observables decay with a rate $\sim 2 \Im[p]$. In all cases, $P_w+2P_s+P_e=1$ for all times.} \label{fig:decay}
\end{figure*}

Recall that the scattering parameters $t_m$, $r_m$ and $e_m(k)$ can be found by solving the eigenvalue equation $H \ket{E_k} = E_k \ket{E_k}$. Since $H$ depends on $\theta$ {($H$ depends on $\Omega$ and $L$ independently)}, so do the scattering parameters, and consequently so do $P_e(t)$, $P_s(t)$ and $\mathcal{P}(x,t)$.  This $\theta$-dependence can be seen in Fig. \ref{fig:decay} for  $\theta=\pi/6$, $\pi/3$, $\pi/2$ and $\pi$. We note that since $v_g=1$, both $x$ and $t$ are plotted with units $J_0^{-1}$. The poles are also  given for each case, where in all cases the third pole $p_3$ has no residue contribution \footnote{The absence of a residue contribution for $p_3$ is highly intriguing and a possible reason for it will be considered further in Section \ref{sec:symmetry}. Our analytical calculations show that the $p_3$ pole can only contribute if the state of the collective system no longer has mirror symmetry w.r.t. the center atom ($x=0$).}  and has therefore been omitted. {The imaginary part of the poles dictates the atom decay, as was pointed out by \cite{tsoi2008quantum}, which become the real part of the decay rates via the relation $\Gamma= 2 i p$.}

The early behavior of the system is dictated by the (superradiant) decay rate with the largest real component, whereas the late behavior is dictated by the (subradiant) one having the smallest real component. 
In Fig. \ref{fig:decay}, the immediate decay of excitation probabilities just after $t=0$ confirms  the early behavior ({superradiant} decay rate is dominant), while the slow decaying behavior at $t=x_0=10J_0^{-1}$ confirms the late behavior (subradiant decay rate is dominant). This is also the case for the photon density profile, but is less obvious from the figure. Additionally, comparing \ref{fig:decay}(a,d) with \ref{fig:decay}(b,c) shows that the system exhibits oscillatory behavior for $\theta$ values close to $\pi/2$. This is expected, since the separation in the imaginary part of the decay rates (e.g. the difference between collective energy levels) is responsible for oscillations and is highest for $\theta=\pi/2$. The early (superradiant) decay rates are higher and late (subradiant) decay rates are slower for $\theta$ values that are far away from $\pi/2$. 

For the special cases $\theta=\pi$ and $\pi/2$, interesting properties occur. We discuss these in detail next.

\subsubsection{Analytic solution for quarter-resonance-wavelength atom spacing ($\theta=\pi/2$)}
In this case, the separation between two neighboring atoms is $L =\lambda / 4$, where $\lambda$ is the resonant wavelength. This means that a non-interacting photon propagating between the two ends of the system acquires an overall phase of $\pi/2$. In this special case, the side atoms never emit to the waveguide, rather only into the system itself with mirror-symmetric initial conditions. 

To see why, consider that when the middle atom emits the photon, it  excites a superposition of side atoms ($m=-1$ and $1$) at later times. The side atoms  become excited equally due to mirror symmetry for times $t>0$. The atoms then emit a photon in superposition, which suggests that the superposition terms are in-phase at  time $t$ of emission, whereas they acquire an overall phase of $\pi/2$ due to propagation inside the system before they leave the system. Since the propagation time of the superposition terms inside the system can be neglected in the Markovian limit, the superposition terms of the photon emitted by the two side atoms (non-interacting with the central atom) interfere destructively and only the photon emission from the central atom leaves the system and gets radiated to the waveguide. For this case,  $P_e(t)$, $P_s(t)$ and $\mathcal{P}(x,t)$ reduce to\clearpage
\begin{widetext}
\begin{subequations}
\begin{align}
\begin{split}
P_e(t) ={}&\frac{1}{7} e^{-J_0t} \Big(3 \cos \left(\sqrt{7}J_0t\right)-\sqrt{7} \sin \left(\sqrt{7}J_0 t\right)+4\Big) \Theta(t), 
\end{split}\\
P_s(t) ={}& \frac{4}{7} e^{-J_0t} \sin ^2\left(\frac{\sqrt{7} J_0 t}{2}\right) \Theta(t),\\
\begin{split}
\mathcal{P}(x,t) ={}&\frac{J_0}{7} e^{-J_0(t-|x|)} \Big(3 \cos \left(\sqrt{7}
   J_0(t-|x|)\right)
   -\sqrt{7} \sin \left(\sqrt{7} J_0(t-|x|)\right)+4\Big)\Theta(t-|x|),
   \end{split}
\end{align}
\end{subequations}
\end{widetext}
where $\Theta(x)$ is the Heavy-side function.

We notice that $\mathcal{P}(x,\tau)|_{(\tau-|x|)=t}= J_0 P_e(t)$, with the proportionality factor $J_0$, suggesting that a photon leaving the linear chain system is indeed emitted by the center atom, as the probability density function of a photon emitted at time $t$ is proportional to the excitation probability of the center atom at time $t$. In Appendix \ref{sec:appendixa}, this proportionality is discussed for a single qubit inside a waveguide. 

Another important result for $\theta=\pi/2$ is that the existence of side-atoms changes the decay rate. An atom on its own has a decay rate of $\gamma_0$, whereas the collective decay rates become $\gamma_0 (0.5\pm i \phi_{\pm})$, where $\phi_{\pm}$ is some phase factor. The decrease in the exponential decay (the real part of the decay rate) and the existence of oscillations (imaginary part of the decay rate) can be explained by the partial reflection of the photon from the two side atoms. Moreover, the { first} local minima of $P_e(t)$ and $\mathcal P(x,\tau)|_{(\tau-x)=t}$ coincide with the { first} local maxima of $P_s(t)$, implying that the side atoms are excited with highest probability when the center atom is in its ground state (where the photon emission is also zero) and vice versa. 

{While we considered the symmetric excitation where the middle qubit is excited, it is also interesting to consider the case where the side qubits are excited with equal weights. Then, by the virtue of the fact that the side qubits do not radiate outside of the system for $\theta=\pi/2$, the outgoing pulse shape looks less like a decaying exponential and has a more symmetric shape. We believe that it could be interesting to consider potential pulse-production applications of this configuration ($\theta=\pi/2$) for increasing number of qubits ($N$) as a future work.}

\subsubsection{Analytic solution for half-resonance-wavelength atom spacing ($\theta=\pi$)}
In this case, the separation between two neighboring atoms is $L =\lambda / 2$ and the system exhibits super- and subradiant modes. The superradiance phenomenon is manifested mathematically in the first pole where $p_1=-i3J_0$. This phenomenon is an ``enhancement effect due to collective dipoles" as pointed out in \cite{tsoi2008quantum}. Additionally, the second pole becomes $p_2=0$, implying that one of the decay modes has zero decay rate, e.g. a finite survival probability is obtained as $t \to \infty$. In this case, $P_e(t)$, $P_s(t)$ and $\mathcal{P}(x,t)$ reduce to
\begin{subequations} \label{eq:pi}
\begin{align}
P_e(t) &=\frac{1}{9} \left(e^{-3 J_0 t}+2\right)^2\Theta(t), \\
P_s(t) &= \frac{1}{9} \left(e^{-3 J_0 t}-1\right)^2 \Theta(t),\\
\mathcal{P}(x,t) &=J_0 e^{- 6 J_0 (t-|x|)}\Theta(t-|x|). \label{eq:pi-c}
\end{align}
\end{subequations}
Interestingly, the integral of (\ref{eq:pi-c}) over $x$, does not tend to one as $t \to \infty$. This suggests that some amplitude of the photon will always remain inside the system. We can also see this from the atom excitation probabilities.  
As $t \to \infty$,  the center atom remains excited with probability $\frac{4}{9}$, and each one of the side atoms remains excited  with probability $\frac{1}{9}$. With  probability $\frac{1}{3}$, all atoms decay and photon-emission takes place. This is in contrast with other values of $\theta$,  where photon-emission is certain for large enough times. This is a consequence of the existing dark states in the system, which are signalled by the fact that two of the three collective decay rates become zero when $\theta=\pi$. In fact, we can obtain this result without any contour integration by first realizing that
\begin{equation}
    \ket{\alpha(0)}=\ket{e_0} = \frac{2}{\sqrt{6}} \ket{D_2}- \frac{1}{\sqrt{3}} \ket{B}.
\end{equation}
Then, the state evolves as
\begin{equation}
    \ket{\alpha(t)}=\frac{2}{\sqrt{6}} \ket{D_2}e^{-i\Omega t}  - \frac{1}{\sqrt{3}} \ket{B(t)},
\end{equation}
where as $t \to \infty$, $\braket{e_m}{B}=0$ for any $m$ due to  superradiance. Then, we find that $P_s=|\frac{2}{\sqrt{6}}\braket{e_{-1}}{D_2}|^2 = |\frac{2}{\sqrt{6}}\braket{e_{1}}{D_2}|^2 =\frac{1}{9}$ and $P_e=|\frac{2}{\sqrt{6}}\braket{e_{0}}{D_2}|^2=\frac{4}{9}$. The probability of complete decay can also be found as $|\frac{1}{\sqrt{3}} \braket{B}{\alpha(0)}|^2=\frac{1}{3}$.

This property may prove fruitful for future quantum memory applications using identical qubits. In fact, as \cite{sato2012strong} has shown,  cavity-waveguide coupling can be controlled, which opens up the possibility of exciting BICs via time-dependent control of cavities using only single-photon scattering states, in addition to exploiting the delayed quantum feedback and multi-photon scattering as discussed in \cite{calajo2019exciting}.

\subsection{Pulse scattering for single-photon states} \label{sec:pulsescat}

To study scattering of single pulses, we will focus on two properties: first on the shape of the transmitted and reflected pulse and second on the excitation probability  $P_m$ of the $m$th  atom. We assume that the initial pulse is situated at $x=-x_0$ ($x_0>0$) far away from origin with an average momentum $k_0>0$ and the atoms are initially in ground state.

To start our discussion, we  define two functions $f(x)$ and $\tilde{f}(k)$---where $\tilde{f}(k)$ is the Fourier transform of $f(x)$---whose standard deviations $\Delta x$ and $\Delta k$ satisfy $\Delta x\ll x_0$ and $\Delta k \ll k_0$. Then, the most general form for the scattering state  $\ket{S(t)}$  at time $t=0$, with one-sided excitation from the left, can be written as
\begin{align} \label{eq:pulsein}
\ket{S(0)} = \int_{-\infty}^\infty \diff x f(x+x_0) e^{ik_0x} C_R^\dag(x) \ket 0,
\end{align}
where we compute $\braket{k}{S(0)}= \tilde f(k-k_0) \allowbreak e^{i(k-k_0)x_0}$. This state represents a pulse initially located at $x=-x_0$ and moving to the right with an average momentum $k_0$. 

We assume $J_0\ll \Omega$, and as before, we  find the time evolution of this state by first projecting it onto energy eigenstates and then time-evolving each part independently:
\begin{equation} \label{eq:timeevolved}
\ket{S(t)} =  \int_{-\infty}^\infty  \frac{\diff k}{2\pi}\braket{E_k}{S(0)} \ket{E_k} e^{-iE_k t}.
\end{equation}

\subsubsection{Shape of Transmitted and Reflected Pulses }
The shape of the photon pulse at time $t$ is
\begin{equation} \label{eq:Sxt}
S(x,t) =   \int_{-\infty}^\infty  \frac{\diff k}{2\pi}\braket{x}{E_k} \braket{E_k}{S(0)} e^{-iE_k t},
\end{equation}
where $S(x<0,t)$ represents the reflected pulse and $S(x>0,t)$ the transmitted pulse at time $t$. We define the functions $\tilde S (k,0)$, $\tilde{S}_+ (k,t)$ and $\tilde{S}_- (k, t)$ as the Fourier transform of the initial, transmitted and reflected pulses.

In Appendix \ref{sec:appendixb}, we show that $|\tilde S_+(k,2x_0)|^2=|t_3 \tilde S(k,0) |^2$ and $|\tilde{S}_- (-k,2x_0)|^2 = |r_1\tilde S(k,0) |^2$. (Here, we pick $t=2x_0$ such that the field no longer interacts with the qubits.) This suggests that, each mode of the input pulse inside the waveguide is modulated independently via the stationary state transmission and reflection coefficients found using the Bethe Ansatz. This property for the specific case of a single qubit and two qubits inside a waveguide was pointed out in  \cite{chen2011coherent} and \cite{liao2016photon}. 

The main advantage of the real-space approach is that the asymptotic scattering calculations do not assume any information on the pulse shape or the internal degrees of freedom of the system. Any pulse that is localized around $x_0$ and has a narrow band in the frequency space can be shown to modulate according to the scattering parameters $t_3$ and $r_1$. Furthermore, as long as $J_0\ll \Omega$, this finding can apply to other quantum networks (see \cite{wang2016dynamics} for an example of a different setting) in addition to the linear chain explored in this paper, since the derivations performed in Appendix \ref{sec:appendixb} only require the overall transmission and reflection coefficients, $t_3$ and $r_1$. Here, the internal interactions of the system are only important for how they affect the output field. 

\subsubsection{Atom excitation probability} 
The excitation probability of individual atoms can be found by using the Born rule for the time evolved state given in (\ref{eq:timeevolved}) as
\begin{align}
P_m(t)
= \left|  \int_{-\infty}^\infty  \frac{\diff k}{2\pi}e_m(k) \braket{E_k}{S(0)} e^{-i\Delta_k t} \right|^2, \label{eq:13}
\end{align}
where we recall that for negative $k$ values we simply set $e_{-1}(-k)= e_1(k)$ from the symmetry of the system. In Appendix \ref{sec:appendixa}, we show the correspondence between this formula and the results of \cite{chen2011coherent} for a single atom inside a waveguide.

\begin{figure*}
\centering
\includegraphics[width=\textwidth]{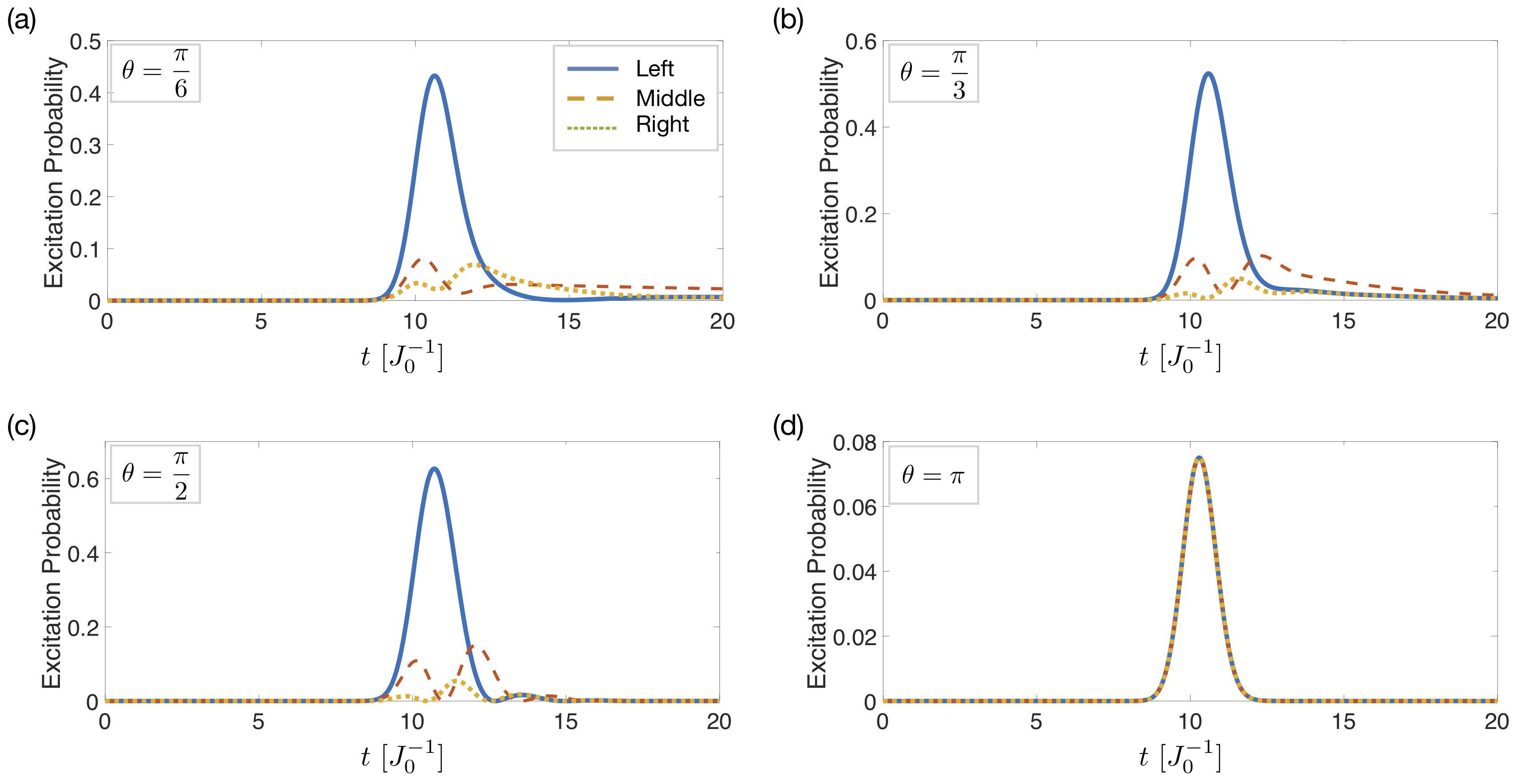}
\caption{Singe-photon scattering in a system of three qubits in the Markovian limit. The excitation probability of each atom  during interaction with a single photon pulse shaped as a Gaussian  centred at $x_0=10/J_0$. For each value of $\theta$, the parameter $\xi$ was optimized to maximise the excitation probability for the left atom. In  d), the linearization assumption causes all three excitation probabilities to be the same, which is accurate as long as $L\sim O(\Omega^{-1})$ and $J_0/\Omega \ll 1$. } \label{fig:exc}
\end{figure*}

\subsubsection{Example: Gaussian pulse} 

Here, we consider a specific example of a resonant Gaussian pulse with $k_0=\Omega$ and $\Delta k = J_0$ incident from the far left (we consider decaying and rising exponential pulse shapes in Appendix \ref{sec:appendixpulsemarkovian}). A  Gaussian-shaped photon is not a regular state and the residue theorem cannot be applied. The integral in (\ref{eq:13}) can still be computed analytically for most cases (for a Gaussian pulse, always) by applying the convolution theorem for Fourier transforms as  discussed for the single-qubit case in Appendix \ref{sec:appendixa}. In summary, the time evolution of a Gaussian state in (\ref{eq:13}) comes down to an integral of a Gaussian function and the scattering parameters. Since the Fourier transform of scattering parameters is a weighted sum of decaying exponentials, a final analytical result can always be obtained by taking the convolution of a Gaussian function with decaying exponentials. As a comparison, we note that this analytical result could not be obtained via the existing methods, where numerical methods were employed \cite{liao2015single,liao2016photon}. As our purpose in this section is the illustration of the pulse scattering, we compute this integral numerically for simplicity.

The projection of the input pulse onto a scattering eigenstate is (see Appendix \ref{sec:appendixb})
\begin{equation}
\braket{E_k}{S(0)} =\sqrt{2\pi} \tilde f(k-k_0) e^{i(k-k_0)x_0} \Theta(k).
\end{equation}
Then, the excitation probability of the $m^{th}$ atom is
\begin{equation}
P_m = \left| \int_{0}^\infty  \frac{\diff E_k}{\sqrt{2\pi}} e_m(k)  \tilde f(k-k_0) e^{-i\Delta_k (t-x_0)} \right|^2,
\end{equation}
as $E_k=k$ for $k>0$. 
For the Gaussian input with $\Delta k = J_0$, we choose $\tilde f(k-k_0)$ such that $\int_{-\infty}^\infty \diff k |\tilde{f}(k-k_0)|^2=1$ to ensure normalization:
% \begin{equation} \label{eq:gaussian}
% \tilde f(k-k_0) = \frac{ \exp(-(\Delta_k-\chi)^2/(4 J_0^2))}{\sqrt{\sqrt{2\pi} J_0}},
% \end{equation}
\begin{equation} \label{eq:gaussian}
\tilde f(k-k_0) = \frac{ e^{-\frac{(\Delta_k-\chi)^2}{4 J_0^2}}}{\sqrt{J_0\sqrt{2\pi} }},
\end{equation}
where, for a resonant Gaussian pulse,  $\chi=0$.

Then, the integral becomes
\begin{equation}
P_m = \left| \int_{-\Omega}^\infty \diff \Delta_k e_m  \frac{ e^{-\frac{(\Delta_k-\chi)^2}{4 J_0^2}}}{\sqrt{2\pi J_0\sqrt{2\pi}} } e^{-i\Delta_k (t-x_0)} \right|^2,
\end{equation}
where by $e_m$ we mean $e_m(|k|)$. Now, setting $\Delta_k \to y J_0$ and assuming $J_0 \ll \Omega$, we obtain
\begin{equation} 
P_m =  \left| \int_{-\infty}^\infty \diff y \, \frac{\sqrt{J_0}e_m  e^{-\frac{y^2}{4}}}{\sqrt{2\pi\sqrt{2\pi}}}  e^{-iy J_0 (t-x_0)} \right|^2. 
\end{equation}

The excitation probability of each atom for the Gaussian pulse scattering is shown for different $\theta$ values in Fig. \ref{fig:exc}.  For the most part, we see similar behavior for the Gaussian pulse as for the decaying exponential pulse in Appendix \ref{sec:appendixpulsemarkovian}, where the left atom is excited with higher probabilities than others (except for $\theta=\pi$, then the excitations are the same) and the excitation probabilities are higher for $\theta$ close to $\pi/2$.
Our numerical calculations show that, for a Gaussian pulse with $\Delta k=J_0$, the maximum probability of excitation for the first atom is achieved for $\theta=\pi/2$, where the separation between the imaginary parts of two poles ($p_1$ and $p_2$) is zero. For this case, the maximum probability  $P^{\mathrm{max}}_{-1}(t)\simeq 0.6266$ is achieved at $t\simeq x_0+0.713/J_0$. Similarly the (maximum) excitation probability is lowest for $\theta=\pi$, with $P^{\mathrm{min}}_{-1}(t)\simeq 0.075$ at $t=x_0+0.289/J_0$ and the separation between the imaginary parts of two poles ($p_1$ and $p_2$) are largest with $3J_0$.

Fig. \ref{fig:exc} reveals a very important property of this system: the excitation probability for a chain of atoms can exceed the maximum excitation probability $P^{\mathrm{max}}(t)=0.5 $ for a single atom. To understand this property, let us consider the excitation of a single qubit by a one-sided excitation pulse. The excitation probability of a single qubit upon one-sided excitation is bounded by $0.5$ \cite{liao2016photon}. This can be understood by decomposing the incident pulse to its even and odd modes, where only even modes can excite the single qubit. For a one-sided pulse, the decomposition is as follows
\begin{equation}
    \ket{S(k)}=\frac{1}{\sqrt{2}} \left( \ket{S_e(k)}+\ket{S_o(k)} \right),
\end{equation}
where all states are unit normalized and $\braket{S_e(k)}{S_o(k)}=0$. Then, the excitation probability can be given as
\begin{subequations}
\begin{align}
    P_{\rm single}(t)={}& |\braket{e_{\rm single}}{S(k)}|^2 \\
    ={}& \frac{1}{2} |\braket{e_{\rm single}}{S_e(k)}|^2,
\end{align}
\end{subequations}
where we note that a single qubit does not couple to odd modes. Here, $\ket{e_{\rm single}}$ is the excited state for a single atom. The term $|\braket{e_{\rm single}}{S_e(k)}|^2\leq 1$ corresponds to excitation of a qubit via an even pulse, where the upper-bound is achieved for a rising exponential in the even basis \cite{rephaeli2010full}. This leads to the bound $P_{\rm single}(t) \leq 0.5$. {Consequently, when three qubits are present, the collective system couples to the odd modes as well, leading to the increased excitation probability of the first qubit for certain cases.}

Numerical optimization of the integral in  (\ref{eq:13}) for a resonant Gaussian pulse reveals that the maximum excitation probability of the first atom is $\simeq 0.6356$ for $\Delta k = 1.175J_0$, $\theta=\pi/2$ and $t=x_0+0.685 J_0^{-1}$. 

\section{Time dynamics of observables in the non-Markovian regime} \label{sec:nonmarkgen}
In this section, we consider the non-Markovian regime, where the Markovian approximation is no longer valid. In this regime, the propagation time of photons within the multi-qubit system is no longer neglected, since the qubits are separated by large distances $L\sim O(J_0^{-1})$. Consequently, the time evolution dynamics of single excitation states can no longer be simplified to a residue calculation. Rather, a single integral must be calculated to obtain the time evolution. We show that the simplicity of this approach gives access to previously unexplored physics.

While it is possible to find fully analytical solution using the real-space formalism for the non-Markovian case, it requires a  longer discussion. For the scope of this paper, we  focus on numerical simulations and discussion of the underlying new physics coming from the non-Markovian behavior. We plan to introduce the analytical method in future work.

As in Section \ref{sec:Markovian}, we consider a chain of three qubits. We first study the time dynamics of atom excitation probabilities and emitted single-photon pulses, and then study the scattering of a single photon pulse from a system of multiple qubits. 

\subsection{Spontaneous emission} \label{sec:nonmarkspontem}
As we did in the Markovian case, here we consider spontaneous emissions with the initial condition where the central qubit ($m=0$) is excited, and begin by defining the relevant observables.

\subsubsection{Observables}

The observables of interest are, as in Section \ref{sec:Markovian}, the survival probability of the middle qubit $P_e(t)$, the excitation probability of the side qubits $P_s(t)$, the probability density of emitted photons $\mathcal{P}(x,t)$, the probability that photons are radiated outside of the system $P_w(t)$ and the probability that photons are trapped inside the system $P_b(t)$. 

\begin{figure*}
    \centering
    \includegraphics[width=\textwidth]{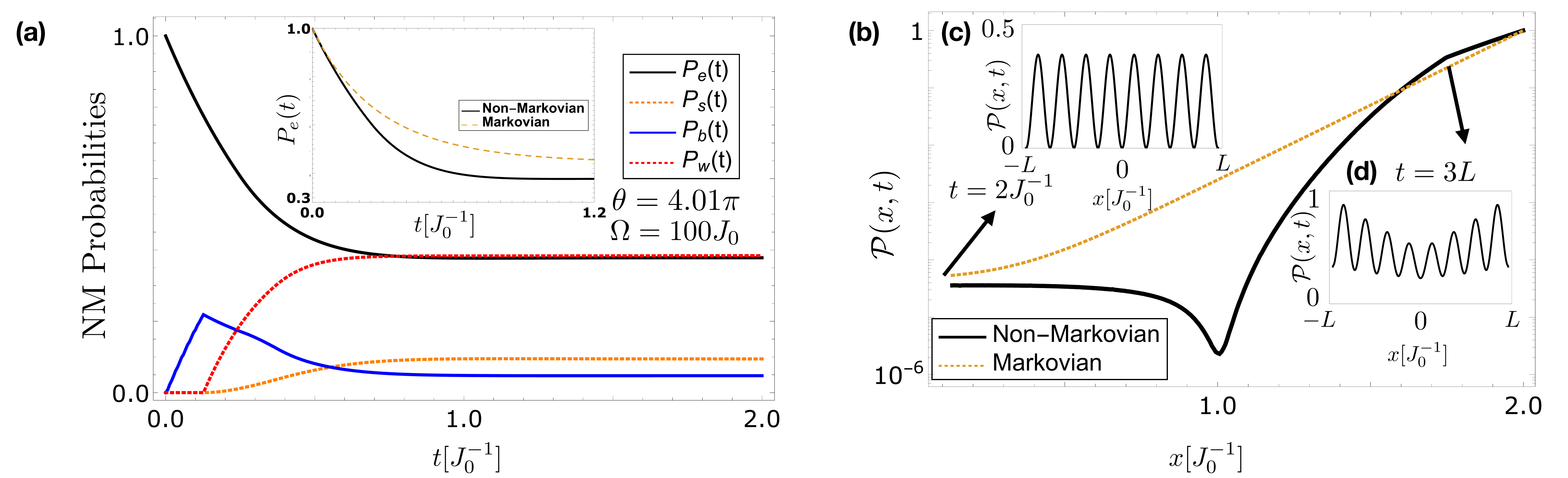}
    \caption{Spontaneous decay of an initially excited middle qubit for $\theta=4.01\pi$. (a) The survival $P_e(t)$ and side-atom excitation $P_s(t)$ probabilities, as well as the probability that the photon stays bounded within the interval $[-L,L]$ ($P_b(t)$) and the probability that the photon gets radiated to the waveguide ($P_w(t)$) are shown for the non-Markovian (exact (NM)) case. The inset figure shows the comparison between the survival probability in the Markovian limit and the exact non-Markovian probability. (b) The comparison between the emitted photon probability density in the Markovian limit and the exact non-Markovian counterpart at $t=2J_0^{-1}$. (c) and (d) illustrate the probability densities of the emitted photon within $[-L,L]$ and the formation of the quasi-bound state. For all figures, $\theta=\Omega L =4.01\pi$ and $\Omega=100J_0$. }
    \label{fig:non-mark-spont}
\end{figure*}

Using the time evolution operator and the Born rule, the observables can be computed, analogously to the Markovian case:
\begin{subequations}
\begin{align}
    P_e(t) &= \left| \int_{0}^\infty \frac{\diff k}{\pi} |e_0|^2 e^{-i\Delta_k t} \right|^2, \\
    P_s(t) &= \left| \int_{0}^\infty \frac{\diff k}{2\pi} |e_1 + e_{-1}|^2 e^{-i\Delta_k t} \right|^2, \\
    \mathcal{P}(x,t) &= \left| \int_{-\infty}^\infty  \frac{\diff k}{2\pi}\braket{x}{E_k} e_0^* e^{-i\Delta_k t} \right|^2, \\
    P_w(t) &= \left(\int_{-\infty}^{-L} + \int_L^\infty\right) \diff x \mathcal{P}(x,t), \\
    P_b(t) &= \int_{-L}^{L}\diff x \mathcal{P}(x,t).
\end{align}
\end{subequations}
Here, $e_m$ are the excitation coefficients computed for $k>0$ values, i.e. for scattering eigenstates where the field is initially incident from far left. It is important to note that the phase is no longer linearized and we keep the definition of $\theta=\Omega L \neq kL$ for convenience. 

\subsubsection{Numerical results}

First, let us consider the time evolution of observed quantities as shown in Fig. \ref{fig:non-mark-spont}. In this figure, $\Omega = 100 J_0$ and $\theta=4.01\pi$ (correspondingly, $L=4.01\pi \Omega^{-1}\simeq 12.6 \Omega^{-1}$). This value of $\theta$ is chosen specifically so that we can observe the formation of the quasi-bound states, where the excitation is trapped inside the system for longer times due to highly subradiant decay modes. (Recall that for $\theta=n\pi$, there are BIC in this system as discussed in Section \ref{sec:energyeigenstate}.) 

Fig. \ref{fig:non-mark-spont}(a) illustrates the time evolution of observable quantities. Since $\theta=4.01\pi$ includes both superradiant and subradiant decay modes, the survival probability $P_e(t)$ decays quickly for earlier times. Similarly, $P_b(t)$ increases rapidly until time $t=L$, where the kink at $t=L$ implies that the excitation is transferred to $P_w(t)$ as the photon leaves the system. There is an interesting feature happening both at $P_w(t)$ and $P_s(t)$. They remain zero until $t\geq L$, which is a manifestation of the causality principle. We plan to  discuss causality in the real-space formalism in detail in future work, for now it is important to note that there has been a long discussion about whether the rotating-wave approximation leads to non-causal behavior \cite{fermi1932quantum,hegerfeldt1994causality,milonni1995photodetection}. A quick recap on the causality behavior of the single qubit case can be found in Appendix \ref{sec:appendixa}. 

For later times in Fig. \ref{fig:non-mark-spont}(a), the survival probability almost saturate around non-zero values and decays extremely slowly such that the excitation stays within the system even for longer times. This marks the formation of the quasi-bound state, where the highly subradiant pole, which was also shown in Fig. \ref{fig:non-mark-col-dec}, dictates the collective behavior. Similarly, $P_b(t)$ stays non-zero for longer times, meaning that some portion of light is trapped between the qubits. The inset figure shows the inability to describe the system in the Markovian limit, as the survival probability calculated using the Markovian approximation can no longer match the exact dynamics. 

The inset figure in Fig. \ref{fig:non-mark-spont}(a) confirms our surprising finding about SSR related to  Fig. \ref{fig:non-mark-col-dec}, where the non-Markovian decay rates can be larger than the sum of the individual qubit decay rates (which is the maximum for the Markovian collective decay rates).  We believe that this may be a case of a self-stimulation process due to time-delayed coherent quantum feedback (\cite{calajo2019exciting} exploits this feedback mechanism to excite a BIC through multi-photon scattering), where after $t=2L$, the initially emitted photon density travels back to the middle qubit and stimulates further emission. {This explanation becomes more likely  when we notice the kink at $x=t_f-2L\simeq 1.75J_0^{-1}$ ($t_f=2J_0^{-1}$ is the time of the snap-shot shown in Fig. \ref{fig:non-mark-spont}(b)) for the radiated photon density $\mathcal{P}(x,t_f)$ as shown in Fig. \ref{fig:non-mark-spont}(b). For $1.75J_0^{-1} \leq x \leq 2J_0^{-1}$, the emitted photon density has a monotonic behavior, where the decay in $\mathcal{P}(x,t_f)$ is less than the Markovian case. Since the self-stimulation process can occur only after $t=2L$, we observe a change in the behavior of $\mathcal{P}(x,t_f)$ at $x=t_f-2L\simeq 1.75J_0^{-1}$ and the decay becomes more rapid than the Markovian case.} In any case, the more rapid decay in the non-Markovian case is linked to the interference between the emitted photon that has travelled back to the qubit position and the initially excited qubit itself. 

The Figs. \ref{fig:non-mark-spont}(c-d) illustrate the quasi-bound photon probability density at times $t=3L$ and $t=2J_0^{-1}$. The arrows mark the radiated probability density corresponding to those times. For $t=2J_0^{-1}$, the quasi-bound state is formed and the photon probability density takes the form of an (almost) sinusoidal function and is nearly zero at the qubit positions. For $t=3L$, the quasi-bound state is not completely formed but is in superposition with the portion of light that is to radiate outside of the waveguide. For $t=2J_0^{-1}$, the quasi-bound state is formed and leaks to the continuum with a subradiant decay rate. The two most subradiant (NM) collective decay rates for this case are $\Gamma_1 \approx (0.000057-i0.02)\gamma_0$ and $\Gamma_2\approx(0.001-i0.05)\gamma_0$. The nonzero imaginary components of the collective decay rates corresponds to the slight detuning in the frequency of the quasi-bound state and the real component corresponds to the actual exponential decay. As $\theta \to 4 \pi$, both collective decay rates tend to zero, where the actual bound state is formed. To describe the exact case of bound-state formation, one could either use the usual time evolution operator for $\theta \neq n\pi$ given in (\ref{eq:timeevolutionop}) and take the limit as $\theta \to 4 \pi$ or use the time evolution operator given in (\ref{eq:timeevol2}) that includes bound state contributions. 

\begin{figure}[h!]
    \centering
    \includegraphics[width=\columnwidth]{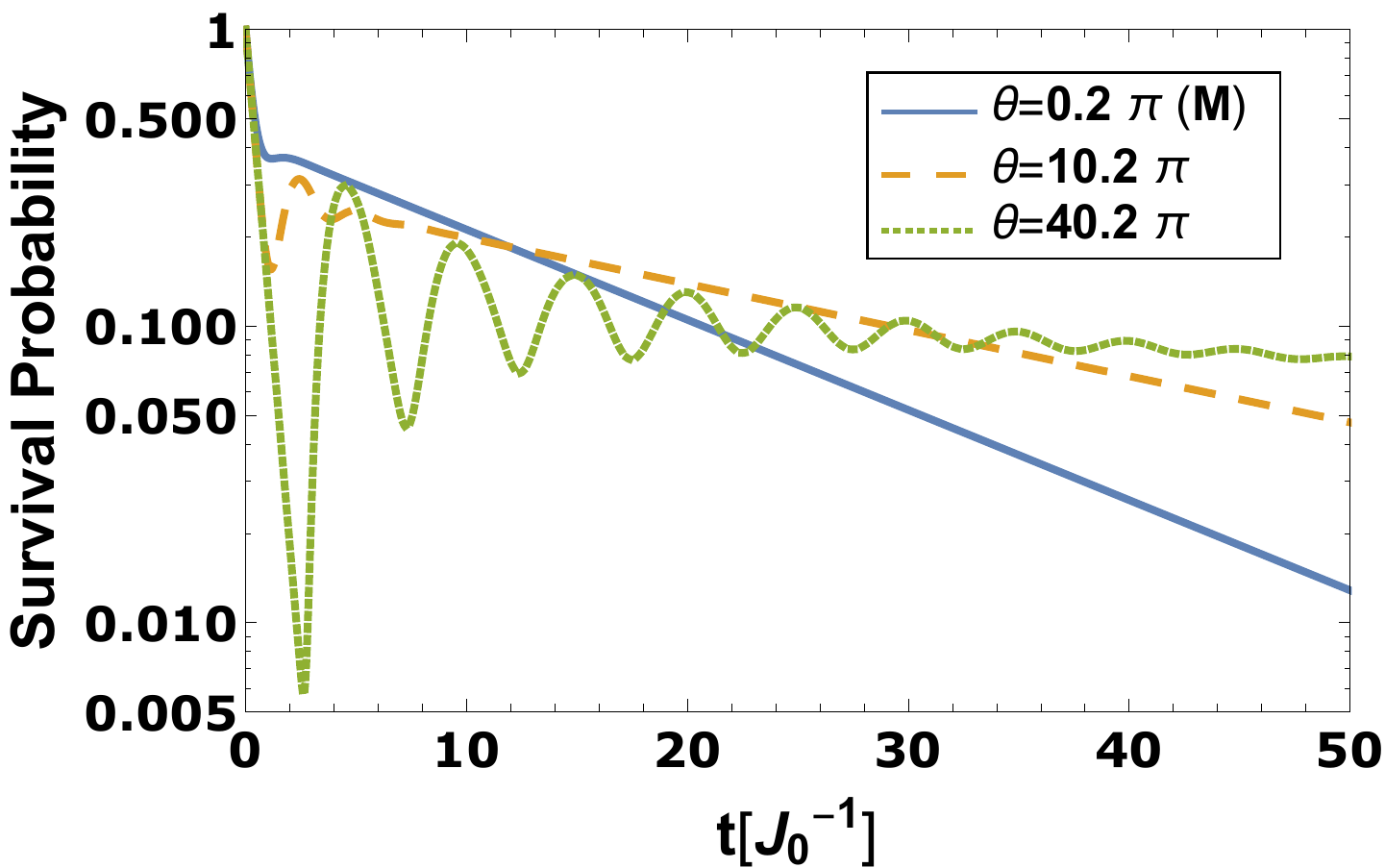}
    \caption{Exact survival probabilities of an initially excited middle qubit for $\theta=0.2\pi + n \pi$, where $n=0,10,40$. In the Markovian limit, all three cases are equivalent, whereas only $\theta=0.2\pi$ case agrees with the Markovian limit. For larger $\theta$, the survival probabilities show non-Markovian characteristics more dominantly. Here, $\Omega=100J_0$.}
    \label{fig:non-mark-surv}
\end{figure}

Non-Markovian behavior is usually associated with oscillatory behavior and beating effects \cite{zheng2013persistent,valente2016non}. This is easily understandable for the multi-qubit case that we consider here. As the time delay introduced by inter-system propagation becomes more significant, the survival probability oscillates rapidly and the oscillation period grows more and more linearly with $L$. While the oscillation period and the strength of oscillations grow, the overall decay rate decreases due to the non-Markovian effects as we have discussed in Section \ref{sec:coldecratnonmark}. This leads to a situation depicted in Fig. \ref{fig:non-mark-surv}. For this system, as the qubits are close to each other such that $\theta=0.2\pi$, the survival probability agrees with the results obtained via the Markovian approximation. For $\theta=10.2\pi$ and $\theta=40.2\pi$, the Markovian approximation is no longer valid, although the survival probabilities would be equivalent in the Markovian limit. As the distance between the qubits increases, the survival probability gains oscillatory behavior and the excitation is trapped within the system for longer times, as we have suspected. 

In this section, we used the real-space formalism to describe non-Markovian behavior of a spontaneously decaying multi-qubit system. While previous work either resorted to modifications of their Markovian formalism or drawing analogies to/taking limits of similar physical models,
we used only textbook quantum theory to describe complicated non-Markovian behaviour. This is only possible because the real-space formalism we introduce in this work is elementary and requires little prior knowledge to perform calculations or generalize to other multi-qubit/multi-waveguide systems. Next, we consider a scattering problem for completeness. 

\subsection{Pulse scattering for single-photon states}\label{sec:nonmark}
In Section \ref{sec:pulsescat}, we  considered scattering problems where $k L \simeq \Omega L$ is a valid assumption. In this section, we investigate a scattering experiment where the qubit separation is large, $L \sim O(J_0^{-1})$ instead of $L\sim O(\Omega^{-1})$, and therefore the linearization assumption is no longer valid. Here, we find that the back-and-forth exchange of the photon between atoms becomes more apparent and echoes are observed in the transmitted and reflected pulses with intervals $\simeq 2L$, which can be interpreted as the delayed feedback time following the approach of \cite{calajo2019exciting}, implying that some portion of light is trapped inside the system. Some fraction of the trapped light is released  upon collusion with the side atoms and we observe it as echoes. 

Unlike in the Markovian limit, the description of time evolution in the non-Markovian regime requires knowledge of BICs, and numerical integration must be employed to describe  time evolution of both regular and irregular states (unless a more complicated re-summation method is used, which we will introduce in a future work). In such cases, however, computing the evolution governed by (\ref{eq:timeevol2}) boils down to a single numerical integral; just as it does for irregular states in the Markovian limit. In the past, scattering of irregular states in both regimes was studied using complicated numerical methods such as the finite-difference time-domain (FDTD) method \cite{liao2015single,liao2016photon}. The numerical approach proposed here is much simpler. 

\subsubsection{Observables}

The excitation probabilities of individual atoms can be found using (\ref{eq:13}). Since the atomic separation is $O(J_0^{-1})$, the interior field between the atoms can no longer be neglected. To distinguish between the right-moving and left-moving modes between  the atoms, we need to identify their respective contributions to the probability density function, $\mathcal{P}(x,t)$. The transmitted (right moving) and reflected (left moving) pulse probability density functions are $\mathcal{P}_{R}(x,t) = |\bra 0 C_R(x) \ket{S(t)}|^2$ and $\mathcal{P}_{L}(x,t) = |\bra 0 C_L(x) \ket{S(t)}|^2$. The total probability density function is therefore $\mathcal{P}(x,t) =\mathcal{P}_R(x,t)+\mathcal{P}_L(x,t)\hspace{1mm}+\hspace{1mm}$``highly oscillating terms that average to zero'', which is properly normalized\footnote{To understand how dividing the probability density $\mathcal{P}(x,t)$ into right- and left-moving pulses preserves the normalization, we can write the most general waveform for the pulse as $\psi(x,t)=\psi_R(x,t)e^{ik_0x} +\psi_L(x,t)e^{-ik_0x}$, where $\psi_{R/L}(x)$ is the waveform for the right-/left-moving pulse with zero average momentum and $k_0$ is the average momentum of the initial pulse. The probability density function of the general pulse is $\mathcal{P}(x,t) = |\psi(x,t)|^2=|\psi_R(x)|^2 + |\psi_L(x)|^2 + 2 Re[\psi_R(x) \psi_L^*(x) e^{2ik_0x}]$. The interference term oscillates rapidly and can be neglected (its integral averages to $0$) as long as $k_0 \Delta x \gg 1 $. Since $J_0\ll k_0$ and $\Delta x \sim O(J_0^{-1})$, the normalization is preserved. } in the limit $J_0\ll \Omega$ . 

\subsubsection{Numerical results}

\begin{figure*}
\centering
\includegraphics[width=16cm]{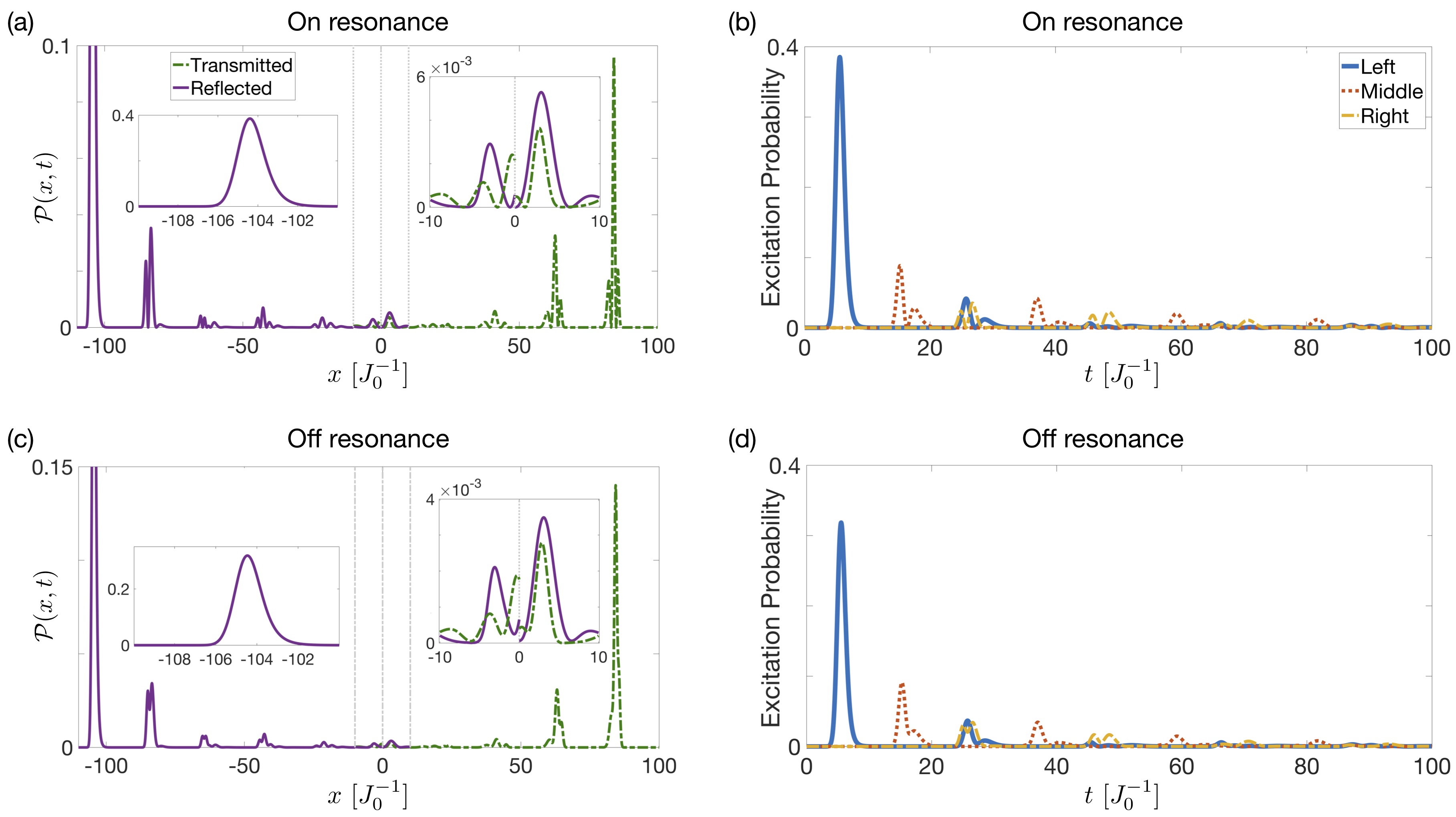}
\caption{Scattering of a single-photon Gaussian-shaped pulse by a three qubit system with a large atomic separation $L=10/J_0$, where the atoms are either on or off resonance with the photon. In (a) and (c), we show emitted photon probability densities for reflected $\mathcal{P}_L(x,t)$ and transmitted $\mathcal{P}_R(x,t)$ modes. The vertical dotted lines indicate the positions of the qubits.  Note that there are periodic photon emissions (echoes) with intervals $\sim 10/J_0$ due to the trapped light inside the system. Moreover, the first peak of reflected light and the excitation probability of the first atom are identical to the ones in the single-atom system explored in Appendix \ref{sec:appendixa}, hinting that the atomic separation is large enough such that the pulse can completely pass through the first atom before reaching another. For both cases, $t=100J_0^{-1}$ and $x_0=15J_0^{-1}$. The insets are zoomed-in portions of the larger plots. Note that while the transmitted and reflected parts of the field are discontinuous at $x=0$, the entire field is actually continuous. In (b) and (d) we show the corresponding excitation probabilities of the individual atoms as a function of time. } \label{fig:final}
\end{figure*}

Fig. \ref{fig:final} shows the single-photon pulse echoes,  as well as  oscillations in atomic excitation probabilities for resonant and non-resonant Gaussian pulse scattering with $L=10/J_0$. The echo interval is $\simeq 2L$ as expected. For illustration purposes, we choose $x_0=15/J_0$, $t=100/J_0$, $\Omega=100J_0$ and $\chi=J_0$ for the non-resonant Gaussian pulse (See Equation (\ref{eq:gaussian})). Since the qubit separation is larger than the pulse-width in real-space, the excitation of the qubits can be considered as a local process \footnote{While we do not show it here explicitly, each pulse shape shown in Fig. \ref{fig:final} can be computed by considering local transmission and reflection of the Gaussian from each qubit. For example, the pulse localized at $x=-85J_0^{-1}$ in Fig. \ref{fig:final}(a) can be obtained by multiplying the frequency spectrum of the Gaussian pulse with two transmission and one reflection coefficients for a single qubit, thus illustrating the locality of the interactions.}. As explained in the previous section, for local interactions between a qubit and one-sided excitation, the atomic excitation probability is bounded by $0.5$ due to odd modes not coupling to the qubit. This shows another difference between the Markovian and non-Markovian regime, as for pulses confined within $\Delta x \ll L$, the atomic excitation cannot exceed $0.5$ in the non-Markovian regime.

\section{Beyond  conventional problems}\label{sec:beyond}
Up to this point, we focused on conventional waveguide QED problems, namely spontaneous emission and scattering in one-dimensional chains. 
In this section, we {sketch out} new problems that are easily accessible only through the machinery of the real-space approach. 

\subsection{Relationship between collective decay rates and symmetry}\label{sec:symmetry}

In this section, we touch upon a possible causal relation between the collective decay rates of a system and its symmetric and anti-symmetric coherent excitations. Our preliminary investigations suggest that certain states cannot have access to the complete decay rate spectrum. First, we will discuss what we mean by (anti-) symmetry for a linear chain of $N$ qubits. As a motivation for this complex problem, we will then discuss the case of two qubits and set up the big picture for future work. 

It is important to note that the symmetric and anti-symmetry coherent excitations of qubits form a basis for the qubit subspace. For even $N$, the basis states can be given as
\begin{subequations}
\begin{align}
    \ket{E_j}&=\frac{1}{\sqrt{2}} \left( \ket{e_j}+ \ket{e_{N+1-j}}  \right), \\
    \ket{O_j}&=\frac{1}{\sqrt{2}} \left( \ket{e_j}- \ket{e_{N+1-j}}  \right),
\end{align}
\end{subequations}
where $\ket{E_j}$ and $\ket{O_j}$ (each with dimension $N/2$ for even $N$ and $(N-1)/2$ for odd $N$) correspond to symmetric and anti-symmetric basis states. For odd $N$, there is an additional even state $\ket{e_{(N+1)/2}}$ that completes the basis. As we argue in this section, these basis states are more physical than we first realize, where we conjecture that they divide the collective decay rates into two subsets: symmetric and anti-symmetric collective decay rates. The symmetric (anti-symmetric) states have access only to symmetric (anti-symmetric) collective decay rates. 

To motivate this, let us consider  two identical qubits coupled to a 1D waveguide, where the qubits are situated at $x=\{-L/2,L/2\}$. The symmetric and anti-symmetric qubit states for the $N=2$ are
\begin{subequations}
\begin{align}
    \ket{E} = \frac{1}{\sqrt{2}} \left( \ket{e_1} + \ket{e_2} \right), \\
    \ket{O} = \frac{1}{\sqrt{2}} \left( \ket{e_1} - \ket{e_2} \right).
\end{align}
\end{subequations}
{Moreover, the collective decay rates of the two-qubit system (calculated in (\ref{eq:nonsymdecay}) for two non-identical qubits)} become
\begin{align} 
    \Gamma_{1/2} &= \gamma_0 \left(1 \pm e^{i\theta} \right).
\end{align}
Let us consider the time evolution of $\ket{\psi(0)}=\ket{E}$. Recall from (\ref{eq:arbitrary2}), where we take {$p=-i\Gamma/2$, its time evolution is
\begin{align}
   \ket{\psi(t)} &= \sum_{j=1}^2 \underset{\Delta_k = -i\Gamma_j/2}{\text{Res}} \left[ \ket{g(k)} \ee^{-i\Delta_kt} \right]\,,
\end{align}}
where $\ket{g(k)}=\frac{1}{\sqrt{2}} \left(e_1^* + e_2^*\right) (\ket{E_k}+\ket{E_{-k}})$ with $e_{1/2}$ corresponding to excitation coefficients for the first and second qubits. After straightforward algebra, we find the time evolved state:
\begin{align} \label{eq:forappendixd}
\begin{split}
        &\ket{\psi(t)}={} \frac{1}{\sqrt{2}} e^{-\left(\frac{\Gamma_1}{2} +i\Omega\right)t} (\ket{e_1} + \ket{e_2}) \\
        &- i \int_{0}^t \frac{ \diff x\Gamma_1}{2\sqrt{2J_0}} e^{-\left(\frac{\Gamma_1}{2}+i\Omega\right)\left(t-x+\frac{L}{2}\right)}C_R^\dag(x) \ket{0}\\
        &-i \int_{-t}^{0}  \frac{\diff x\Gamma_1}{2\sqrt{2J_0}} e^{-\left(\frac{\Gamma_1}{2} + i \Omega\right) \left(t+x+\frac{L}{2}\right)} C_L^\dag(x) \ket 0 ,
\end{split}
\end{align}
where we note $J_0/\Omega \ll 1$ implies that $ \Gamma_1 L \ll 1$. Moreover, the position ($x$) and time ($t$) have units $J_0^{-1}$. It is important to note that $\hat P \ket{\psi(t)}=\ket{\psi(t)}$ such that the state stays symmetric for later times $t$. This is expected, as the parity is a conserved quantity since it commutes with the Hamiltonian. We show that $\ket{\psi(t)}$ satisfies the Schr\"odinger equation in the Markovian limit in Appendix \ref{sec:appendixd}. As can be seen from this expression, the symmetric coherent excitation of the qubits decay through only one of the decay modes, corresponding to the decay rate $\Gamma_1$, whereas it does not have access to the anti-symmetric decay mode $\Gamma_2$. A similar calculation shows that $\ket{O}$ does not have access to $\Gamma_1$ either. This observation agrees with our findings for $N=3$  in Section \ref{sec:Markovian} and the findings of \cite{tsoi2008quantum}, where for $N=5$, the survival probability of the initially excited middle qubit depends only on $3$ decay rates {and not $5$ of them\footnote{Note that number of symmetric collective decay rates is $N/2$ for $n$ even and $(N+1)/2$ for odd $n$, inline with the dimensionality of symmetric states. Similarly, the number of anti-symmetric decay rates is also consistent with the dimensionality of the anti-symmetric subspace}.} Consequently, we conclude this discussion with the following conjecture
\begin{conj}[Symmetric and Anti-Symmetric Collective Decay Rates]
  The space of collective decay rates is divided into two subspaces called  symmetric and anti-symmetric collective decay rates. The symmetric (anti-symmetric) states, that have $\pm 1$ eigenvalues under the general parity transformation $\hat P$, have access to only symmetric (anti-symmetric) collective decay rates.
\end{conj}
We leave the proof of this conjecture as an open problem. We believe the alternative method of finding collective decay rates, that we discuss in Appendix \ref{sec:appendixalternativemethod}, can be beneficial for proving this conjecture. {Further evidence for the conjecture can be found by computing the eigenvectors of the coupling matrix discussed in this appendix and checking that for any eigenvector $\ket{\xi}$, $\hat P \xi= \pm \xi$ holds.}

\subsection{Towards pulse-shaping with quantum emitters: exploiting collective decay rates} \label{sec:pulseshaping}

In Section \ref{sec:pulsescat} (and Appendix \ref{sec:appendixa}) we saw that the shape of the single photon strongly affects the excitation probability of a given system. 
To ensure high-efficiency coupling to different systems, it is therefore important to prepare pulses with different shapes.

To explore this possibility,  we first write down the most general form for the symmetric emitted pulse probability density of a black box system, that is coupled to a 1D waveguide, by generalizing (\ref{eq:emittedphoto}) for a general transmission and reflection coefficients $t_b$ and $r_b$ (and symmetric excitation coefficient $e_b$ for a coherent single excitation of qubits)
\begin{equation} \label{eq:pulseshaping}
      \mathcal{P}(x,t) = \left| \underset{\Delta_k = p}{\text{Res}} \left[(t_b + r_b ) e_b^* e^{-i\Delta_k (t-|x|)}\right] \right|^2,
\end{equation}
\noindent
where we realize that $\braket{x}{E_k}=t_b e^{ikx} \theta(k) + (r e^{ikx} + e^{-ikx})\Theta(-k)$  for $x>0$. Terms containing $e^{ikx}$ vanishes due to causality arguments. Considering (\ref{eq:pulseshaping}), we can clearly see that the shape of the emitted pulse is dictated by the poles (hence, collective decay rates) of the system. Therefore, the decaying exponential shape emitted from the single qubit can be traced back to the single pole of $-iJ_0$ of the scattering parameters. 

\begin{figure}[h!]
    \centering
    \includegraphics[width=\columnwidth]{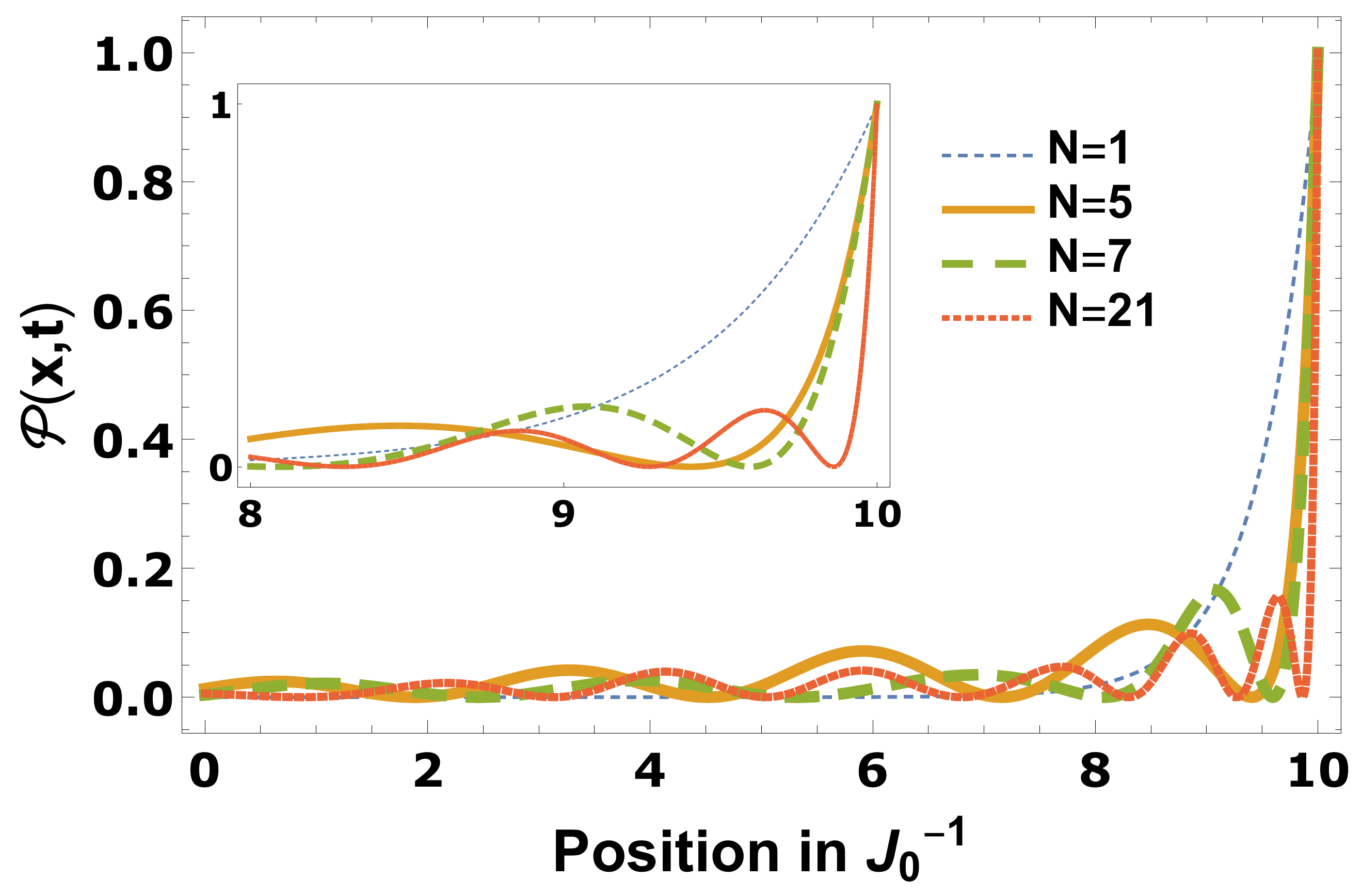}
    \caption{The emitted photon probability density $\mathcal{P}(x,t)$ (normalized w.r.t. $J_0$) for $N=1,5,7,21$ identical qubits in a linear chain, where $\theta=\pi/2$ and $t=10J_0^{-1}$.}
    \label{fig:pulsehaping2}
\end{figure}

This opens up the possibility for  pulse-shaping using quantum emitters. By adjusting the collective decay rates of the complete system, one can prepare pulses that might be beneficial for various applications in quantum information processing. This adjustment can be made through any of the following:
\begin{itemize}
    \item Increasing the number of qubits in the system
    \item Introducing qubits with different energy separation $\Omega_j$ and coupling energies $J_j$ 
    \item Using multiple waveguides and $d>1$ dimensional topologies of waveguide/qubit systems
\end{itemize}
To see the effects of changing the qubit number on the emitted photon probability density, see Fig. \ref{fig:pulsehaping2}. For larger $N$, the initial decay is faster, whereas the tail of the exponential remains finite for longer times. We plan to discuss this further in  future work.

\subsection{Breaking bounds: near-perfect coherent excitation with a one-sided pulse} \label{sec:beast-mode-on}
In this section, we demonstrate the 
power and efficiency of the real-space formalism by considering pulse scattering from a system of up to $500$ identical qubits coupled to a 1D waveguide. For comparison,  \cite{liao2015single} considered  time evolution in the presence of $N=5$ identical qubits and \cite{albrecht2019subradiant} considered the case of $N=10$ qubits in the subradiant regime. We pick the qubits to be identical for simplicity, but modelling non-identical qubits is equally computationally efficient. The results obtained in this section are computed via a standard personal computer and not using a cluster.

\begin{figure}[h!]
    \centering
    \includegraphics[width=\columnwidth]{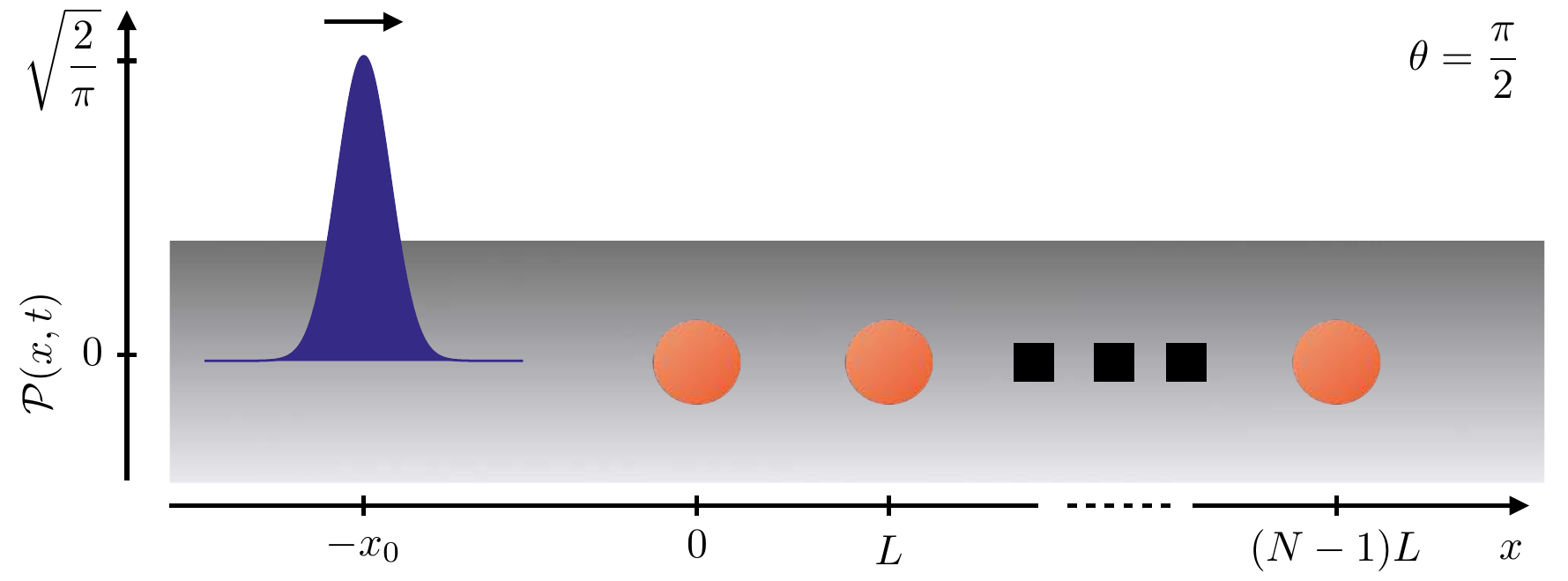}
    \caption{A Gaussian pulse, initially located at $x=-x_0$, is incident upon a linear chain of $N$ qubits. The qubits are initially in the ground state and are separated by quarter (resonance) wavelength ($\theta=\pi/2$).}
    \label{fig:opt3}
\end{figure}

Similar to Section \ref{sec:pulsescat}, let us consider a pulse that is incident from far left (see Fig. \ref{fig:opt3}). The initial state can be given as in (\ref{eq:pulsein}), where $k_0=\Omega$ for a resonant pulse. We consider two types of pulses: Gaussian  and Rising Exponential.

\subsubsection{Example: Gaussian pulse} 
Since the Gaussian pulse is an irregular state, the observables can be expressed by a simple integral, which can be computed via numerical integration methods.

For the Gaussian pulse,
\begin{subequations}
\begin{align}
      f(x) &= \frac{\sqrt{\sigma\sqrt{2}}e^{-\sigma^2 x^2}}{\sqrt{\sqrt{\pi}}}, \\
      \tilde f(k) &= \frac{e^{-\frac{k^2}{4\sigma^2}}}{\sqrt{\sigma\sqrt{2\pi}}},  
\end{align}
\end{subequations}
where we call $\sigma$  the pulse width (or frequency deviation)  in units of $J_0$. Then, the excitation probability of the $m$th qubit, where $m=1,\ldots,N$, becomes
\begin{equation}
    P_m(t)=\left| \int_{-\Omega}^\infty \diff \Delta_k e_m  \frac{ e^{-\frac{\Delta_k^2}{4 \sigma^2}}}{\sqrt{2\pi\sigma\sqrt{2\pi}} } e^{-i\Delta_k (t-x_0)} \right|^2,
\end{equation}
\noindent
where $e_m$ is the excitation coefficient of the $m$th qubit obtained with the initial conditions $t_1=0$ and $r_{N+1}=0$ in (\ref{eq:non-id-field-amp}). 

\begin{figure}[t!]
    \centering
    \includegraphics[width=\columnwidth]{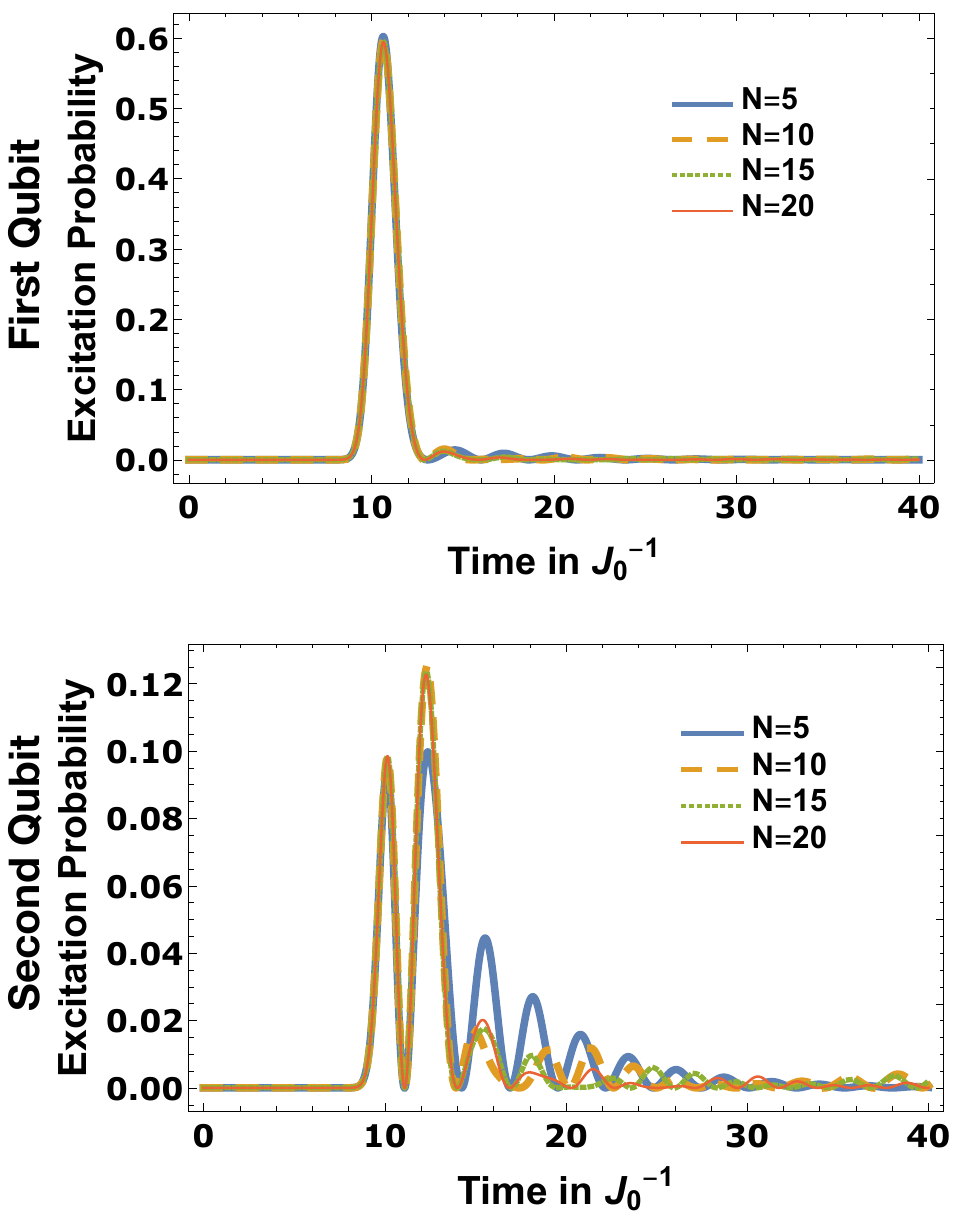}
    \caption{Excitation probabilities of first and second  qubits in a linear chain of $N=5,10,15,20$ qubits upon a resonant Gaussian pulse incidence. The pulse is situated at $x_0=10J_0^{-1}$ and has pulse width (frequency deviation) $\sigma=J_0$. Note that quantities are normalized with respect to $J_0$.}
    \label{fig:opt1}
\end{figure}

The excitation of the first and second qubit upon pulse incidence is shown in Fig. \ref{fig:opt1} for $\sigma=J_0$, $x_0=10J_0^{-1}$, $\theta=\pi/2$ and $N=5,10,15,20$ total qubits. For $N=5$, this reproduces the results of \cite{liao2015single}. As apparent from the figure, the excitation probability of the first qubit does not change significantly for various $N$ values, whereas the maximum excitation probability is somewhat saturated around $\simeq 0.60$. The excitation probability of the second qubit oscillates within time intervals $\sim O(J_0^{-1})$, which is caused by photon exchange between the many qubits inside the waveguide. Note that similar oscillations can be observed for the first qubit, although they are not as visible due to the large initial excitation probability.

\begin{figure}[t!]
    \centering
    \includegraphics[width=\columnwidth]{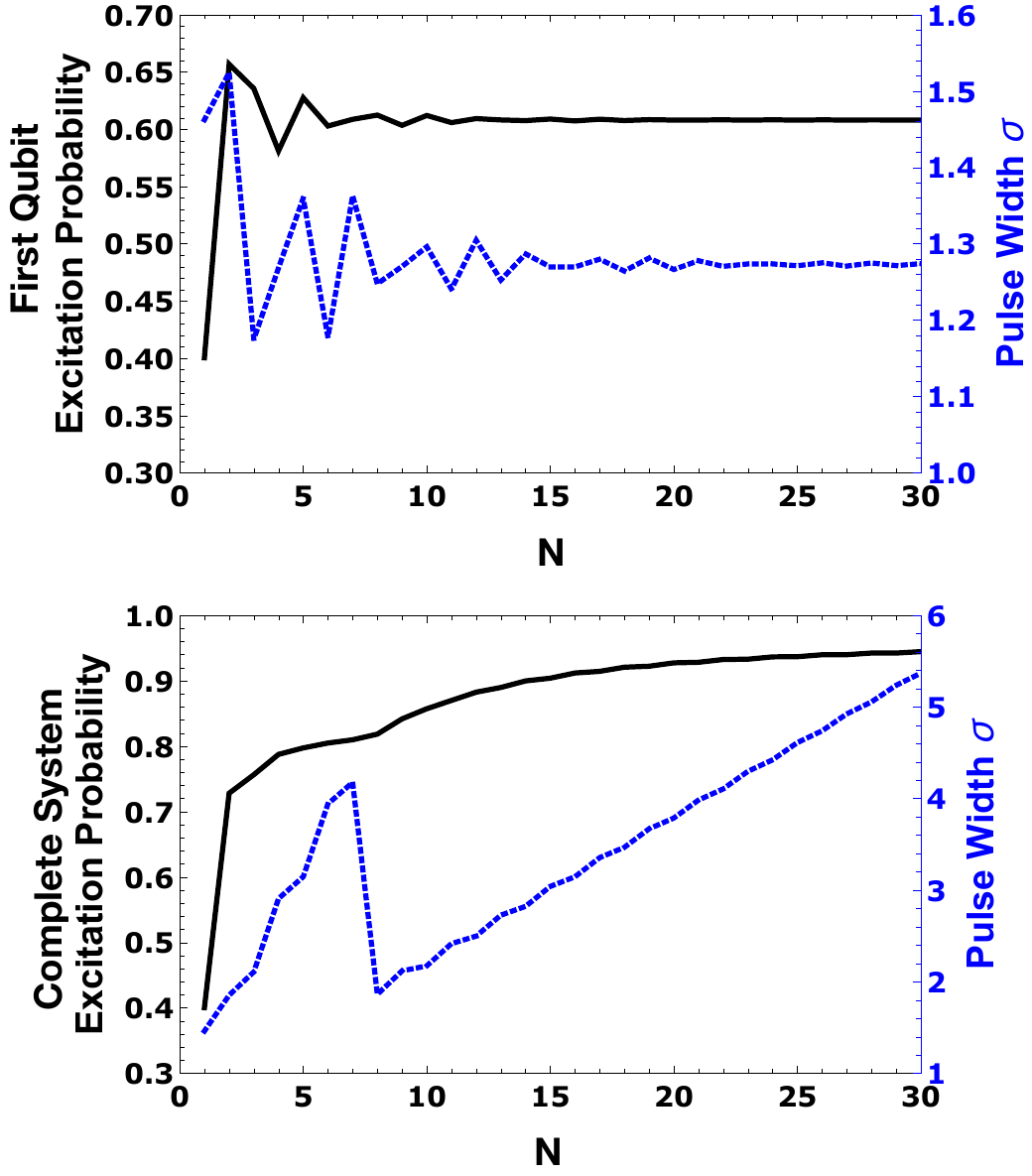}
    \caption{ The optimum excitation probability and corresponding $\sigma$ for the first qubit  and the complete system (all qubits). Note that quantities are normalized with respect to $J_0$.}
    \label{fig:opt2}
\end{figure}

As another observable, we consider the total excitation inside the multi-qubit system, which is defined as
\begin{equation}
    P_{\rm tot}(t) = \sum_{m=1}^N P_m(t).
\end{equation}
In Fig. \ref{fig:opt2}, we maximize both $P_1(t)$ and $P_{\rm tot}(t)$ for increasing qubit number $N$. We denote the maximum quantities as $P_1^{\rm max}$ and $P_{\rm tot}^{\rm max}$, respectively. Since both quantities are given by simple integrals, the numerical maximization can be carried out in Mathematica by using the NMaximize function. The saturation of $P_1^{\rm max}$ around $0.60$ is clear for $N\geq 10$, which has also been observed in Fig. \ref{fig:opt1}. Moreover, the total excitation probability increases monotonically and is $P_{\rm tot}^{\rm max}\simeq 0.9445$ for $N=30$. 

\subsubsection{Example: rising exponential}
For the rising exponential, we can obtain fully analytical expressions since the state is regular. Consequently, we can compute observables much more efficiently than the Gaussian case. For example, $P_{\rm tot}(t)$ can be computed for $N= 500$ qubits using a personal computer within 24 hours. 

For a rising exponential,
\begin{subequations}
\begin{align}
        f(x) &= \sqrt{2\sigma} e^{-\sigma x} \Theta(x),\\
        \tilde f(k) &= \frac{-\sqrt{\sigma}i}{\sqrt{\pi} (k-i\sigma)}.
\end{align}
\end{subequations}
Using these, the excitation probability for the $m$th qubit is
\begin{equation} 
    P_m(t)=\left| \int_{-\infty}^\infty  \frac{\diff \Delta_k}{2\pi}\frac{\sqrt{2 \sigma} e^{-i \Delta_k (t-x_0)}}{\Delta_k-i\sigma} e_m \right|^2.
\end{equation}
We note that the integrand of this expression has only one pole in the upper half plane (UHP) and $N$ poles (corresponding to $e_m$) in the LHP. The contour is closed from the UHP for $t<x_0$ and LHP vice versa. $P_e(t)$ is continuous at $t=0$ by construction. So, if we are clever about it, we can maximize this quantity by simply considering only the single UHP pole. First, let us realize that for $t<x_0$, $P_m(t)$ is described by a rising exponential, whereas for $t>x_0$, it is described by a decaying exponential due to the residue theorem. Then, $P_m(t)$ is maximized for $t=x_0$ regardless of $\sigma$. Then, we can calculate $P_m(t)$ for $t<x_0$ by using the only pole at $\Delta_k = i \sigma$, take the limit $t \to x_0$ and maximize $\sum_{m=1}^N P_m(0)$ over all possible $\sigma$ values. The result gives  the maximum total (coherent) excitation probability $P_{\rm tot}^{\rm max}$.

\begin{figure}[b!]
    \centering
    \includegraphics[width=\columnwidth]{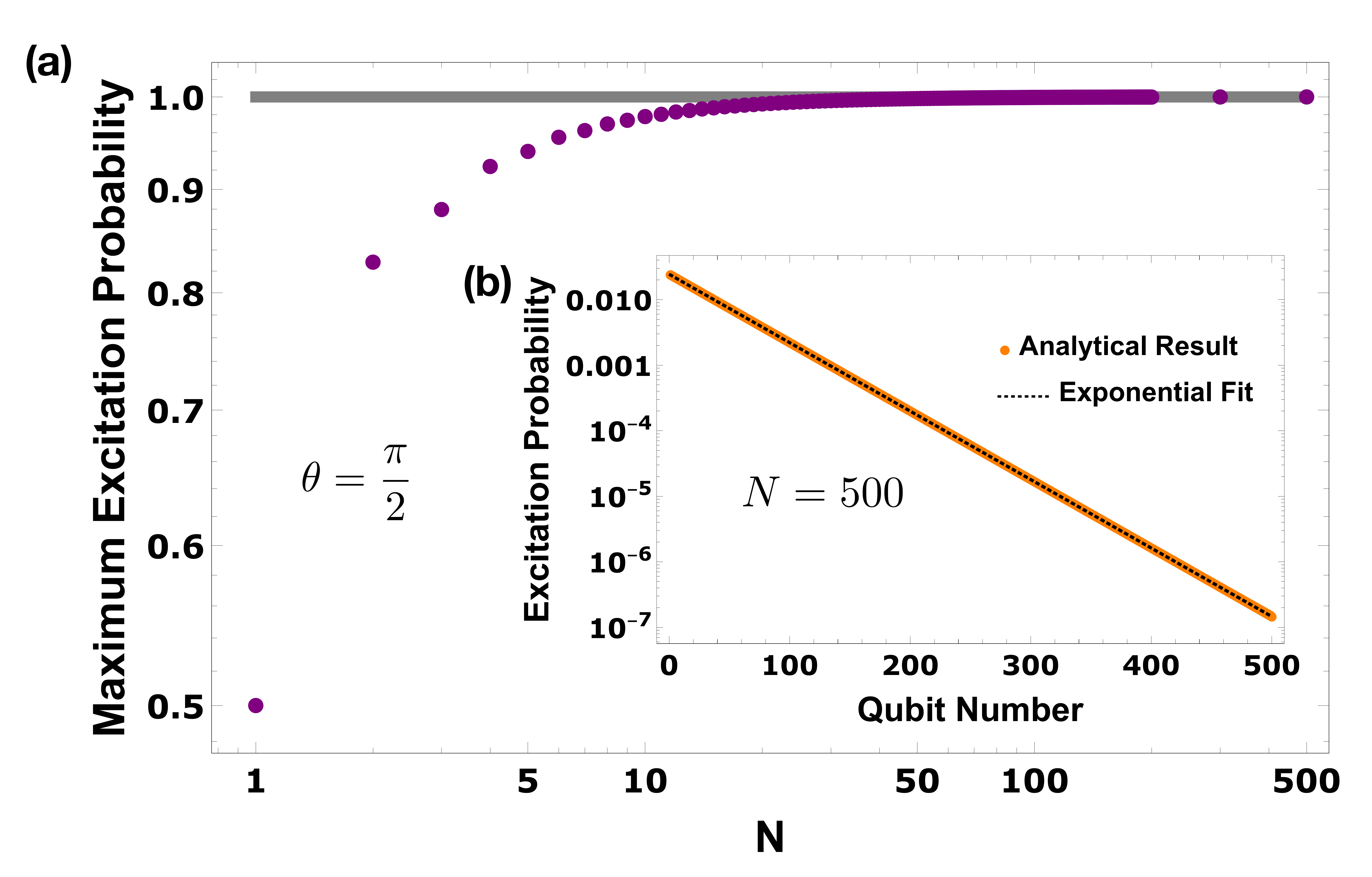}
    \caption{(a) The maximum total excitation probability, $P_{\rm tot}^{\rm max}$, upon incidence of a rising exponential pulse for $\theta=\pi/2$. The gray line marks the unity excitation probability. (b) The individual excitation probabilities of each qubit for $N=500$. Note that for $N=500$, $P_{\rm tot}^{\rm max}\simeq 0.99996$. }
    \label{fig:nearperfect}
\end{figure}

In Fig. \ref{fig:nearperfect}(a), we plot $P_{\rm tot}^{\rm max}$ that is obtained through maximizing over $\sigma$ values for each $N$. As $N$ increases, the total coherent excitation increases monotonically, where  $P_{\rm tot}^{\rm max}\simeq 0.99996$ for $N=500$. This shows that using a rising-exponential, one can achieve effectively a unity collective excitation probability using only a one-sided pulse, which is impossible for a single qubit as we show in Section \ref{sec:pulsescat}. Moreover, the excitation probabilities of each qubit follow an exponential fit as shown in Fig. \ref{fig:nearperfect}(b). We note that this exponential decay of excitation probability is not unique for $N=500$. In fact, we observe that the qubit excitations follow an exponential fit even for $N=10$, while the exponential fit becomes more fitting as $N$ increases. The reason behind this phenomenon is currently not known to us and is a subject for future work. 

While previous sections focused mainly on generalizing the theory and observing new physical phenomena, this section has shown the advantages of the real-space approach for modelling time evolution of single-photon states, where one can obtain fully analytical expressions for  time dynamics in the presence of even $500$ qubits. 

First, we reproduced some results from the literature and then considered multi-qubit systems that are two orders of magnitude larger than the ones considered so far in the literature. In doing so, we also showed that nearly-perfect coherent excitation can be achieved for a linear chain of qubits via a one-sided rising-exponential pulse. { Similar effects have been observed in the literature for cascaded setups \cite{cirac1997quantum,stannigel2011optomechanical}}

\section{Discussion} \label{sec:sum}

The virtue of the real-space approach lies in its elementary\footnote{Following the definition by Richard Feynman, elementary does not mean easy to understand. Elementary means that \emph{very little knowledge is required ahead of the time to understand something except to have an infinite amount of intelligence.}} nature. In developing the formalism in this work, we just had to find the energy eigenbasis, use the spectral theorem to construct the time evolution operator, and apply the Born rule to find observable quantities.  We did not introduce any approximations beyond those already established in the literature.  This is precisely the abstraction level taught in elementary  quantum theory  \cite{griffiths2010introduction}.

So, why can't we achieve the same with the momentum space approach? Because we can't diagonalize the Hamiltonian exactly in the momentum basis. To see why, consider the Heaviside function $\Theta(x)$ and its Fourier transform $\tilde \Theta(k)= \sqrt{\frac{\pi}{2}} \delta(k)+ \frac{i}{\sqrt{2\pi}k}$. Whenever we see discontinuities, we can identify the existence of $\Theta(x)$ and thus it is easier to construct superpositions of $\Theta(x)$ multiplied with complex superpositions. This is precisely how the Bethe Ansatz for the energy eigenstates are constructed for the real-space Hamiltonian. The same approach is not as trivial for the momentum space Hamiltonian, as there is not a clear physical intuition behind why $\tilde \Theta(k)$ would emerge in the scattering problem. 

One might be cautious about working with photon densities in position space, due to potential causality issues associated with photon wavefunctions. We have not encountered these yet. We see no violation of causality, nor have we seen it appear in previous work on the real-space approach. Furthermore, our results that overlap with those predicted using other methods match exactly. We plan, however, to explore this in future work. 

The real-space approach is  popular  for solving steady-state waveguide QED problems. But it might be dismissed  (see for example \cite{chen2011coherent})  because it requires calculating the stationary eigenstates, which might be considered too arduous if one's goal is  ``only'' steady-state solutions.  In this paper, we argued that finding the stationary eigenstates is a worthwhile pursuit, since they reveal much more than  the steady-state behaviour of a waveguide QED system. 

We demonstrated this approach for $N=3$ identical qubits in a linear waveguide, and later showed that it can be generalized to quantum networks with multiple emitters and multiple waveguides, as well as non-identical systems. We then used this approach to study some interesting physics. In particular, we studied spontaneous emission and scattering from multi-qubit systems. We also studied subradiance, superradiance and bound states in continuum. We  discussed new phenomena such as  subdivision of collective decay rates into symmetric and anti-symmetric subsets and non-Markovian superradiance effects that can lead to  collective decay stronger than Dicke superradiance. And we  discussed possible  applications such as pulse-shaping and coherent absorption. 

Our analyses led to several new insights about these systems: (1) the specific way in which collective decay rates dictate the collective behaviour of the system, (2) that exact causality emerges for the real-space approach, (3) that the non-Markovian behavior of multi-qubit systems is extremely easy to obtain using the real-space approach, where we simply omit the linearization of $kL$, (4) that the real-space approach explains the black box behavior we see in waveguide QED systems, and (5) that the real-space approach makes it possible to calculate exact non-Markovian dynamics. These insights are discussed in Appendix \ref{sec:int}. 

We also showed that this approach can be used to study time dynamics, of a photon interacting with up to 500 qubits, using  a personal computer. We therefore expect the formalism presented in this paper to enable the study of complicated quantum networks such as multi-dimensional waveguide arrays \cite{unpublished}. Furthermore, we expect that applications such as quantum logic \cite{sipahigil2016integrated}, quantum memory \cite{lvovsky2009optical}, quantum photon routing \cite{shomroni2014all,zhou2013quantum}, as well as quantum sensing \cite{degen2017quantum} and communication \cite{chang2007single,kimble2008quantum} will also benefit from analysis using the real-space approach for scattering phenomena. 

\section*{Acknowledgements}
\.{I}lke Ercan  acknowledges generous resources provided by the Electrical \& Electronics Engineering Department at Bo\u{g}azi\c{c}i University. Research at  Perimeter Institute is supported by the Government of Canada through the Department of Innovation, Science and Economic Development Canada, and by the Province of Ontario through the Ministry of Research and Innovation. This research was supported in part by the Natural Sciences and Engineering Research Council of Canada (RGPIN-2016-04135).

\appendix
\onecolumngrid
\newpage

\section{A single qubit inside a waveguide} \label{sec:appendixa}
In this appendix, we shall investigate  photon emission from a single qubit inside a waveguide. The scattering energy-eigenstates for a single qubit have been found in \cite{shen2005coherent} and the scattering parameters $t$, $r$ and $e_k$ are given as follows
\begin{subequations} \label{eq:fan}
\begin{align}
t&= \cos b e^{ib}, \\
r&= i \sin b e^{ib}, \\
e_k &= - \frac{1}{\sqrt{J_0}} \sin b e^{ib}
\end{align}
\end{subequations}
where $b = \arctan(-J_0/\Delta_k)$ is the phase shift. We first find the survival probability of the atom upon excitation:
\begin{align}
P_{se}(t) = \left| \int_{-\infty}^\infty  \frac{\diff k}{2\pi}|e_k|^2 e^{-i\Delta_k t}\right|^2 = e^{-2 J_0 t}.
\end{align}
{ Here, we find that $\gamma_0=2J_0$ is indeed the single emitter decay rate.}
Moreover, we can find the probability density function of the emitted photon with respect to $x$ by using (\ref{eq:appendixa}) and the scattering parameters given in (\ref{eq:fan}):
\begin{equation}
      \mathcal{P}(x,t)= \left| \int_{-\infty}^\infty  \frac{\diff k}{2\pi}\braket{x}{E_k} e_k^* e^{-i\Delta_k t} \right|^2 = J_0 e^{-2J_0(t-|x|)} \Theta(t-|x|).
\end{equation}
\begin{figure*}[h!]
    \centering
    \includegraphics[width=12cm]{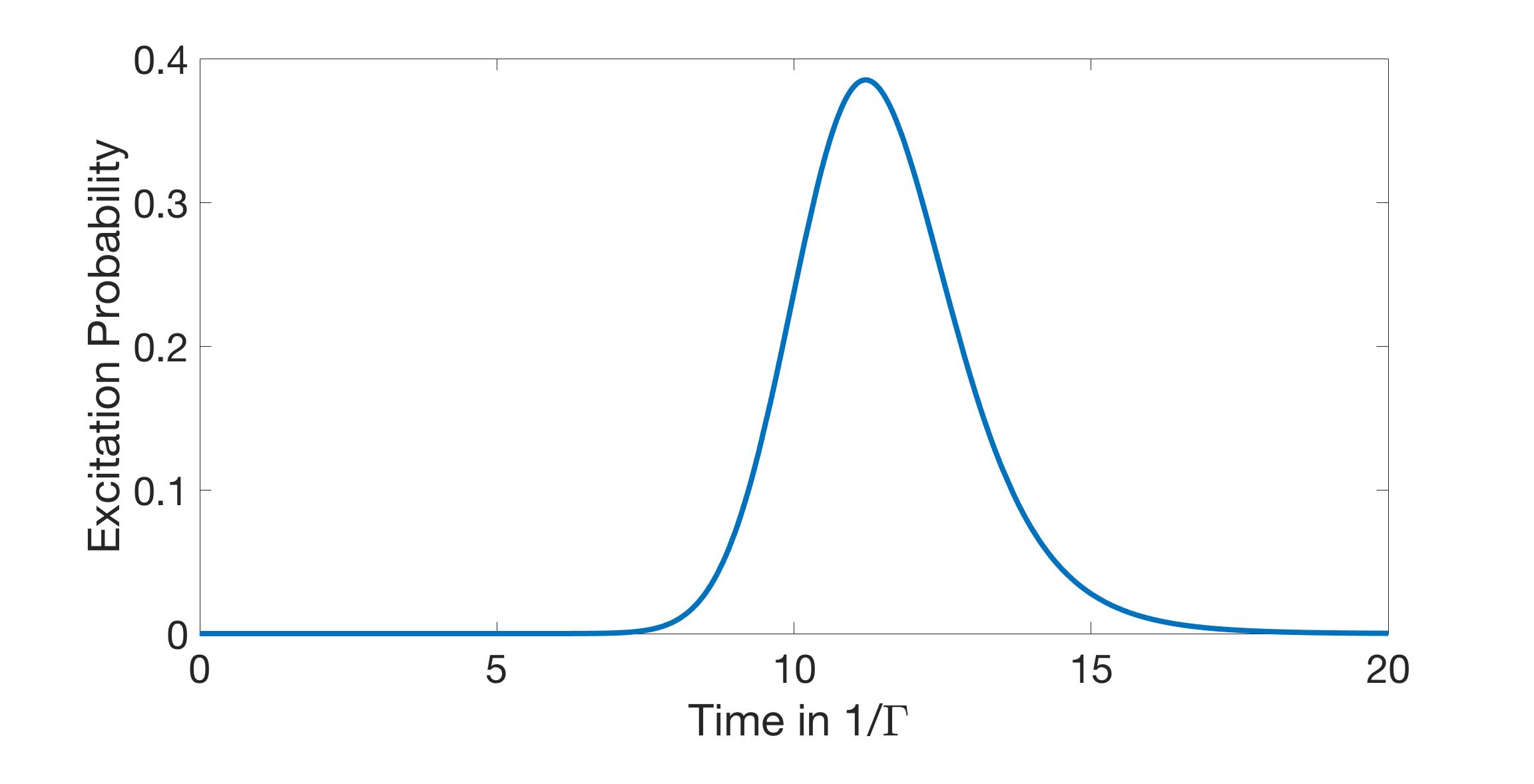}
    \caption{The excitation probability for the Gaussian pulse with $k_0=\Omega$ and $\Delta k = J_0$, where $x_0=10/\Gamma$ and $\Gamma=2J_0$. Our formalism reproduces the excitation probability found in Fig. 2b of \cite{chen2011coherent}.}
    \label{fig:my_label}
\end{figure*}

Here, we can realize that the probability density function for photon emission taking place between time $t$ and $t+\diff t$ can be found as by simply changing $\mathcal{P}(t)=2\mathcal{P}(x,\tau)|_{(\tau-|x|)=t}$, as a photon at distance $|x|$ from the center is emitted at a time $t = \tau-|x|$, where $\tau$ is the current time and $2$ comes from the symmetry. To probe this, we can find the probability that a photon is emitted until time $t$ as
\begin{equation}
\int_0^t \diff t' \mathcal{P}(t') \diff t' = \int_0^t \diff t' 2J_0 e^{-2J_0 t'} = (1- e^{-2J_0t}),
\end{equation}
which is indeed equal to $1-P_{se}(t)$ as expected. This suggests that for a single atom inside a waveguide, the probability of photon emission is linearly proportional to the probability of atomic excitation. Assuming that a photo-detection device is situated at a distance $x=L_p$, we can find the detection probability density function as
\begin{equation}
P_d(t)= J_0 e^{-2J_0 (t-L_p)} \Theta(t-L_p),
\end{equation}
which is consistent with the literature \cite{grynberg2010introduction} (See Complement 5.B).

\begin{figure*}[h!]
    \centering
    \includegraphics[width=12cm]{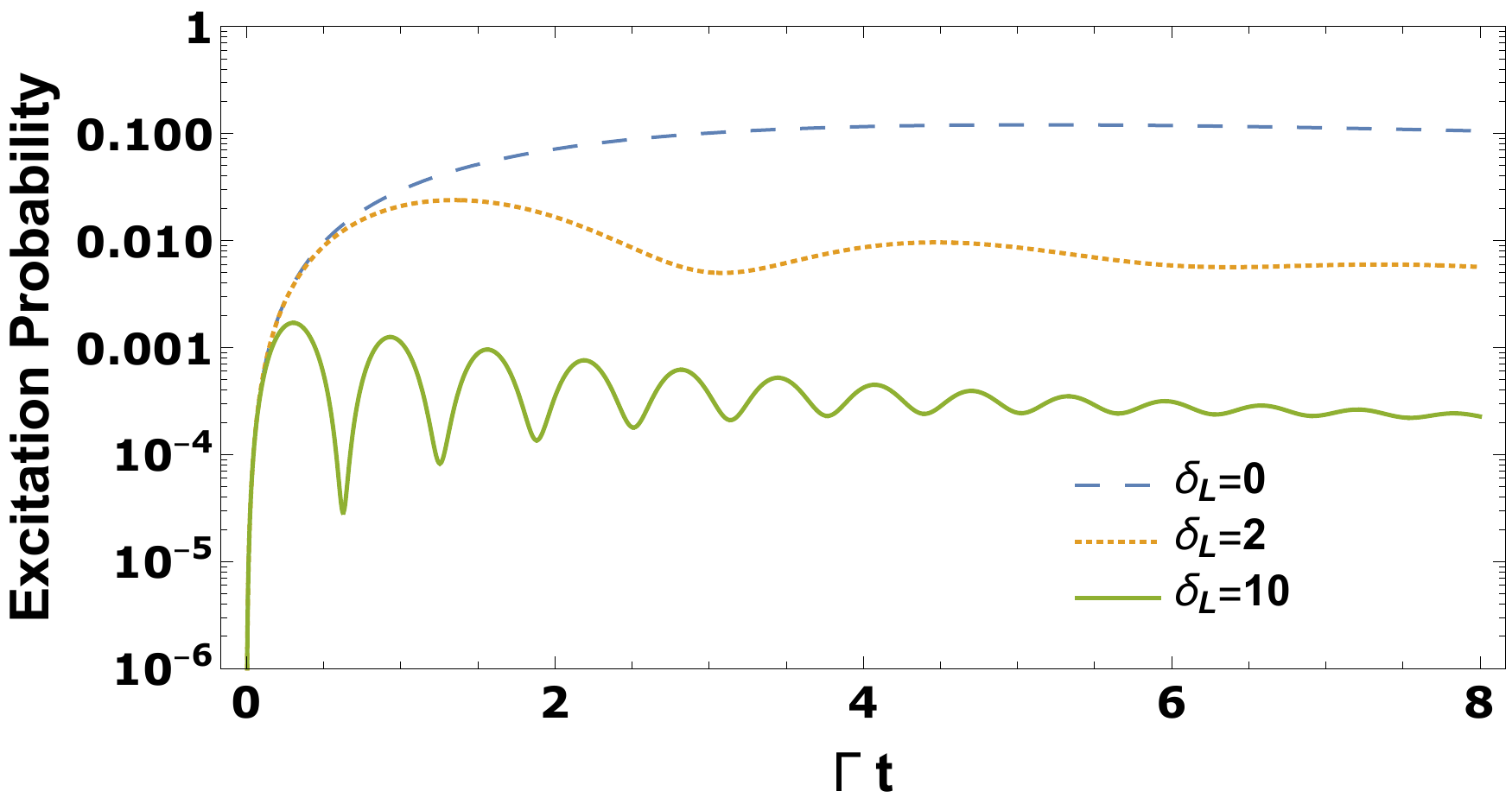}
    \caption{The excitation probability of the single qubit upon incidence of a decaying exponential. $\delta_L$ is given in terms of $\Gamma=2J_0$ and $\xi=0.05 \Gamma$ to   reproduce Fig. 2(a) of \cite{valente2016non}.}
    \label{fig:valente}
\end{figure*}

Now, we can focus on the excitation probability of the two-level system upon  incidence of a decaying pulse as in (\ref{eq:decayingpulse}). The excitation probability is
\begin{equation}
    P_e(t)=\left| \text{Res}_{\Delta_k =-iJ_0} \left[ \frac{\sqrt{2 \xi} e^{-i \Delta_k (t-x_0)}}{\xi - i \Delta_k} e_k\right]\right|^2=\left| \text{Res}_{\Delta_k =-iJ_0} \left[ \frac{\sqrt{2}J_0 e^{-i \Delta_k (t-x_0)}}{(J_0 - i \Delta_k)^2} \right]\right|^2 = 2 J_0^2 \tau^2 e^{-2J_0\tau}.
\end{equation}
where we set $\xi=J_0$ and $\tau=t-x_0$ to allow comparison with \cite{wang2011efficient}. Now that we have the analytical expression for $P_e(t)$, we can find its maximum value and corresponding $t_{\rm max}$:
\begin{equation}
    t_{\rm max}=x_0+J_0^{-1}, \quad P_e(t_{\rm max})= \frac{2}{e^2} \simeq 0.27.
\end{equation}
Note that this is half of the value found in \cite{wang2011efficient}, since we consider a waveguide with left and right moving modes. Similar computation for a single-direction waveguide results in $P_e(t_{\rm max})=\frac{4}{e^2} \simeq 0.54$ as found in \cite{wang2011efficient}. A similar calculation with the rising exponential yields $0.5$ for $\tau=0$, the theoretical maximum for a one-side excitation of a single qubit \cite{liao2016photon}. Moreover, we can also reproduce Fig. 2(a) of \cite{valente2016non} by using a non-resonant decaying exponential as shown in Fig. \ref{fig:valente}. In this figure, we define $\Gamma = 2J_0$ and $\delta_L=\Omega_P-\Omega$, where $\Omega_P$ is the average momentum of the pulse. This figure clearly shows that the real-space approach can be used to capture single qubit non-Markovianity quite easily. 

Finally, we show that using (\ref{eq:13}), we can find the excitation probability of the atom inside the waveguide for an incident pulse. For a Gaussian pulse with $k_0=\Omega$ and $\Delta k = J_0$, we can find the excitation probability of the two-level atom as shown in Fig. \ref{fig:my_label} through numerical integration, reproducing Fig. 2b of \cite{chen2011coherent}. Since $e_k$ is Lorentzian for the single qubit case, we can even find an analytical solution. In fact, for a more general Gaussian with $\Delta k=J_0 \sigma$, we can find the excitation probability analytically as
 \begin{subequations}
\begin{align}
    P_{\rm gaus}(\tau)={}& \frac{1}{2 \pi \sigma \sqrt{2\pi}} \left|\int_{-\infty}^\infty \diff y \frac{\exp(-\frac{y^2}{4\sigma^2}-iy\tau)}{y+i}\right|^2\\\label{eq:gausexact}
    ={}& \frac{\sqrt{2\pi}}{4 \sigma} \exp(\frac{1}{2 \sigma^2}-2J_0\tau) \text{erfc}^2\left(\frac{1-2\sigma^2 J_0 \tau}{2 \sigma}\right).
\end{align}
    \end{subequations}
This is consistent with the findings of \cite{chen2011coherent} in the limit $x_0 \to -\infty$ such that the pulse does not interact with the qubit initially (See Eq. B3 from \cite{chen2011coherent}, which becomes the same integral as in our (\ref{eq:gausexact})). Note that the excitation probability depends on $\tau$, and not on $t$ or $x_0$ independently. This is again a manifestation of the causality principle: the pulse has to travel a distance $\simeq x_0$ before it can have an effect on the excitation probability. Increasing $x_0$ increases this distance, hence the overall excitation probability depends only on $t-x_0$. We stress that this is true for asymptotically free pulses, where, initially, the qubit is in the ground state and the field amplitude is effectively zero at the qubit position.

For a more general Gaussian pulse with width $\Delta k$, we can use numerical maximization to find that the maximum excitation probability of the atom is $\simeq 0.40$, for $\Delta k = 1.46J_0$ and $t=x_0+(2J_0)^{-1}$ ($\tau= 0.5 J_0^{-1}$), which is in agreement with the literature \cite{stobinska2009perfect} (Up to a factor of two, which comes from the fact that we consider waveguides that are non-chiral, hence the excitation probability from one-sided pulses are divided by two). Hence, we have shown that our approach agrees perfectly with the existing literature on single qubit excitation. 

Before we finish this discussion, we write down the time-evolved state for the case where the qubit decays spontaneously to the waveguide, which is calculated using the time-evolution in (\ref{eq:arbitrary}). The time-evolved state $\ket{\psi(t)}$, that describes the spontaneous emission of the single qubit to the 1D waveguide can be written as
\begin{equation} \label{eq:newappendix1}
    \ket{\psi(t)} = \ket{\psi_q(t)}+\ket{\psi_R(t)} + \ket{\psi_L(r)},
\end{equation}
where each component is
\begin{subequations}\label{eq:newappendix2}
\begin{align}
    \ket{\psi_q(t)}&=e^{-(J_0+i\Omega)t}\ket{e}, \\
    \ket{\psi_R(t)}&=-i\sqrt{J_0} \int_{-\infty}^\infty \diff x e^{-(J_0+i\Omega)(t-x)} [\Theta(x)-\Theta(x-t)] C_R^\dag(x) \ket{0}, \\
    \ket{\psi_L(t)}&=-i\sqrt{J_0} \int_{-t}^0 \diff x e^{-(J_0+i\Omega)(t+x)} [\Theta(x+t)-\Theta(x)] C_L^\dag(x) \ket{0}.
\end{align}
\end{subequations}

Now, we show that the time evolved state satisfies the Schr\"odinger equation 
\begin{equation}
    i \frac{\partial}{\partial t} \ket{\psi(t)}= (H_0 + H_I) \ket{\psi(t)},
\end{equation}
where the free and interaction Hamiltonian are
\begin{subequations}
\begin{align}
    H_0 &= \Omega \ket{e} \bra{e} + i \int_{-\infty}^\infty \diff x \left( C_L^\dag(x) \frac{\partial}{\partial x} C_L(x) -C_R^\dag(x) \frac{\partial}{\partial x} C_R(x)  \right), \\
    H_I &= \sqrt{J_0}\left( \sigma^\dag [C_L(0) + C_R(0)] + \sigma [C_L^\dag(0) + C_R^\dag(0)] \right).
\end{align}
\end{subequations}

Let us start with the time derivatives
\begin{subequations}
\begin{align}
    i\partial_t \ket{\psi_q(t)}&=- i(J_0 + i\Omega) \ket{\psi_q(t)}, \\
    i\partial_t \ket{\psi_R(t)}&=- i(J_0 + i \Omega) \ket{\psi_R(t)}+ \sqrt{J_0} C_R^\dag(t) \ket 0, \\
   i \partial_t \ket{\psi_L(t)}&=-i (J_0 + i \Omega) \ket{\psi_R(t)}+\sqrt{J_0} C_L^\dag(-t) \ket 0.
\end{align}
\end{subequations}
Next, let us find the action of the free Hamiltonian ($H_0$) on each part of $\ket{\psi(t)}$
\begin{subequations}
\begin{align}
    H_0 \ket{\psi_q(t)}&=\Omega \ket{\psi_q(t)}, \\
    H_0 \ket{\psi_R(t)}&= -i(J_0 + i\Omega) \ket{\psi_R(t)} - \sqrt{J_0} e^{-(J_0+i\Omega)t} C_R^\dag(0) \ket 0 +\sqrt{J_0} C_R^\dag(t) \ket{0}, \\
    H_0 \ket{\psi_L(t)}&=-i(J_0 + i\Omega) \ket{\psi_L(t)} - \sqrt{J_0} e^{-(J_0+i\Omega)t} C_L^\dag(0) \ket 0 +\sqrt{J_0} C_L^\dag(-t) \ket{0}.
\end{align}
\end{subequations}
Finally, we consider the action of the interaction Hamiltonian ($H_I$) on each part of $\ket{\psi(t)}$
\begin{subequations}
\begin{align}
    H_I \ket{\psi_q(t)}&=\sqrt{J_0} e^{-(J_0 +i\Omega) t} [C_R^\dag(0) + C_L^\dag(0)] \ket 0 , \\
    H_I \ket{\psi_R(t)}&= -i J_0 \Theta(0) e^{-(J_0+i\Omega)t}  \ket{e}, \\
    H_I \ket{\psi_L(t)}&= -i J_0 e^{-(J_0+i\Omega)t}  (1-\Theta(0)) \ket{e}.
\end{align}
\end{subequations}
From our calculations, we see that $i\partial_t \ket{\psi_q(t)}=H_0 \ket{\psi_q(t)}+ H_I \ket{\phi_R(t)}+H_I \ket{\phi_L(t)}$. Similarly, $i\partial_t (\ket{\psi_R(r)}+\ket{\psi_L(t)})=H_I \ket{\psi_q(t)}+H_0(\ket{\psi_R(t)}+\ket{\psi_L(t)})$. Consequently, the time evolved state given in (\ref{eq:newappendix1}-\ref{eq:newappendix2}) satisfies the Schr\"odinger equation for $t>0$ and describes the spontaneous decay of an initially excited qubit. As can be seen from this expression, the excited qubit component of the state decays with a decay rate $J_0$ and the position component consists of a superposition of two decaying exponentials travelling away from the qubit, as we argued at the beginning of this section. This state has an interesting feature: it is exactly a solution to the Schr\"odinger equation, whereas in evaluation of (\ref{eq:arbitrary}), approximations are performed. This is due to the fact that all these approximations are exact in the limit $J_0/\Omega \to 0$, and the time evolution of single qubit states has similar behavior, i.e. follows a similar physical pattern, for all $J_0/\Omega$ values within the accuracy of the rotating-wave approximation. This property will be  discussed further in  future work, where we plan to show that the cure of causality violations in the RWA by the extension of energy integrals to negative energies (as explained in \cite{milonni1995photodetection} and had been employed by Fermi \cite{fermi1932quantum}) is not a trick, but rather has a physical basis. 

\section{Waveguide QED with networks of quantum emitters}  \label{sec:appendixgeneral}
In this appendix, we extend the  analysis beyond three qubits in a 1D waveguide. We re-derive the results for quantum networks with various emitters that can be modelled as identical two-level systems. We show that, in the end, the generalization comes down to simply summing over all possible stationary states, in analogy to the case of three qubits. This demonstrates the power of the real-space approach. At the end of this section, we  argue that this approach can also be used for non-identical quantum emitters. 

For the general network, the Hamiltonian can be written as $H=H_0+H_{\textsc{i}}$, 
where $H_0$ contains the self energies and $H_{\textsc{i}}$ is the interaction Hamiltonian that includes point-like interactions. In  real space, the Ansatz for such a system can be written as
\begin{equation}
\begin{split}
        |E_k^{(i)}\rangle&= \sum_{\Xi \in \{\rm WG\}} \int_{-\infty}^\infty \diff x \phi_\Xi^{(i)}(x) C_\Xi^\dag(x) \ket 0 + \sum_{s \in \{\rm qubits\}} e_s^{(i)} \ket{e_s},
\end{split}
\end{equation}
where, $\Xi$ represents all possible modes of all 1D waveguides inside the network, $\phi_\Xi^{(i)}(x)$ includes the field amplitudes with the initial condition $i$, $C_\Xi^\dag(x)$ is the creation operator corresponding to the mode $\Xi$ and $e_s^{(i)}$ is the excitation coefficient for the qubit denoted by $s$. If the interactions are point interactions, then the field amplitudes $\phi_\Xi^{(i)}(x)$ change only at the atomic positions while behaving as free fields inside the waveguide. 

Without loss of generality, we assume that the energy eigenstates are normalized as $\langle E_k^{(i)}|E_p^{(j)}\rangle=2 \pi \delta(k-p) \delta_{ij}$.
Here, there is one aspect that requires further attention. When finding the energy eigenbasis, we usually consider the scattering of light that is initially incident from one side of the system. It is not always clear whether the (degenerate) eigenstates corresponding to the same momentum value found via this method are orthogonal. \cite{tsoi2008quantum} doesn't mention this for the linear chain of qubits, but one can prove this  using the transfer matrix properties. For the general case, we keep in mind that this normalization is only valid in the limit $J_0/\Omega \to 0$ such that the left and right moving particle subspace does not mix. This ensures the orthogonality condition for the degenerate eigenstates found by considering one-sided excitations. 

\subsection{Number of Markovian collective decay rates}
In a network with $N$ qubits, the maximum number of Markovian collective decay rates can be at most $N$. To illustrate this fact, we first consider a generalization of Equation (6c) from \cite{tsoi2008quantum}:
\begin{equation}\label{eq:deltak}
    \sum_\Xi \sqrt{J_\Xi} \phi_\Xi(x_s) - \Delta_k e_s^{(i)}=0,
\end{equation}
where $x_s$ is the position of the qubit denoted by $s$ and $J_\Xi$ is the coupling energy of the $\Xi$ mode to the qubits. This equation needs to be satisfied at each atomic position, such that $|E_k^{(i)}\rangle$ is an energy eigenstate. 

Now, let us construct a coefficient matrix $A$, such that $Ax=b$, with $x$ including the scattering parameters and $b$ including the initial conditions. The coefficient matrix $A$ includes the coefficient of the scattering parameters in the generalized versions of the equations (6a-c) from \cite{tsoi2008quantum}. The scattering parameters, $x$, diverge for $\Delta_k$ values, for which $A$ is singular. As a result, the characteristic equation for the poles of scattering parameters can be given as $\det(A(\Delta_k))=0$. Since $A$ has only $N$ distinct $\Delta_k$ values, each at different rows according to (\ref{eq:deltak}), and the phase is linearized in the Markovian limit; the characteristic polynomial for the poles is of $N$th order for $\Delta_k$ and can have only $N$ distinct solutions $p'$. Then, let us define $p$ equal to $p'$ if $p'$ is in the lower-half plane or on the real axis; or as $p^{'*}$ if $p'$ is in the higher-half plane. In analogy with the three qubits case, these solutions relate to the decay rates via a {$\Gamma = 2i p$.} In a future work, we will prove that the poles $p'$ of the scattering parameters are indeed in the LHP or on the real axis, consequently $p=p'$. This theorem requires a long discussion of the causality principle in waveguide QED and is therefore left as a future work.

\subsection{Time evolution}
As before, we start by assuming that there are no BICs present in the system. Then, the time evolution operator is
\begin{equation}
    U(t) = \sum_{i \in \{\rm I.C.\}} \int_0^\infty \frac{\diff k}{2\pi} |E_k^{(i)}\rangle\langle E_k^{(i)}| e^{-i E_k t}.
\end{equation}
Here, $i$ represents the initial conditions (I.C.), two for each 1D waveguide present in the system. Then, the time evolution of any regular state is
\begin{subequations}
\begin{align}
    \ket{\psi(t)}&= \sum_{i \in \{\rm I.C.\}} \int_0^\infty \frac{\diff k}{2\pi} |E_k^{(i)}\rangle\langle E_k^{(i)}|\psi(0)\rangle e^{-i E_k t}, \\
    &= \int_{-\Omega}^\infty \frac{\diff \Delta_k}{2\pi} \ket{g(k)} \exp{-i \Delta_k t}, \\
    &\simeq \int_{-\infty}^\infty \frac{\diff \Delta_k}{2\pi} \ket{g(k)} \exp{-i \Delta_k t}, \\
    &= \sum_p \underset{\Delta_k = p}{\text{Res}} \left[ \ket{g(k)} \exp{-i\Delta_k t} \right],
\end{align}
\end{subequations}
where in the second line, we substitute $k \to \Delta_k$ and define $\ket{g(k)}=\sum_{i \in \{\rm I.C.\}}\langle E_k^{(i)}|\psi(0)\rangle |E_k^{(i)}\rangle$. The rest is analogous to the case of three qubits considered earlier, with the final step being meaningful only in the Markovian limit as we have discussed in Section \ref{sec:formalism}. Here, $p$ are the lower half plane poles of  $|E_k^{(i)}\rangle$ and $\langle E_k^{(i)}|$, and consequently of scattering parameters. As in the case of three qubits, the poles contain complete information of the collective decay rates of the system.

As a result, the collective decay rates in a quantum network consisting of 1D waveguides can be read-off from the scattering parameters. The collective decay rates govern the time evolution of the overall system in the Markovian limit. This intuitive approach to the time evolution in waveguide QED can be employed as long as the coupling energy between the qubits and light is smaller than the resonance frequency and the distance between the emitters is $L \sim O (\Omega^{-1})$ such that any exponential in the characteristic equation can be linearized:
\begin{equation}
   kL_{ij} = (\Delta_k + \Omega) L_{ij} \simeq  \Omega L_{ij} = \theta_{ij}, 
\end{equation}
where $L_{ij}$ is the distance between the two emitter and $\theta_{ij}$ is the corresponding phase acquired by light. 

Networks with non-identical emitters and/or with non-radiative decay can also be described in the Markovian limit, as long as the resonance energies of the emitters differ only $\sim O(J_0)$, with $J_0 = \min\{J_\Xi\}$, so that the phase linearization is valid. We will do this for a linear chain in Appendix \ref{sec:appendixnonidentical}. The non-Markovian limit is trivial to describe by calculating the final step of time evolution as a numerical integral. Overall, adding more qubits or waveguides to the system changes only the scattering eigenstates $|E_k^{(i)}\rangle$ and not the theory itself. Since the scattering parameters can be found via solving a linear set of equations, the real space approach is easily scale-able to large systems.

\subsection{Single-photon pulse scattering}
In section \ref{sec:pulsescat}, we discussed  pulse scattering in the Markovian limit for a linear chain of three qubits and found  that the transmission and reflection amplitudes are determined by the $k$-mode scattering eigenstates, inline with the findings of \cite{liao2016photon}. The calculations in Appendix \ref{sec:appendixb} present a direct proof that this property is more general than the specific example of a linear quantum emitter chain and are not specific to the Markovian limit. The scattering by any ``black box'' system can be described by the external degrees of freedom, such that the scattered light amplitudes vary via the scattering parameters of the stationary states. In such a case, the asymptotic field amplitudes for each $k$-mode  can be written as
\begin{equation}
    |\psi_s(k,t \to \infty)|^2 = |\psi(k,0) t_s(k)|^2\,,
\end{equation}
where $|\psi_s(k,t \to \infty)|^2$ is the output field amplitude corresponding to the (normalized) scattering parameter $t_s(k)$ and $|\psi(k,0)|^2$ is the input field amplitude. 

We thus see that the real-space approach can be used to find collective decay rates in complicated systems with multiple quantum emitters and multiple waveguides. In doing so, the approach can be used to study subradiance and superradiance, BICs, single-photon pulse scattering and time evolution of system observable such as excitation probabilities. 

\begin{figure}
    \centering
    \includegraphics[width=8cm]{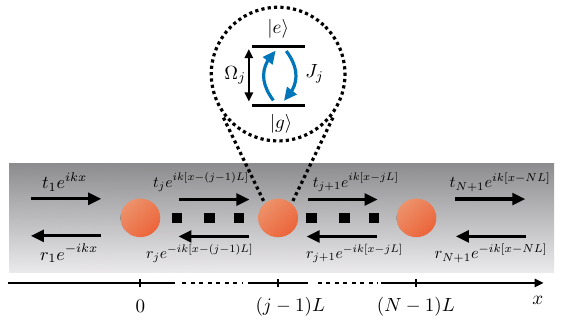}
    \caption{The Bethe ansatz for a linear chain of non-identical qubits with various energy separation $\Omega_j$ and coupling energies $J_j$.}
    \label{fig:non-identical}
\end{figure}

\section{Waveguide QED with  non-identical emitters: a general theory} \label{sec:appendixnonidentical}

In this appendix, we set up the theory for studying non-identical emitters. We do this for a single waveguide coupled to $N$ qubits, but the derivations are more general and can be employed for multi-waveguide systems.

An illustration of the linear chain consisting of $N$ non-identical qubits is given in Fig. \ref{fig:non-identical}. The real-space Hamiltonian for this system is
 $H = H_{\textsc{q}}+H_{\textsc{f}}+H_{\textsc{i}}$, 
where 
\begin{align}
     H_{\textsc{q}}={}&  \sum_{j=1}^{N} \Omega_j \ket{e_j}\bra{e_j}
     \end{align}
     is the free Hamiltonian of the qubits, where $\ket{e_j}$ is the excited state of  the $j^{th}$ qubit and  $\Omega_j$ is  the energy separation for qubit $j$, 
     \begin{align}
     \begin{split}
   H_{\textsc{f}}&= i \hbar v_g \int_{-\infty}^\infty \diff x\left( C_L^\dag(x)\frac{\partial}{\partial x} C_L(x) -C_R^\dag(x)\frac{\partial}{\partial x} C_R(x)  \right)   
           \end{split}
           \end{align}
           is the free Hamiltonian of the field,  where $v_g$ is the group velocity of photons inside the waveguide and $C_{R/L}(x)$ are annihilation operators for right/left moving photons,  and
           \begin{align}
           \begin{split}
         H_{\textsc{i}}={}&\sum_{j=1}^N \sqrt{J_j} \int_{-\infty}^\infty \diff x \delta(x-(j-1)L) \left( (C_R^\dag(x)+C_L^\dag(x)) \sigma_j + \text{H.c.} \right)\,,
           \end{split}
\end{align}
is the interaction Hamiltonian between the systems, where $J_j$ is the coupling energy between  qubit $j$ and light. From now on, we set $\hbar=v_g=1$ for algebraic simplicity, as usual. 

Similar to the identical qubit case, we start by writing a Bethe Ansatz for a general eigenstate
\begin{equation}
\begin{split}
    \ket{E_k}&= \sum_{\Xi \in \{WG\}}\int_{-\infty}^\infty \diff x  \phi_\Xi(x) C_\Xi^\dag(x)\ket 0+ \sum_{j=1}^N e_j \ket{e_j},
\end{split}
\end{equation}
where $e_j$ is the excitation coefficient for the $j$th qubit and $\phi_{R/L}(x)$ are field amplitudes for left/right moving photons. Note that $e_j$ and $\phi_{R/L}(x)$ depend on $\theta$ implicitly. Owing to delta-function interactions, the field becomes free for $x\neq (j-1)L$. Hence, we can write an Ansatz for the field amplitudes $\phi_{R/L}(x)$ as:
\begin{subequations} \label{eq:non-id-field-amp}
\begin{align}
    \phi_R(x)&= 
    \begin{cases}
    t_1e^{ikx} &x<0,\\
    t_{j+1} e^{ik[x-jL]} &(j-1)L<x<jL, \\
    t_{N+1} e^{ik[x-NL]} & x>(N-1)L,
    \end{cases}\\
    \phi_L(x)&= 
    \begin{cases}
    r_1e^{-ikx} &x<0,\\
    r_{j+1} e^{-ik[x-jL]} &(j-1)L<x<jL, \\
    r_{N+1}e^{-ik[x-NL]} & x>(N-1)L,
    \end{cases}
\end{align}
\end{subequations}
where $t_1$, $r_1$, $t_{N+1}$ and $r_{N+1}$ are picked for different initial conditions. For a photon incident from the far left ($\ket{E_k}$): $t_1=0$, $r_{N+1}=0$ and for non-radiating conditions (for BIC, $\ket{D_i}$): $t_1=r_1=t_{N+1}=r_{N+1}=0$. For $\ket{E_{-k}}$, one can mirror the state $\ket{E_k}$ w.r.t. the center of the linear chain, as in Section \ref{sec:formalism}. For now, we suppress the implicit $k$-dependence for brevity and discuss the Markovian limit and corresponding additional conditions. The non-Markovian regime calculations can be performed analogous to Section \ref{sec:nonmarkgen}.

Having set an Ansatz for the energy eigenstates, we apply the condition $H \ket{E_k}=E_k \ket{E_k}$, with $E_k=|k|$, to obtain the equation of motion for the scattering parameters:
\begin{subequations} \label{eq:sec2eom}
\begin{align}
    t_{j+1}e^{-i\theta}-t_{j}+ i \sqrt{J_j}e_j &=0, \\
    r_{j+1} e^{i\theta}-r_j - i \sqrt{J_j}e_j &=0, \\
    \sqrt{J_j} (t_j + r_j) - (\Delta_k+\delta_j) e_j &= 0,
\end{align}
\end{subequations}
where we linearize the propagation phase as usual $\theta \simeq \Omega_1 L$, and define $\Delta_k = E_k-\Omega_1$ and $j=1,...,N$. Moreover, for $j\neq 1$, we define $\delta_j = \Omega_1-\Omega_j$ such that $E_k - \Omega_j=\Delta_k + \delta_j$. For the case of non-identical emitters, the Markovian limit has an additional constraint, apart from $J_j /\Omega_j \ll 1$. So that the interacting frequencies are confined within $|\Delta_k|\leq O(J_i)$, the detuning of the qubit separation frequencies should satisfy $|\delta_j|\sim O(J_0)$. Consequently, the linearization of $\theta$ around $\Omega_j$ are all equivalent. If this is not satisfied, the linearization assumption of the phase is no longer valid. 

There are two possible methods that could be used to solve (\ref{eq:sec2eom}):
\begin{enumerate}
    \item By writing the equations as a matrix equation with $3N$ unknowns and solving this linear system. We used this method in Section \ref{sec:formalism}, and it is reasonable to do so here as long as the system is small. But for large systems, this method requires unnecessarily high computation and is therefore inefficient.
    \item By using the transfer matrix method and obtaining a recursive algorithm to find each scattering parameter. This method preserves the polynomial shape of the scattering parameters and leads to the  fully analytical results that we desire. We will use this method to find the scattering parameters.
\end{enumerate}
There is also an additional method considered in \cite{tsoi2008quantum}. We could use the transfer matrix method and draw parallels to one-dimensional photonic crystals \cite{photoniccrystal}. But apart from the fact that this method can only be used for identical emitters, even in the identical qubit case, the numerator and denominator of scattering parameters are no longer polynomial in $\Delta_k$, hence the poles of the system cannot be obtained easily. As we have seen so far, the complex analysis plays an important role in time evolution of single-photon states. Therefore, this method is also not preferred for obtaining fully analytical results. 

To use the transfer matrix method, we shall first reshape (\ref{eq:sec2eom}) and eliminate the $e_j$ degree of freedom:
\begin{subequations}
\begin{align}
    t_{j+1}e^{-i\theta}&=\left( 1- i \frac{J_j}{\Delta_k+\delta_j}\right)t_j - i \frac{J_j}{\Delta_k+\delta_j} r_j , \\
    r_{j+1} e^{i\theta}&= i \frac{J_j}{\Delta_k+\delta_j} t_j + \left( 1+ i \frac{J_j}{\Delta_k+\delta_j}\right)t_j.
\end{align}
\end{subequations}
Writing this in matrix form and re-arranging terms, we obtain the matrix equation
\begin{align}
\begin{split}
    \begin{pmatrix}
    t_j \\
    r_j
    \end{pmatrix}
    ={}&
    \begin{pmatrix}
    1 + i\frac{J_j}{\Delta_k+\delta_j} & i \frac{J_j}{\Delta_k+\delta_j} \\
    - i \frac{J_j}{\Delta_k+\delta_j} &  1 - i\frac{J_j}{\Delta_k+\delta_j}
    \end{pmatrix}
    \begin{pmatrix}
    e^{-i\theta} & 0 \\
    0 & e^{i\theta}
    \end{pmatrix}
    \begin{pmatrix}
    t_{j+1} \\
    r_{j+1}
    \end{pmatrix}.
    \end{split}
\end{align}
Then, we find the transfer matrix for a unit cell, which contains the scattering from the qubit $j$ and a propagation phase $\theta$, as
\begin{equation}
\begin{split}
    T_j&=
    \begin{pmatrix}
    1 + i\frac{J_j}{\Delta_k+\delta_j} & i \frac{J_j}{\Delta_k+\delta_j} \\
    - i \frac{J_j}{\Delta_k+\delta_j} &  1 - i\frac{J_j}{\Delta_k+\delta_j}
    \end{pmatrix}
    \begin{pmatrix}
    e^{-i\theta} & 0 \\
    0 & e^{i\theta}
    \end{pmatrix}  \implies \begin{pmatrix}
    t_j \\
    r_j
    \end{pmatrix}= 
    T_j 
    \begin{pmatrix}
    t_{j+1} \\
    r_{j+1}
    \end{pmatrix}.
\end{split}
\end{equation}
By recursively relating them, we can eliminate the internal scattering parameters and obtain the relation between the external fields for the initial condition, where the photon is incident from far left, as
\begin{equation}
    \begin{pmatrix}
    1 \\
    r_1
    \end{pmatrix}= 
    T
    \begin{pmatrix}
    t_{N+1} \\
    0
    \end{pmatrix},
\end{equation}
where $T=T_1 T_{2}...T_N=\prod_{j=1}^N T_j$. From this equation, we can find the external transmission and reflection coefficients as
\begin{equation}
    t_{N+1} = 1/(T)_{11}, \quad r_1 = (T)_{21}/(T)_{11}. 
\end{equation}
Here, $(A)_{ij}$ represents the $ij$th element of the matrix $A$. Then, the rest of the transmission and reflection coefficients can be found via
\begin{equation}
    \begin{pmatrix}
    t_j \\
    r_j
    \end{pmatrix}
    =
    S_j
    \begin{pmatrix}
    1\\
    r_1
    \end{pmatrix},
\end{equation}
where we define $S_j=T_{j-1}^{-1}...T_2^{-1} T_1^{-1}$. Finally, the excitation coefficients can be obtained from (\ref{eq:sec2eom}) as
\begin{equation}
    e_j = \sqrt{J_j} \frac{(t_j+r_j)}{\Delta_k+\delta_j}.
\end{equation}

As an example, let us consider the two qubit case and find the collective decay rates. A perturbative result for the decay rates has been obtained in \cite{muller2017nonreciprocal}, here we find the complete expressions. Since the collective decay rates can be read-off from any scattering parameter, let us find the reflection coefficient $r_1$ by employing the transfer matrix approach:
\begin{equation}
    r_1=-\frac{i \left(J_1 (\delta_2+\Delta_k+i J_2)+J_2 e^{2 i \theta} (\Delta_k-i J_1)\right)}{(\Delta_k+i J_1) (\delta_2+\Delta_k+i J_2)+J_1 J_2 e^{2 i \theta}}.
\end{equation}
Then, the two collective decay rates are
{
\begin{align} \label{eq:nonsymdecay}
    \Gamma_{1/2} &=  \left(J_1+ J_2-i\delta_2\pm \Delta\right),
\end{align}}
where $\Delta=\sqrt{(J_1+i\delta_2-J_2)^2+4 J_1 J_2 e^{2 i \theta}}$. For $J_1=J_2$ and $\theta=\pi$, expanding around small $\delta$ leads to the perturbative results obtained in \cite{muller2017nonreciprocal}. Furthermore, we emphasize that for a photon incident from far right, exchanging $J_1\iff J_2$, setting $\delta_2 \to - \delta_2$ results in the same expressions for the collective decay rates up to a global imaginary shift of $i\delta_2$, which is simply due to the redefinition of $\Delta_k$ and is just a re-normalization of energy levels. This shows clearly that collective decay rates are indeed independent of initial conditions, as expected.

\begin{figure}
    \centering
    \includegraphics[width=8cm]{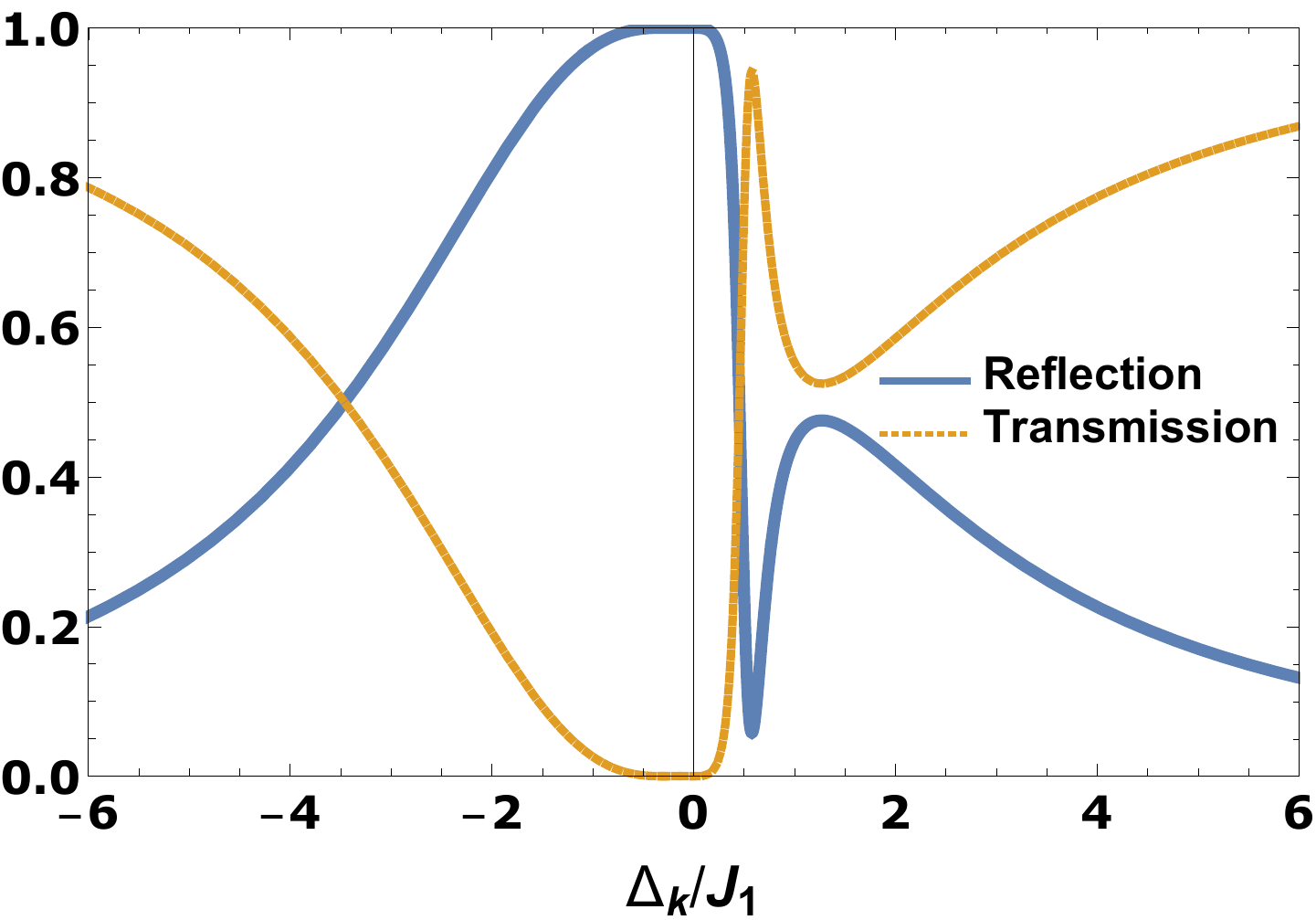}
    \caption{The transmission and reflection intensities, $|t_2|^2$ and $|r_1|^2$ respectively, for a photon incident from far left to a two qubit system with $J_2/J_1=2$, $\delta_2=0.3J_1$ and $\theta=0.85\pi$. The intensities show Fano type line-shapes and the system no longer becomes completely transparent for the Fano minimum.}
    \label{fig:non-identical-two-qubit}
\end{figure}

Moreover, the two-qubit system with identical emitters becomes transparent for a Fano minimum (\cite{fano}) such that $r_1=0$, where the transmission becomes unity \cite{tsoi2008quantum}. By the same logic, for the general two-qubit system, we can find the $\Delta_k$ value corresponding to the zero reflection as 
\begin{equation} \label{eq:fano}
   \Delta_k= -\frac{J_1 \left(\delta_2-i J_2 e^{2 i \theta}+i J_2\right)}{J_1+J_2 e^{2 i \theta}}.
\end{equation}
However, this frequency detuning, for which $r_1=0$, is complex, whereas only real $\Delta_k$ values are physical. This means that the expression in (\ref{eq:fano}) is not the Fano dip and the reflected pulse intensity may not be zero for the general case. Consequently, the two-qubit system is only transparent if the qubits are identical. In fact, the minimum value of the reflection intensity takes a very complicated form, which is different than the real part of (\ref{eq:fano}). The transmission and reflection intensities are illustrated in Fig. \ref{fig:non-identical-two-qubit} for a specific example. As apparent from the figure, the intensities show Fano type line-shapes and the system no longer becomes completely transparent for the Fano minimum. We emphasize that (\ref{eq:fano}) reduces to the result found in \cite{tsoi2008quantum} for $J_1=J_2=J_0$ and $\delta_2=0$, which is $\Delta_k = -J_0 \tan(\theta)$. The two-qubit system becomes completely transparent for this special case. 

We note that while we have not considered the non-radiative decay, the non-radiative modes can be modelled as waveguides (as long as the dispersion relation can be linearized) and interactions between the qubits can be described via this formalism by using multiple waveguides, whose only effect is to increase the dimensions of the transfer matrices. 

\section{Parity (mirror) symmetry} \label{sec:appendixparity}We define the general parity operator, $\hat P$, that acts on the combined Hilbert space of the single-excitation states. This operator simply mirrors the state w.r.t. the center of the multi-qubit system. When the multi-qubit system is centered at $x=0$, its action on creation operators is defined as
\begin{subequations}
\begin{align}
    \hat P \sigma_j^\dag \hat P^\dag = \sigma_{N-j+1}^\dag &\implies \hat P \ket{e_j}= \ket{e_{N-j+1}}, \\
    \hat P C_{R/L}^\dag(x) \hat P^\dag = C_{L/R}^\dag(-x) &\implies \hat P \ket{x} = \ket{-x}.
\end{align}
\end{subequations}
Then, the symmetry states are defined as the eigenvectors of $\hat P$ that have $+1$ eigenvalues, whereas anti-symmetric states correspond to $-1$ eigenvalues. For example, for odd $N$, the state where the initial qubit is excited is a symmetric state since $\hat P \ket{e_{(N+1)/2}}=\ket{e_{(N+1)/2}}$. Note that the general parity operator leads to the natural definition of even (symmetric) and odd (anti-symmetric) basis operators, as discussed in the literature (for example \cite{shen2007strongly})
\begin{subequations}
\begin{align}
    C_e^\dag(x) &= \frac{1}{\sqrt{2}} \left( C_R^\dag(x) + C_L^\dag(-x) \right), \\
     C_o^\dag(x) &= \frac{1}{\sqrt{2}} \left( C_R^\dag(x) - C_L^\dag(-x) \right),
\end{align}
\end{subequations}
such that $\hat P C_{e/o}^\dag(x) \hat P^\dag = \pm C_{e/o}^\dag(x)$. Since a single qubit couples to light symmetrically, odd (e.g. anti-symmetric) states cannot excite single qubits, which will be discussed in details in Section \ref{sec:pulsescat}. 

Let us now use the parity operator to find the energy eigenstates. First, it is straightforward to show that the parity operator commutes with the Hamiltonian $[H,\hat P]=0$. Then, let $\ket{E_k}$ be an eigenstate of the Hamiltonian but not an eigenstate of the parity operator. (Any scattering state, where the photon is initially incident from one side, fits this definition.) Then, $\hat P \ket{E_k}$ is also an eigenstate of the Hamiltonian since:
\begin{equation}
     H (\hat P \ket{E_k})= \hat P H \ket{E_k}= \hat P E_k \ket{E_k}= E_k (\hat P \ket{E_k}).
\end{equation}
Consequently, $\ket{E_{-k}}=\hat P \ket{E_k}$. Throughout the paper, we refer to this procedure simply by stating that $\ket{E_{-k}}$ can be found via symmetry considerations.

\section{An example for how residues contain information on BIC} \label{sec:appendixfinal}

For the bright state $\ket{B}$, the system completely couples with the light and the residues corresponding to subradiant poles are zero. The reason why the subradiant residues are non-zero for cases, where the initial state has non-zero overlap with subradiant states, can be explained as follows. Consider $|\theta - n \pi| \simeq \delta$. In this case, the subradiant state couples to the light slightly and decay is described solely by scattering eigenstates, since there are no BICs. Any regular state $\ket \psi$ can be written as
\begin{equation}
\ket{\psi}=\sum_i \alpha_i \ket{D_i} + \beta \ket{B}.    
\end{equation}
For any $\delta>0$, no matter how small, the residue has a finite value for $t=0$, which is the complex coefficients corresponding to states $\ket{D_i}$ and $\ket{B}$ (which is, for example, $\alpha_1$ for the coefficient of $\ket{D_1}$). Taking the limit $\delta \to 0$ then results in $\alpha_i$ for $\ket{D_i}$, since $\alpha_i$ is obtained for each value of $\delta>0$. The exponential term corresponding to dark states become unity, since the subradiant poles become zero. 

To illustrate how a state approaches the dark state as $\theta\rightarrow\pi$, let us compute the overlap $|\braket{D_2}{\psi(t)}|$, where  $\ket{\psi(0)}=\ket{e_0}$. This overlap is important, because it is a continuous function of $\theta\neq \pi$ for any $t$ for the time evolution operator presented in (\ref{eq:timeevolutionop}), whereas it becomes discontinuous when $\theta \neq \pi$. In this case, the time evolution given in (\ref{eq:timeevol2}) should be used. Now, of course, the immediate question is whether this discontinuity is a physical one, that is whether we can remove it by using (\ref{eq:timeevolutionop}) asymptotically at $\theta \rightarrow \pi$ and omit the definition of (\ref{eq:timeevol2}). The overlap is discontinuous at $\theta =pi$, since the scattering part, which constitutes (\ref{eq:timeevolutionop}), does not contribute to this overlap exactly at $\theta =\pi$, whereas they do for any other value $\theta \neq \pi$. 

\begin{figure}[!h]
    \centering
    \includegraphics[width=0.5\textwidth]{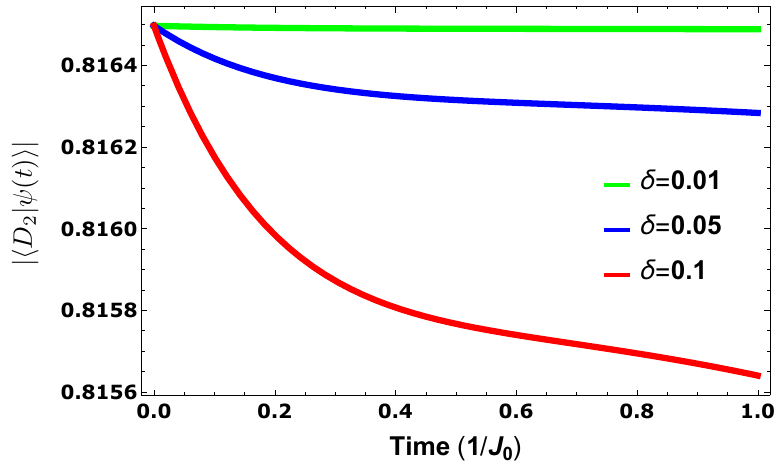}
    \caption{The overlap $|\braket{D_2}{\psi(t)}|$ where $\ket{\psi(0)}=\ket{e_0}$ for various $\delta = \pi-\theta $. As $\delta$ gets smaller, the system approaches the regime where BICs appear. Taking the limit $\delta \to 0$, the information on BIC is stored in the residues of the poles.}
    \label{fig:fig1}
\end{figure}

Defining $\delta = \pi-\theta $, we compare the overlap as a function of time $t$ (with units $J_0^{-1}$) for various $\delta$. As $\delta$ smaller, this illustrates taking the limit $\theta\rightarrow\pi$. Since $\ket{D_2}$ does not decay as $\delta \to 0$, the coefficient $|\braket{D_2}{\psi(t)}|$ is also expected to be non-zero and non-decaying in this limit. For $\delta=0$, this term can only be obtained from the BIC contribution in (\ref{eq:timeevol2}), where the residue obtained from the scattering states would not contribute to this coefficient since $\ket{D_2}$ is a dark state.  Fig. \ref{fig:fig1} shows that the residue obtained from the scattering parameter is non-zero in the limit $\delta\rightarrow 0$. Thus, the time evolution in the limit $\delta \to 0$ can be described by taking the limit of the residues found using only the scattering eigenstates.

This shows that the proposed strategy yields the correct results, because the limit of the residue converges to the coefficient that would have been obtained via the complete time evolution operator with the BIC contribution. 

\section{An alternative method for finding collective decay rates} \label{sec:appendixalternativemethod}
In this appendix, we show an alternative method for finding collective decay rates for the special case of a linear chain of $N$ qubits. This method is less efficient than simply finding the poles of the scattering parameters, but it provides further insight into the physics of the system. Specifically, this method shows  that each collective decay rate corresponds to a specific coherent excitation of the system. This means that one can always prepare initial states that excited only one of the $N$ collective decay rates. 

Let us start with the following Ansatz for an initially excited system following the approach by \cite{liao2015single}
\begin{equation}
    \ket{\psi(t)}= \sum_{j=1}^N \alpha_j(t) e^{-i\Omega t} \ket{e_j} + \ket{\chi(t)},
\end{equation}
where $\ket{\chi(t)}$ includes the photon contribution with the initial condition {$\ket{\chi(0)}=0$}. Then, the relation between $\alpha_j(t)$ is found in \cite{liao2015single} and can be given in terms of the variables defined in this paper as
\begin{equation}\label{eq:liao}
    \dot \alpha_j(t) = - \sum_{k=1}^N  J_0 e^{i\theta |j-k|} \alpha_k(t-L|j-k|),
\end{equation}
where $\dot \alpha_j(t)$ denotes the time derivative of $\alpha_j(t)$. \cite{liao2015single} uses a more general form of this equation to calculate observables (upon pulse incidence). Here, we will take another approach and show that this equation can yield the collective decay rates. We also note that here we are not interested in the time evolution of $\ket{\chi(t)}$.

Since we are interested in the collective decay rates, we take the Markovian limit such that $\alpha_k(t-L|j-k|) \simeq \alpha_k(t)$ since $L\sim O(\Omega^{-1})$ and $t\sim O(J_0^{-1})$. Consequently, (\ref{eq:liao}) can be written as a matrix equation
\begin{equation}
    \dot x(t) = -J x(t),
\end{equation}
where $x(t) = [\alpha_1(t), \ldots, \alpha_N(t)]$ includes the qubit excitation coefficients and $(J)_{jk}=J_0 e^{i\theta |j-k|}$ is the collective coupling matrix. The solution to this equation is trivial (as long as $J$ is non-singular, which it is for $\theta \neq n\pi$; for $\theta=n\pi$, the existence of BIC changes the time evolution as  discussed in Section \ref{sec:timecoldec}) and can be given as
\begin{equation}
    x(t) = \sum_{l=1}^N \beta_l e^{-0.5\Gamma_l t} \xi_l,
\end{equation}
where $\xi_l$ and $\Gamma_l/2$ are eigenvectors and eigenvalues of the matrix $J$, and $\beta_l$ are some complex constants that can be determined by the initial conditions \footnote{Here, we assumed, for simplicity, that $J$ has a non-degenerate spectrum, which does not affect any of the arguments that follows.}. This expression ties back to (\ref{eq:arbitrary2}), where $\Gamma_l$ are indeed the collective decay rates.

This derivation reveals an important property of collective decay rates: they have one-to-one correspondence with states ($\ket{\xi_l}$) in the qubit subspace. It is important to note that these states are not guaranteed to be orthogonal, although they are distinct and span the qubit subspace. While \cite{liao2015single} discusses the eigenvalues and eigenvectors of the coupling matrix and associates, in passing, eigenvalues with collective decay rates when considering a pulse scattering problem, here we proved this relationship, which is only accurate for the Markovian limit. 

The eigenstates of the matrix are the basis states of the $N$ distinct decay modes. Consequently, each collective decay rate corresponds to a physical decay mode. This leads to the following phase space picture: the initial coherent excitation of the qubits can be written in terms of a linear combination of decay mode basis states. Consequently, a decay mode can only be accessed if the overlap is nonzero\footnote{It is important to clarify that the eigenstates are not necessarily orthogonal, so the overlap is not taken as orthogonal projections, but according to the angles between eigenstates. This leads to the fact that a system prepared in a certain eigenmode has non-zero probability to be observed at another mode. Nonetheless, this is not an interference effect of different modes, since any observed quantity decays only with the decay rate of the initially prepared mode.}. Hence, there is always a specific coherent excitation (i.e. one that overlaps perfectly with the corresponding eigenvector) of qubits that can excite a single decay mode only. We discuss this property in the next section in the context of  pulse shaping by engineering  collective decay rates. Moreover, this property also explains how decay rates can signal the existence of BIC and why the dimensionality ($N-1$) of BIC  is linked to the number ($N-1$) of zero collective decay rates. Since for $\theta=n\pi$, the $J$ matrix has $N-1$ zero eigenvalues, the corresponding subspace has dimensionality $N-1$ and can be constructed with orthogonal basis states. We also infer that subradiant states become BIC in a continuous manner as $\theta$ approaches $n\pi$.

On another note, combining the one-to-one correspondence of decay modes and $\ket{\xi_l}$ with the symmetric and anti-symmetric collective decay rates conjecture, we realize that the states $\ket{\xi_l}$ have either even or odd parity such that $P \ket{\xi_l} = \pm \ket{\xi_l}$. The even (odd) parity states correspond to symmetric (anti-symmetric) states. The implication is easy to prove via proof by contradiction. Assume $\ket{\xi_1}$ has both symmetric and anti-symmetric parts. Then, it can be decomposed into both parts and hence can excite certain symmetric and anti-symmetric modes (that have non-zero overlap with the symmetric/anti-symmetric part of $\ket{\xi_1}$), which contradicts  the conjecture. The fact that the decay mode states, $\ket{\xi_l}$, are either symmetric or anti-symmetric, and not a mixture of two, is intriguing. For now, we do not have a conclusive proof for this, although we believe that the highly special shape of the $J$ matrix might be the first step towards understanding this phenomenon. 

We emphasize that finding the collective decay rates via this method is inefficient, since it requires diagonalization of a $N\times N$ matrix. Using the transfer matrix method is efficient, since it eliminates the internal degrees of freedom and deals with only $2\times 2$ matrices, as  shown in Appendix \ref{sec:appendixgeneral}. For example, using the transfer matrix method, the collective decay rates for $N=30$  can be found almost instantly, whereas it is nearly impossible to diagonalize the coupling matrix. This phenomenon illustrates clearly why real-space approach outperforms the existing methods \cite{liao2015single} by a large margin. We also emphasize that this approach works only in the Markovian limit. To find non-Markovian collective decay rates, one needs to use the machinery of the real-space approach.

\section{Pulse scattering examples for the Markovian regime}  \label{sec:appendixpulsemarkovian}

Here we consider two more  distinct pulse shapes: 1) Decaying Exponential, i.e. the shape of a photon emitted from a two-level quantum emitter \cite{grynberg2010introduction} and 2) Rising Exponential, which is known to give maximum excitation for a single qubit \cite{wang2011efficient} (see Appendix \ref{sec:appendixa}).
\subsubsection{Decaying exponential} 
A decaying exponential pulse corresponds to a regular state, therefore we can obtain analytical expressions for each atomic excitation probability, as well as the transmitted and reflected pulse shapes. For a (resonant) decaying exponential incident from the left, $f(x)$ in (\ref{eq:pulsein}) takes the form
\begin{equation} \label{eq:decayingpulse}
    f(x) = \sqrt{2 \xi} \ee^{\xi x} \Theta(-x).
\end{equation}
Then, the state of the pulse at time $t$ is
\begin{equation}
    \begin{split}
      &\ket{S(t)}= \int_{-\infty}^\infty  \frac{\diff \Delta_k}{2\pi}\frac{\sqrt{2 \xi} e^{-i \Delta_k (t-x_0)}}{\xi - i \Delta_k} \ket{E_k} =\sum_{p'} \underset{\Delta_k =p'}{\text{Res}} \left[ \frac{\sqrt{2 \xi} e^{-i \Delta_k (t-x_0)}}{\xi - i \Delta_k} \ket{E_k} \right],
    \end{split}
\end{equation}
where $p'$ includes, in addition to three poles of the system, an additional pole ($-i\xi$) introduced by the decaying exponential. It is important to note that this expression has no UHP poles, hence the interaction between the photon and the system occurs only for $t\leq x_0$, which is a consequence that photons travel with a speed $v_g=1$. Consequently, the causality principle manifests itself in this scattering problem by the absence of poles in the upper-half plane. The derivation for the single photon case (\cite{wang2011efficient}) can be found in Appendix \ref{sec:appendixa}. The emitted photon probability density $\mathcal{P}(x,t)$ can also be found analytically, but we don't show it here as it doesn't offer any more insight into the formalism.

\begin{figure}
    \centering
    \includegraphics[width=\textwidth]{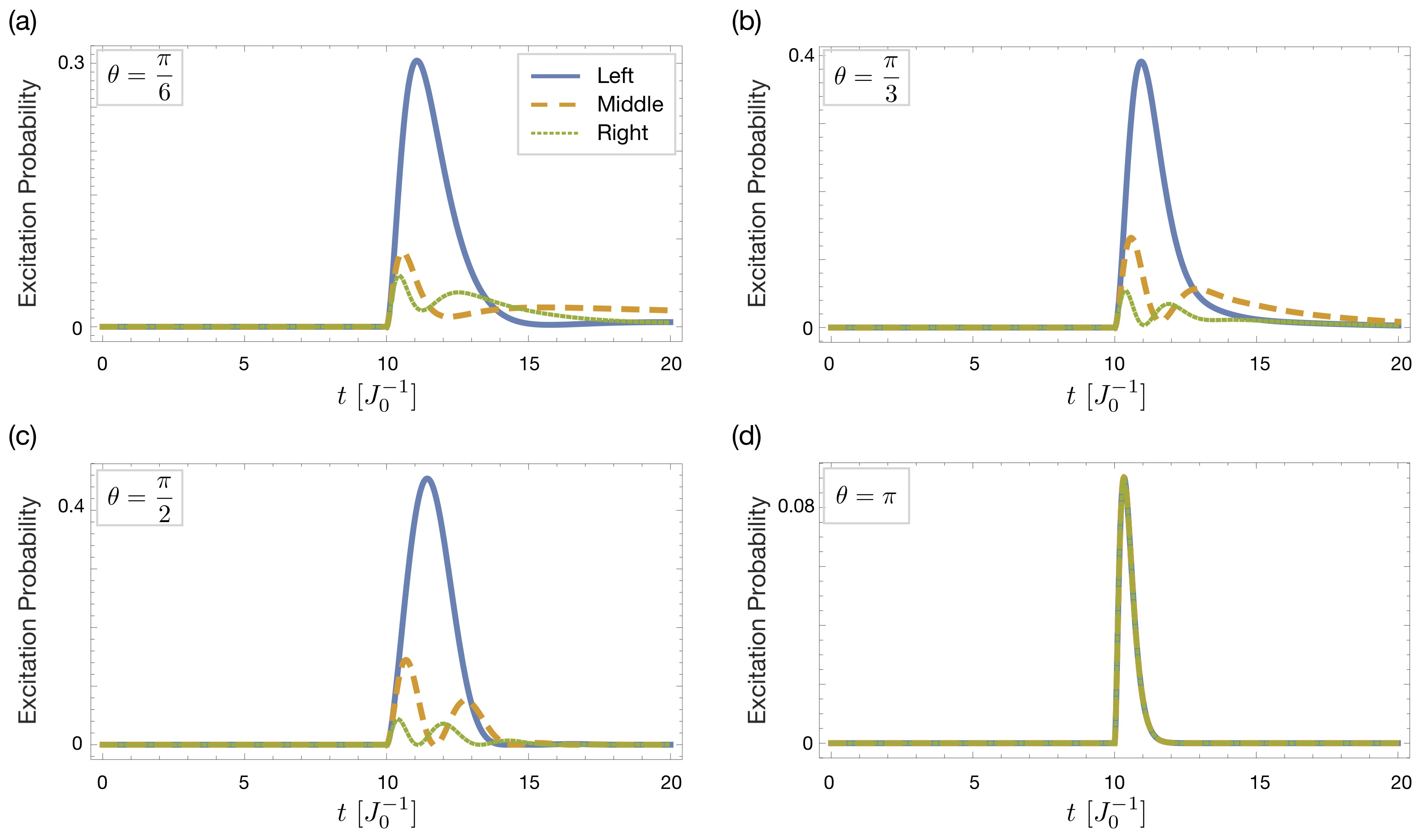}
    \caption{Single-photon scattering in a system of three qubits in the Markovian limit. The excitation probability of each atom  during interaction with a single photon pulse shaped as a decaying exponential centred at $x_0=10/J_0$. For each value of $\theta$, the parameter $\xi$ was optimized to maximise the excitation probability for the left atom; the optimal values are a) $\xi =0.94J_0$, b) $\xi =1.12J_0$, c) $\xi =0.73J_0$ and d) $\xi =3J_0$. In  d), the linearization assumption causes all three excitation probabilities to be the same, which is accurate as long as $L\sim O(\Omega^{-1})$ and $J_0/\Omega \ll 1$.}
    \label{fig:decayingexponential}
\end{figure}

We do, however, study the atom excitation probabilities. In particular, we optimize the decaying exponential and $\theta$ to maximize or minimize the (maximum) excitation probabilities. 

The maximum excitation probability for the left atom is $P^{\mathrm{max}}_{-1}(t)\simeq 0.454$, and occurs when the system is parameterized by $\theta=\pi/2$, and the pulse is parameterized by $\xi=0.73J_0$ and $t=x_0+1.43/J_0$. 

The fact that the maximum excitation is obtained for $\theta=\pi/2$ can be explained as follows. The imaginary parts of the poles correspond to decaying exponential. If the decay is large, then the coupling between the light and the atom is strong and the atom can be excited by the incoming pulse more easily. Nonetheless, an excited atom can decay faster through high decay rate modes, leading to a trade-off. Therefore, from an heuristic point of view, it is expected that the highest probability of excitation is achieved for the case where the relevant decay modes have similar decay rates, which is when $\theta=\pi/2$. 

In the other extreme, the minimum (maximum) excitation probability  is $P^{\mathrm{min}}_{-1}(t)=\frac{2}{3e^2} \simeq 0.09$, and occurs when the system is parameterized by $\theta=\pi$, and the pulse is parameterized by $\xi=3J_0$ and  $t=x_0+(3J_0)^{-1}$. This can be understood as follows. When  $\theta=\pi$, out of three decay rates, two of them become zero, leading to dark states that do not couple to the incoming pulse. Therefore, the collective system behaves as a two-level system between the ground state and the bright state $\ket{B}$. This explains the excitation probability $\frac{2}{3e^2}$ of individual atoms, since the overall qubit excitation of $\frac{2}{e^2}$ (See Appendix \ref{sec:appendixa}) is divided equally between them. 

We can also optimize the system when a subset of the parameters are fixed. Fig. \ref{fig:decayingexponential}, show excitation probabilities for which $P_{-1}(t)$ is maximized for different $\theta$. 

\subsubsection{Rising exponential} A rising exponential photon corresponds to a regular state. Therefore, the discussion on rising exponential pulse is identical to the decaying exponential. For brevity, we report directly the result. The rising exponential pulse results in a maximum excitation of $\simeq 0.6808$ for the first atom for $\xi=0.97J_0$, $\tau=0$ (with $x_0 \to \infty$) and $\theta=\pi/2$. This shows that the rising exponential is a more suitable one-sided pulse to achieve a maximum excitation probability in comparison to decaying exponential and Gaussian pulses. 

\section{Derivation of scattering pulse shape} \label{sec:appendixb}
In this appendix, we carry out the derivations of $S(x<0,2x_0)$ and $S(x>0,2x_0)$ from the general formula of $S(x,t)$ given in (\ref{eq:Sxt}). Here, the time is chosen as $t=2x_0$ so that the qubits are practically in the ground state. We shall start by dividing the integral into two parts
\begin{align}
S(x,t)= \frac{1}{2\pi} \left( \int_{0}^\infty \diff k \braket{x}{E_k} \braket{E_k}{S(0)} e^{-iE_k t} +  \int_{0}^\infty \diff k \braket{x}{E_{-k}} \braket{E_{-k}}{S(0)} e^{-iE_k t}\right),
\end{align}
where by $E_{-k}$ we mean the scattering energy-eigenstate for a photon far from right. For the next step, we find
\begin{subequations}
\begin{align} 
\braket{E_k}{S(0)}&\simeq \int_{-\infty}^\infty f(x+x_0)e^{-i(k-k_0)x} \diff x =\sqrt{2\pi} \tilde f(k-k_0) e^{i(k-k_0)x_0}, \\
\braket{E_{-k}}{S(0)}&\simeq 0, 
\end{align}
\end{subequations}
where we make use of the fact that $f(x+x_0)\simeq 0$ for $x>0$ and $[C_R^\dag(x),C_L(x)]=0$. 

Now, we set $t=2x_0$ and assume $x>0$ to find the transmitted signal, then
\begin{subequations}
\begin{align}
S(x,2x_0)&= \frac{1}{\sqrt{2\pi}} \int_0^\infty \diff k \left( t_3 e^{ikx} \right) \tilde f(k-k_0) e^{i(k-k_0)x_0} e^{-2iE_kx_0}, \\
&=\frac{e^{-ik_0x_0}}{\sqrt{2\pi}} \int_0^\infty \diff k \left( t_3 e^{ikx} \right)  \tilde f(k-k_0) e^{-ikx_0}.
\end{align}
\end{subequations}
Here, we recall that $E_k=k$ for $k>0$  and $\tilde f(k-k_0)\simeq 0$ for $k<0$. Then, we have
\begin{align}
S(x,2x_0)&=\frac{e^{-ik_0x_0}}{\sqrt{2\pi}} \int_{-\infty}^\infty \diff k  \,  t_3 \tilde  f(k-k_0) e^{i k(x-x_0)},
\end{align}
where we realize that $\tilde{S}_+(k,2x_0)= e^{-i(k_0+k)x_0} \, t_3 \tilde f(k-k_0) $. Now, as $|\tilde{S}(k,0)|^2 = |\tilde f(k-k_0)|^2$, we have that $|\tilde{S}_+(k,2x_0)|^2 = |t_3\tilde f(k-k_0)|^2 =|t_3\tilde{S}(k,0) |^2$.

For the reflected signal, we assume $x<0$. Then, we find
\begin{subequations}
\begin{align}
S(x,2x_0)&= \frac{1}{\sqrt{2\pi}} \int_0^\infty \diff k \left( e^{ikx} + r_1 e^{-ikx} \right) \tilde f(k-k_0) e^{i(k-k_0)x_0} e^{-2iE_kx_0}, \\
&=\frac{e^{-ik_0x_0}}{\sqrt{2\pi}} \int_0^\infty \diff k \left( r_1 e^{-ikx} \right)  \tilde f(k-k_0) e^{-ikx_0},
\end{align}
\end{subequations}
as the first term of the integral (which includes $e^{ik(x-x_0)}$) corresponds to a function localized at $x=x_0$ and is therefore $\simeq 0$ for $x<0$ and $E_k=k$ for $k>0$. We first realize that $\tilde f(k-k_0)\simeq 0$ for $k<0$ and expand the order of integration to $-\infty$. Then, we change $k\to -k$
\begin{align}
S(x,2x_0)&=\frac{e^{-ik_0x_0}}{\sqrt{2\pi}} \int_{-\infty}^\infty \diff k  \,  r_1 \tilde  f(-k-k_0) e^{i k(x+x_0)},
\end{align}
where we realize that $\tilde{S}_- (k,2x_0)= e^{i(k-k_0)x_0} r_1 \tilde f (-k-k_0) $. Then, $|\tilde{S}_-(-k,2x_0)|^2= |r_1 \tilde f(k-k_0)|^2 = |r_1 \tilde S(k,0)|^2$. The narrow band pulses scatter indeed corresponding to the stationary eigenstate scattering coefficients $t_3$ and $r_1$.

\section{The time-evolved state $\ket{\psi(t)}$ satisfies the Schr\"odinger equation ($N=2$ Qubits)} \label{sec:appendixd}
In this appendix, we show that the time evolved state given in (\ref{eq:forappendixd}) satisfies the Schr\"odinger equation 
\begin{equation}
    i \frac{\partial}{\partial t} \ket{\psi(t)}= (H_0 + H_I) \ket{\psi(t)},
\end{equation}
where the Hamiltonian for the $2$ qubit system is
\begin{subequations}
\begin{align}
    H_0 &= \Omega \sum_{j=\{-1,1\}} \ket{e_i} \bra{e_i} + i \int_{-\infty}^\infty \diff x \left( C_L^\dag(x) \frac{\partial}{\partial x} C_L(x) -C_R^\dag(x) \frac{\partial}{\partial x} C_R(x)  \right), \\
    H_I &= \sqrt{J_0}\sum_{j=\{-1,1\}} \left( \sigma_j^\dag [C_L(jL/2) + C_R(jL/2)] + \sigma_j [C_L^\dag(jL/2) + C_R^\dag(jL/2)] \right),
\end{align}
\end{subequations}
where we pick $x=0$ as the center of the $2$ qubit system.

We re-write (\ref{eq:forappendixd}) once more while scaling $\Gamma_1 \to 2 \Gamma_1$ for convenience:
\begin{equation} \label{eq:forappendixdind}
\begin{split}
        \ket{\psi(t)}= \frac{1}{\sqrt{2}} \Bigg( e^{-\Gamma_1 t -i\Omega t} (\ket{e_1} + \ket{e_2}) &- i \int_{0}^t \diff x \frac{\Gamma_1}{\sqrt{J_0}} e^{-(\Gamma_1+i\Omega)(t-x+L/2)}C_R^\dag(x) \ket{0}\\
        &-i \int_{-t}^{0} \diff x \frac{\Gamma_1}{\sqrt{J_0}} e^{-(\Gamma_1 + i \Omega) (t+x+L/2)} C_L^\dag(x) \ket 0 \Bigg),
\end{split}
\end{equation}
Our calculations are performed in the Markovian limit, where $\Omega L = \theta$, $J_0 L \simeq 0$ and $C_{R/L}^\dag(\pm L/2) \simeq e^{\mp i\theta/2} C_{R/L}^\dag(0)$. The latter can be proven using the definition of $C_{R/L}^\dag(x)$ as a Fourier Transform of momentum mode creation operator $a_R^\dag(k)$. This is consistent with the time evolution in (\ref{eq:arbitrary}), where the Markovian limit is assumed implicitly to obtain a polynomial characteristic equation for the collective decay rates.

First, we divide $\ket{\psi(t)}=\ket{\psi_q(t)}+\ket{\psi_R(t)}+\ket{\psi_L(t)}$ into three parts as
\begin{subequations}
\begin{align}
    \ket{\psi_q(t)}&=\frac{1}{\sqrt{2}} e^{-\Gamma_1 t -i\Omega t} (\ket{e_1} + \ket{e_2}), \\
    \ket{\psi_R(t)}&=\frac{-i}{\sqrt{2}} \int_{-\infty}^\infty \diff x \frac{\Gamma_1}{\sqrt{J_0}} e^{-(\Gamma_1+i\Omega)(t-x+L/2)} [\Theta(x)-\Theta(x-t)]C_R^\dag(x) \ket{0}, \\
    \ket{\psi_L(t)}&=\frac{-i}{\sqrt{2}} \int_{-\infty}^\infty \diff x \frac{\Gamma_1}{\sqrt{J_0}} e^{-(\Gamma_1+i\Omega)(t+x+L/2)} [\Theta(t+x) - \Theta(x)]C_L^\dag(x) \ket{0}.
\end{align}
\end{subequations}
Let us start with the time derivatives
\begin{subequations}
\begin{align}
    i\partial_t \ket{\psi_q(t)}&=- i(\Gamma_1 + i\Omega) \ket{\psi_q(t)}, \\
    i\partial_t \ket{\psi_R(t)}&=- i(\Gamma_1 + i \Omega) \ket{\psi_R(t)}+ \frac{\Gamma_1}{\sqrt{2J_0}} e^{-i(\Gamma_1+i\Omega) L/2} C_R^\dag(t) \ket 0, \\
   i \partial_t \ket{\psi_L(t)}&=-i (\Gamma_1 + i \Omega) \ket{\psi_R(t)}+\frac{\Gamma_1}{\sqrt{2J_0}} e^{-i(\Gamma_1+i\Omega) L/2} C_L^\dag(-t) \ket 0.
\end{align}
\end{subequations}
Now, let us find the action of the free Hamiltonian ($H_0$) on each part of $\ket{\psi(t)}$
\begin{subequations}
\begin{align}
    H_0 \ket{\psi_q(t)}={}&\Omega \ket{\psi_q(t)}, \\
    \begin{split}
    H_0 \ket{\psi_R(t)}={}& -i(\Gamma_1 + i\Omega) \ket{\psi_R(t)} - \frac{\Gamma_1}{\sqrt{2J_0}} e^{-(\Gamma_1+i\Omega)t}e^{-i\theta/2} C_R^\dag(0) \ket 0 \\
    &+\frac{\Gamma_1}{\sqrt{2J_0}} e^{-(\Gamma_1+i\Omega)L/2} C_R^\dag(t) \ket{0}, 
    \end{split}\\
    \begin{split}
    H_0 \ket{\psi_L(t)}={}&-i(\Gamma_1 + i\Omega) \ket{\psi_L(t)} - \frac{\Gamma_1}{\sqrt{2J_0}} e^{-(\Gamma_1+i\Omega)t}e^{-i\theta/2} C_L^\dag(0) \ket 0 \\
    &+\frac{\Gamma_1}{\sqrt{2J_0}} e^{-(\Gamma_1+i\Omega)L/2} C_L^\dag(-t) \ket{0},
    \end{split}
\end{align}
\end{subequations}
where we define $e^{- (\Gamma_1+i\Omega)L/2}\simeq e^{-i\Omega L/2}=e^{-i\theta/2}$.

Finally, we consider the action of the interaction Hamiltonian ($H_I$) on each part of $\ket{\psi(t)}$
\begin{subequations}
\begin{align}
    H_I \ket{\psi_q(t)}&=\frac{\sqrt{J_0}}{\sqrt{2}} e^{-\Gamma_1 t -i\Omega t} [C_R^\dag(L/2) + C_L^\dag(L/2)+ C_R^\dag(-L/2) + C_L^\dag(-L/2))] \ket 0 , \\
    H_I \ket{\psi_R(t)}&= \frac{-i}{\sqrt{2}} \Gamma_1 e^{-(\Gamma_1+i\Omega)t}  \ket{e_2}, \\
    H_I \ket{\psi_L(t)}&= \frac{-i}{\sqrt{2}} \Gamma_1 e^{-(\Gamma_1+i\Omega)t}  \ket{e_1}.
\end{align}
\end{subequations}
From our calculations, we see that $i\partial_t \ket{\psi_q(t)}=H_0 \ket{\psi_q(t)}+ H_I \ket{\phi_R(t)}+H_I \ket{\phi_L(t)}$. Eliminating other components in a similar fashion, we obtain
\begin{subequations}
\begin{align}
   & \frac{\Gamma_1}{\sqrt{2J_0}} e^{-(\Gamma_1+i\Omega)t}e^{-i\theta/2} \left[ C_R^\dag(0) + C_L^\dag(0) \right]\ket{0} \\
    &= \frac{\sqrt{J_0}}{\sqrt{2}} e^{-\Gamma_1t-i\Omega t} [C_R^\dag(L/2) + C_L^\dag(L/2)+ C_R^\dag(-L/2) + C_L^\dag(-L/2))] \ket 0.
\end{align}
\end{subequations}
Now, we use the condition $C_{R}^\dag(\pm L/2) \simeq e^{\mp i\theta/2} C_{R}^\dag(0)$, $C_{L}^\dag(\pm L/2) \simeq e^{\pm i\theta/2} C_{L}^\dag(0)$ (which are consequences of the Markovian approximation) and $\Gamma_1=J_0 (1+e^{i\theta})$ {(with the re-scaling $\Gamma_1\to 2\Gamma_1$ at the beginning taken into account)} to obtain
\begin{align}
    2J_0\cos(\theta) \left[ C_R^\dag(0) + C_L^\dag(0) \right] \ket{0} = 2 J_0 \cos(\theta) [ C_R^\dag(0) + C_L^\dag(0))] \ket 0,
\end{align}
which shows that (\ref{eq:forappendixd}) is a solution to the Schr\"odinger equation in the Markovian limit.

\section{New insights gained}\label{sec:int}

In this Appendix, we discuss various areas where we developed new insights as a result of applying the real-space formalism to the problems considered here.

\subsection{Collective decay rates}
The most important intuition that we gained  is the idea of collective decay rates and how they dictate the collective behavior. When there is an effective coupling between two or more isolated systems, the coupling shifts the energy levels as well as the individual decay rates. In waveguide QED, this effective coupling is mediated by bosonic fields (photons).  The real part of the collective decay rate corresponds to the coupling energy of the corresponding interaction basis state, whereas the imaginary part {(divided by 2)} corresponds to the shift in energies. Here, the interaction basis states, let's call them $\ket{S_i}$, are coherent excitations of single qubit excitation states such that $\ket{S_i}=\sum_j \alpha_j^{(i)} \ket{e_j}$, where $\alpha_j^{(i)}$ can be found by diagonalizing the coupling matrix as described in Appendix \ref{sec:appendixalternativemethod}. Consequently, by finding poles of the scattering parameters (i.e. highly mathematical objects), we gain information about the shifts in energy levels and individual decay rates. 

\subsection{Causality}
Another important aspect we saw is how the causality principle is related to the position of scattering parameter poles (and correspondingly the collective decay rates) in the complex plane. As we saw in the example of the decaying exponential, the causality principle dictates that the poles should be in the LHP for the linear chain of $N=3$ qubits. In fact, this is a more general phenomenon and the scattering parameter poles should be in the LHP for \emph{any} waveguide QED system. We plan to prove this  in future work. Here, we emphasize another important feature of this mathematical connection. The imaginary part of the poles gives  the decay rates, and so the poles should be in the LHP for the decay rate of the multi-qubit system to be positive. If the system had a pole in the UHP, then one of the decay rates would be negative, that is the system would get excited with more than unity probability after some time. This is clearly not physical. What is striking is the fact that both this unphysical behavior and the causality principle are linked by the same mathematical identity, which can be found simply by considering \emph{the steady-state solution} without any time dynamics calculations. We have not seen this intuition discussed in the literature. 

\subsection{The Markovian limit}

In \cite{tsoi2008quantum}, the authors linearize the phase of photons propagating between two adjacent qubits as $kL \simeq \Omega L$. A similar substitution is later referred to as a Markovian approximation in \cite{zheng2013persistent}. In the Markovian limit, the inter-system propagation time of photons is neglected and only the phase acquired by the propagating photon is accounted for. The latter is trivial to see in the linearization process $e^{ikL} \simeq e^{i\Omega L}$ as the acquired phase can be approximated by the phase of the resonant photon. Neglecting the time delay portion, however, is not as straightforward to see in this linearization. As we have seen in Section \ref{sec:formalism}, the $k$- dependence of the phase $e^{ikL}$ shifts the time $t$ by $t-L$ (See, for example, Eq. (\ref{eq:arbitrary})). To see this clearly, let us calculate the excitation probability of a qubit upon pulse scattering:
\begin{equation}
    P_m(t)=\left|  \int_{-\infty}^\infty  \frac{\diff k}{2\pi}e_m S(k,0) \exp(-i\Delta_k t) \right|^2.
\end{equation}
Here, let us assume that the qubit is situated at position $x=L$ instead of $x=0$. Then, $e_m$ would gain a phase of $e^{ikL}$, which would later be picked up by $\exp(-i\Delta_k t)$ term to shift the time $t \to t-L$. This leads to a delay of $L$ in time. If, however, $e^{ikL}$ was to be linearized by replacing $k\simeq \Omega$, the time delay would not be accessible since $e^{i\Omega L}$ does not depend on $k$ and can be taken out of the integral. By linearizing $kL \simeq \Omega L$, we effectively disregard any time delay that is of order $L$. Consequently, the real-space approach shows how this linearization is equivalent to applying the Markovian approximation. Consequently, the non-Markovian behavior of multi-qubit systems is extremely easy to obtain using the real-space approach, where we simply omit the linearization.

\subsection{Black-box behaviour}

The real-space approach also explains the black box behavior we see in waveguide QED systems. When a pulse is incident on a system, the modulation of the transmitted/reflected pulse depends only on the external scattering parameters. Each momentum mode of the initial pulse gets modulated via the external parameters and the asymptotic shape of the pulse at $t\to \infty$ can be found without any time dynamics consideration. We proved this using the real-space approach in Appendix \ref{sec:appendixb} for a specific case, however the general proof is analogous.

\subsection{Non-Markovian dynamics}

Finally, the real-space approach makes it possible to calculate exact and fully analytical non-Markovian dynamics. The fact that non-Markovian behavior for a single qubit system was explored only recently \cite{fang2018non,valente2016non}, means that probing fully analytical multi-qubit non-Markovian dynamics would be a big step forward. We will discuss an approach to this problem in upcoming work. For now, we note that this process is  analogous to scattering from a finite well in introductory quantum mechanics. The elementary nature of the real-space approach guides our intuition about simple problems toward solving more complicated ones. 

\end{document}